\documentclass[a4paper,fleqn,usenatbib]{mnras}

\usepackage{newtxtext,newtxmath}

\usepackage[T1]{fontenc}
\usepackage{ae,aecompl}
\usepackage{fixltx2e}

\usepackage{graphicx}	% Including files
\usepackage{amsmath}	% Advanced maths commands
\usepackage{comment}
\usepackage{ulem}
\usepackage{subcaption}%

\usepackage{xcolor}		% To create colored notes during the process of drafting
\usepackage{mathrsfs} % for fancy cursive fonts
\usepackage{breakcites} %to ensure citations can break over multiple lines
\usepackage{caption} %caption for figures outside of a floating environment (\begin{figure}). Also, to use \ContinuedFloat for breaking figures* over multiple pages
\usepackage{multicol}
\usepackage{rotating} %to rotate tables
\usepackage{threeparttable}
\usepackage{float}

\usepackage{dblfloatfix}

\usepackage{letltxmacro} % This and the next one to make a command to wrap each \citep in an \mbox to prevent the error that occurs when a link spans two pages
\usepackage{xparse}
\newcommand{\todo}{\ifmmode {\color{red}\text{\Huge{\(\bullet\)}}} \else {\color{red} \Huge$\bullet$}\fi}
\newcommand{\tido}{\ifmmode {\bullet} \else $\bullet$\fi}

\newcommand{\E        }[1]{\ifmmode 10^{#1} \else $10^{#1}$\fi}
\newcommand{\tE        }[1]{\ifmmode \times10^{#1} \else $\times10^{#1}$\fi}
\newcommand{\til}{\ifmmode \sim \else $\sim$\fi}
\renewcommand{\~} {\ifmmode \sim \else $\sim$\fi}

\newcommand{\pc}	{\ifmmode {\rm pc} \else pc\fi}
\newcommand{\ld}	{\ifmmode {\rm l.d.} \else l.d.\fi}
\newcommand{\kms}	{\ifmmode {\rm km\,s}^{-1} \else km\,s$^{-1}$\fi}
\newcommand{\Jykms}	{\ifmmode {\rm Jy\,km\,s}^{-1} \else Jy\,km\,s$^{-1}$\fi}
\newcommand{\cc}	{\ifmmode {\rm cm}^{-3}    \else cm$^{-3}$\fi}
\newcommand{\cmii}	{\ifmmode {\rm cm}^{-2}    \else cm$^{-2}$\fi}
\newcommand{\ergs}	{\ifmmode {\rm erg\,s}^{-1} \else erg s$^{-1}$\fi}
\newcommand{\ergcms}	{\ifmmode {\rm erg\,cm}^{-2}\,{\rm s}^{-1} \else erg\,cm$^{-2}$\,s$^{-1}$\fi}
\newcommand{\ergcmsA}	{\ifmmode {\rm erg\,cm}^{-2}\,{\rm s}^{-1}\,{\rm\AA}^{-1}
\else erg\,cm$^{-2}$\,s$^{-1}$\,\AA$^{-1}$\fi}
\newcommand{  \ergcmsHz  }{\ifmmode{\rm erg\,cm}^{-2}\,{\rm s}^{-1}\,{\rm Hz}^{-1}
                       \else ergs\,cm$^{-2}$\,s$^{-1}$\,Hz$^{-1}$\fi}
\newcommand{\kev}	{\ifmmode {\rm keV} \else keV\fi}

\newcommand{\mic}	{\ifmmode {\rm \mu m} \else $\mu$m\fi}
\newcommand{\vFWHM}	{\ifmmode v_{\mbox{\tiny FWHM}} \else $v_{\mbox{\tiny FWHM}}$\fi}
\newcommand{\vBLR}	{\ifmmode v_{\mbox{\tiny BLR}} \else $v_{\mbox{\tiny BLR}}$\fi}
\newcommand{\sigBLR}	{\ifmmode \sigma_{\mbox{\tiny BLR}} \else $\sigma_{\mbox{\tiny BLR}}$\fi}
\newcommand{\vNLR}	{\ifmmode v_{\mbox{\tiny NLR}} \else $v_{\mbox{\tiny NLR}}$\fi}
\newcommand{\tauBLR}	{\ifmmode \tau_{\mbox{\tiny BLR}} \else $\tau_{\mbox{\tiny BLR}}$\fi}

\newcommand{\Hubble}	{\ifmmode {\rm km\,s}^{-1}\,{\rm Mpc}^{-1} \else km\,s$^{-1}$\,Mpc$^{-1}$\fi}
\newcommand{\NDunit}	{\ifmmode {\rm Mpc}^{-3} \else Mpc$^{-3}$\fi}
\newcommand{\LFunit}	{\ifmmode {\rm Mpc}^{-3}\,{\rm mag}^{-1} \else Mpc$^{-3}$\,mag$^{-1}$\fi}
\newcommand{\MFunit}	{\ifmmode {\rm Mpc}^{-3}\,{\rm dex}^{-1} \else Mpc$^{-3}$\,dex$^{-1}$\fi}

\newcommand{\Msun}{\ifmmode M_{\odot} \else $M_{\odot}$\fi}
\newcommand{\Lsun}{\ifmmode L_{\odot} \else $L_{\odot}$\fi}
\newcommand{\Zsun}{\ifmmode Z_{\odot} \else $Z_{\odot}$\fi}
\newcommand{\mpyr}{\ifmmode \Msun\,{\rm yr}^{-1} \else $\Msun\,{\rm yr}^{-1}$\fi}

\newcommand{\qnote}{\ifmmode q_{0} \else $q_{0}$\fi}
\newcommand{\Hnote}{\ifmmode H_{0} \else $H_{0}$\fi}
\newcommand{\hnote}{\ifmmode h_{0} \else $h_{0}$\fi}
\newcommand{\anote}{\ifmmode a_{0} \else $a_{0}$\fi}
\def\gsim{\;\rlap{\lower 2.5pt \hbox{$\sim$}}\raise 1.5pt\hbox{$>$}\;}
\def\lsim{\;\rlap{\lower 2.5pt \hbox{$\sim$}}\raise 1.5pt\hbox{$<$}\;}

\newcommand{  \Halpha   }{\ifmmode {\rm H}\alpha \else H$\alpha$\fi}

\newcommand{  \ha       }{\Halpha}
\newcommand{  \Hbeta    }{\ifmmode {\rm H}\beta \else H$\beta$\fi}

\newcommand{  \hb       }{\Hbeta}
\newcommand{  \Hgamma   }{\ifmmode {\rm H}\gamma \else H$\gamma$\fi}
\newcommand{  \Hdelta   }{\ifmmode {\rm H}\delta \else H$\delta$\fi}
\newcommand{  \Lya      }{\ifmmode {\rm Ly}\alpha \else Ly$\alpha$\fi}
\newcommand{  \Lyb      }{\ifmmode {\rm Ly}\beta \else Ly$\beta$\fi}
\newcommand{  \Pa       }{\ifmmode {\rm P}\alpha \else P$\alpha$\fi}
\newcommand{  \Pb       }{\ifmmode {\rm P}\beta \else P$\beta$\fi}
\newcommand{  \Bra      }{\ifmmode {\rm Br}\alpha \else Br$\alpha$\fi}
\newcommand{  \Brg      }{\ifmmode {\rm Br}\gamma \else Br$\gamma$\fi}
\newcommand{  \hi       }{\ifmmode {\rm H}\,\textsc{i} \else H\,\textsc{i}\fi}
\newcommand{  \HI       }{\ifmmode {\rm H}\,\textsc{i} \else H\,\textsc{i}\fi}
\newcommand{  \hii      }{\ifmmode {\rm H}\,\textsc{ii} \else H\,\textsc{ii}\fi}
\newcommand{  \hei      }{\ifmmode {\rm He}\,\textsc{i} \else He\,\textsc{i}\fi}
\newcommand{  \heii     }{\ifmmode {\rm He}\,\textsc{ii} \else He\,\textsc{ii}\fi}
\newcommand{  \HeIIuv   }{\ifmmode {\rm He}\,\textsc{ii}\,\lambda1640 \else He\,\textsc{ii}\,$\lambda1640$\fi}
\newcommand{  \HeIIop   }{\ifmmode {\rm He}\,\textsc{ii}\,\lambda4686 \else He\,\textsc{ii}\,$\lambda4686$\fi}
\newcommand{  \CII	}{\ifmmode \left[{\rm C}\,\textsc{ii}\right]\,\lambda157.74\,\mu{\rm m} \else [C\,{\sc ii}]\ $\lambda157.74\,\mu{\rm m}$\fi}
\newcommand{  \cii	}{\ifmmode \left[{\rm C}\,\textsc{ii}\right] \else [C\,{\sc ii}]\fi}

\newcommand{  \ciii     }{\ifmmode {\rm C}\,\textsc{iii}\right] \else C\,\textsc{iii}]\fi}
\newcommand{  \CIII     }{\ifmmode {\rm C}\,\textsc{iii}\right]\,\lambda1909 \else C\,\textsc{iii}]\,$\lambda1909$\fi}
\newcommand{  \civ      }{\ifmmode {\rm C}\,\textsc{iv}  \else C\,\textsc{iv}\fi}
\newcommand{  \CIV      }{\ifmmode {\rm C}\,\textsc{iv}\,\lambda1549 \else C\,\textsc{iv}\,$\lambda1549$\fi}
\newcommand{  \nii      }{\ifmmode {\rm N}\,\textsc{ii}  \else N\,\textsc{ii}\fi}
\newcommand{  \niii     }{\ifmmode {\rm N}\,\textsc{iii} \else N\,\textsc{iii}\fi}
\newcommand{  \niv      }{\ifmmode {\rm N}\,\textsc{iv}  \else N\,\textsc{iv}\fi}
\newcommand{  \NIVuv    }{\ifmmode {\rm N}\,\textsc{iv}\,\lambda1486 \else N\,\textsc{iv}\,$\lambda1486$\fi}
\newcommand{  \nv       }{\ifmmode {\rm N}\,\textsc{v}   \else N\,\textsc{v}\fi}
\newcommand{\oi}{\ifmmode \left[{\rm O}\,\textsc{i}\right] \else [O\,{\sc i}]\fi}
\newcommand{\OI}{\ifmmode \left[{\rm O}\,\textsc{i}\right]\,\lambda6300 \else [O\,{\sc i}]$\,\lambda6300$\fi}
\newcommand{\oii}{\ifmmode \left[{\rm O}\,\textsc{ii}\right] \else [O\,{\sc ii}]\fi}
\newcommand{\OII}{\ifmmode \left[{\rm O}\,\textsc{ii}\right]\,\lambda3727 \else [O\,{\sc ii}]\,$\lambda3727$\fi}
\newcommand{\oiii}{\ifmmode \left[{\rm O}\,\textsc{iii}\right] \else [O\,{\sc iii}]\fi}
\newcommand{\OIII}{\ifmmode \left[{\rm O}\,\textsc{iii}\right]\,\lambda5007 \else [O\,{\sc iii}]\,$\lambda5007$\fi}
\newcommand{  \OIIIuv   }{\ifmmode {\rm O}\,\textsc{iii}\,\lambda1663 \else O\,\textsc{iii}\,$\lambda1663$\fi}
\newcommand{  \oiv      }{\ifmmode {\rm O}\,\textsc{iv}  \else O\,\textsc{iv}\fi}
\newcommand{  \OIVuv    }{\ifmmode {\rm O}\,\textsc{iv}\,\lambda1402  \else O\,\textsc{iv}\,$\lambda1402$\fi}
\newcommand{  \OIVIR    }{\ifmmode {\rm O}\,\textsc{iv}\,25.9\,\mu {\rm m} \else O\,\textsc{iv}\,$25.9\,\mu$m\fi}
\newcommand{  \ovi      }{\ifmmode {\rm O}\,\textsc{vi}   \else O\,\textsc{vi}\fi}
\newcommand{  \Ovi      }{\ifmmode {\rm O}\,\textsc{vi}\,\lambda1035 \else O\,\textsc{vi}\,$\lambda1035$\fi}
\newcommand{  \nei      }{\ifmmode {\rm Ne}\,\textsc{i}   \else Ne\,\textsc{i}\fi}
\newcommand{  \neii     }{\ifmmode {\rm Ne}\,\textsc{ii}  \else Ne\,\textsc{ii}\fi}
\newcommand{  \NeiiIR   }{\ifmmode {\rm Ne}\,\textsc{ii}\,12.8\,\mu {\rm m} \else Ne\,\textsc{ii}\,$12.8\,\mu$m\fi}
\newcommand{  \neiii    }{\ifmmode {\rm Ne}\,\textsc{iii} \else Ne\,\textsc{iii}\fi}
\newcommand{  \neiv     }{\ifmmode {\rm Ne}\,\textsc{iv}  \else Ne\,\textsc{iv}\fi}
\newcommand{  \nev      }{\ifmmode {\rm Ne}\,\textsc{v}   \else Ne\,\textsc{v}\fi}
\newcommand{  \NevIR    }{\ifmmode {\rm Ne}\,\textsc{v}\,24.3\,\mu {\rm m} \else Ne\,\textsc{v}\,$24.3\,\mu$m\fi}
\newcommand{  \nevi     }{\ifmmode {\rm Ne}\,\textsc{vi}  \else Ne\,\textsc{vi}\fi}
\newcommand{  \mgi      }{\ifmmode {\rm Mg}\,\textsc{i} \else Mg\,\textsc{i}\fi}
\newcommand{  \mgii     }{\ifmmode {\rm Mg}\,\textsc{ii} \else Mg\,\textsc{ii}\fi}
\newcommand{  \MgII     }{\ifmmode {\rm Mg}\,\textsc{ii}\,\lambda2798 \else Mg\,\textsc{ii}\,$\lambda2798$\fi}
\newcommand{  \sii      }{\ifmmode {\rm S}\,\textsc{ii} \else S\,\textsc{ii}\fi}
\newcommand{  \siii     }{\ifmmode {\rm S}\,\textsc{iii} \else S\,\textsc{iii}\fi}
\newcommand{  \siv      }{\ifmmode {\rm S}\,\textsc{iv} \else S\,\textsc{iv}\fi}
\newcommand{  \sili     }{\ifmmode {\rm Si}\,\textsc{i}   \else Si\,\textsc{i}\fi}
\newcommand{  \silii    }{\ifmmode {\rm Si}\,\textsc{ii}  \else Si\,\textsc{ii}\fi}
\newcommand{  \Siliv    }{\ifmmode {\rm Si}\,\textsc{iv}  \else Si\,\textsc{iv}\fi}
\newcommand{  \SilIVuv  }{\ifmmode {\rm Si}\,\textsc{iv}\,\lambda1400  \else Si\,\textsc{iv}\,$\lambda1400$\fi}
\newcommand{  \AlIII   }{\ifmmode {\rm Al}\,\textsc{iii}\,\lambda1857 \else Al\,\textsc{iii}\,$\lambda1857$\fi}
\newcommand{  \Aliii   }{\ifmmode {\rm Al}\,\textsc{iii} \else Al\,\textsc{iii}\fi}
\newcommand{  \caii     }{\ifmmode {\rm Ca}\,\textsc{ii} \else Ca\,\textsc{ii}\fi}
\newcommand{  \feii     }{\ifmmode {\rm Fe}\,\textsc{ii} \else Fe\,\textsc{ii}\fi}
\newcommand{  \feiii    }{\ifmmode {\rm Fe}\,\textsc{iii} \else Fe\,\textsc{iii}\fi}
\newcommand{  \Kalpha   }{\ifmmode {\rm K}\alpha \else K$\alpha$\fi}

\newcommand{ \Lhb   }{\ifmmode L_{\hb} \else $L_{\hb}$\fi}
\newcommand{ \Lha   }{\ifmmode L_{\ha} \else $L_{\ha}$\fi}
\newcommand{ \fwhb  }{\ifmmode {\rm FWHM}\left(\hb\right) \else FWHM(\hb)\fi}
\newcommand{\sighb  }{\ifmmode \sigma\left(\hb\right) \else $\sigma\left(\hb\right)$\fi}
\newcommand{ \ewhb  }{\ifmmode {\rm EW}\left(\hb\right) \else EW(\hb)\fi}
\newcommand{ \fwha  }{\ifmmode {\rm FWHM}\left(\ha\right) \else FWHM(\ha)\fi}
\newcommand{ \ewha  }{\ifmmode {\rm EW}\left(\ha\right) \else EW(\ha)\fi}
\newcommand{ \Lmg   }{\ifmmode L\left(\mgii\right) \else $L\left(\mgii\right)$\fi}
\newcommand{ \fwmg  }{\ifmmode {\rm FWHM}\left(\mgii\right) \else FWHM(\mgii)\fi}
\newcommand{ \Lciv  }{\ifmmode L\left(\civ\right) \else $L\left(\civ\right)$\fi}
\newcommand{ \fwciv }{\ifmmode {\rm FWHM}\left(\civ\right) \else FWHM(\civ)\fi}
\newcommand{ \fwhm  }{\ifmmode {\rm FWHM} \else FWHM\fi} 
\newcommand{ \voff  }{\ifmmode v_{\rm off} \else $v_{\rm off}$\fi} 
\newcommand{ \vmax  }{\ifmmode v_{\rm max} \else $v_{\rm max}$\fi} 

\newcommand{ \mumg  }{\ifmmode \mu\left(\mgii\right) \else $\mu\left(\mgii\right)$\fi}
\newcommand{ \fmg   }{\ifmmode f\left(\mgii\right) \else $f\left(\mgii\right)$\fi}
\newcommand{ \muciv }{\ifmmode \mu\left(\civ\right) \else $\mu\left(\civ\right)$\fi}
\newcommand{ \fciv  }{\ifmmode f\left(\civ\right) \else $f\left(\civ\right)$\fi}
\newcommand{  \auvo     }{\ifmmode \alpha_{\nu,{\rm UVO}} \else $\alpha_{\nu,{\rm UVO}}$\fi}
\newcommand{  \Ledd     }{\ifmmode L_{\rm Edd} \else $L_{\rm Edd}$\fi}
\newcommand{  \lamLlam  }{\ifmmode \lambda L_{\lambda} \else $\lambda L_{\lambda}$\fi}
\newcommand{  \lLl      }{\ifmmode \lambda L_{\lambda} \else $\lambda L_{\lambda}$\fi}
\newcommand{  \nuLnu    }{\ifmmode \nu L_{\nu} \else $\nu L_{\nu}$\fi}
\newcommand{  \nLn      }{\ifmmode \nu L_{\nu} \else $\nu L_{\nu}$\fi}
\newcommand{  \Luv      }{\ifmmode L_{1450} \else $L_{1450}$\fi}
\newcommand{  \Lop      }{\ifmmode L_{5100} \else $L_{5100}$\fi}
\newcommand{  \lLop     }{\ifmmode \log\left(\Lop/\ergs\right) \else $\log\left(\Lop/\ergs\right)$\fi}
\newcommand{  \Lthree   }{\ifmmode L_{3000} \else $L_{3000}$\fi}
\newcommand{  \lLthree  }{\ifmmode \log\left(\Lthree/\ergs\right) \else $\log\left(\Lthree/\ergs\right)$\fi}
\newcommand{  \Lsix      }{\ifmmode L_{6200} \else $L_{6200}$\fi}
\newcommand{  \lLisx     }{\ifmmode \log\left(\Lop/\ergs\right) \else $\log\left(\Lop/\ergs\right)$\fi}
\newcommand{  \Lxray    }{\ifmmode L_{\rm X} \else $L_{\rm X}$\fi}
\newcommand{  \Lhard    }{\ifmmode L_{\rm 2-10} \else $L_{\rm 2-10}$\fi}
\newcommand{  \Lsoft    }{\ifmmode L_{\rm 0.5-2} \else $L_{\rm 0.5-2}$\fi}

\newcommand{\Fthree}{\ifmmode F_{3000} \else $F_{3000}$\fi}
\newcommand{\fuv}{\ifmmode f_{\lambda}\left(1450{\rm \AA}\right) \else $f_{\lambda}\left(1450 {\rm \AA}\right)$\fi}
\newcommand{\fthree}{\ifmmode f_{\lambda}\left(3000{\rm \AA}\right) \else $f_{\lambda}\left(3000{\rm \AA}\right)$\fi}
\newcommand{\fH}{\ifmmode f_{\lambda}\left(1.65\micron\right) \else
$f_{\lambda}\left(1.65\micron\right)$\fi}

\newcommand{\fbol}{\ifmmode f_{\rm bol} \else $f_{\rm bol}$\fi}
\newcommand{\fbolwv}{\ifmmode f_{\rm bol}\left(\lambda\right) \else $f_{\rm bol}\left(\lambda\right)$\fi}
\newcommand{\fbolopt}{\ifmmode f_{\rm bol}\left(5100{\rm \AA}\right) \else $f_{\rm bol}\left(5100{\rm \AA}\right)$\fi}
\newcommand{\fbolthree}{\ifmmode f_{\rm bol}\left(3000{\rm \AA}\right) \else $f_{\rm bol}\left(3000{\rm \AA}\right)$\fi}
\newcommand{\fboluv}{\ifmmode f_{\rm bol}\left(1450{\rm \AA}\right) \else $f_{\rm bol}\left(1450{\rm \AA}\right)$\fi}

\newcommand{\fobs}{\ifmmode f_{\rm obs} \else $f_{\rm obs}$\fi}

\newcommand{  \mbh      }{\ifmmode M_{\rm BH} \else $M_{\rm BH}$\fi}
\newcommand{  \lmbh     }{\ifmmode \log\left(\mbh/\Msun\right) \else $\log\left(\mbh/\Msun\right)$\fi} 
\newcommand{  \lledd    }{\ifmmode L/L_{\rm Edd} \else $L/L_{\rm Edd}$\fi}
\newcommand{  \mmedd    }{\ifmmode \dot{m}/\dot{m}_{\rm \,Edd} \else $\dot{m}/\dot{m}_{\rm \,Edd}$\fi}
\newcommand{  \Lbol     }{\ifmmode L_{\rm bol} \else $L_{\rm bol}$\fi}
\newcommand{  \lbol     }{\ifmmode L_{\rm bol} \else $L_{\rm bol}$\fi}
\newcommand{  \lLbol    }{\ifmmode \log\left(\Lbol/\ergs\right) \else $\log\left(\Lbol/\ergs\right)$\fi} 
\newcommand{  \Lagn     }{\ifmmode L_{\rm AGN} \else $L_{\rm AGN}$\fi}
\newcommand{  \lagn     }{\ifmmode L_{\rm AGN} \else $L_{\rm AGN}$\fi}

\newcommand{  \tgrow     }{\ifmmode t_{\rm growth} \else $t_{\rm growth}$\fi}
\newcommand{  \tUni      }{\ifmmode t_{\rm Universe} \else $t_{\rm Universe}$\fi}

\newcommand{  \Mindot	}{\ifmmode \dot{M}_{\rm infall} \else $\dot{M}_{\rm infall}$\fi}
\newcommand{  \Mbhdot	}{\ifmmode \dot{M}_{\rm BH} \else $\dot{M}_{\rm BH}$\fi}

\newcommand{  \Maddot	}{\ifmmode \dot{M}_{\rm AD} \else $\dot{M}_{\rm AD}$\fi}

\newcommand{  \Mdot	}{\ifmmode \dot{M} \else $\dot{M}$\fi}

\newcommand{  \as	}{\ifmmode a_{\rm *} \else $a_{\rm *}$\fi}
\newcommand{  \avec	}{\ifmmode \vec{a}_{\rm *} \else $\vec{a}_{\rm *}$\fi}
\newcommand{  \re	}{\ifmmode \eta      	 \else $\eta$\fi}
\newcommand{  \RISCO	}{\ifmmode R_{\rm ISCO}  \else $R_{\rm ISCO}$\fi}
\newcommand{  \rg	}{\ifmmode r_{\rm g}  \else $r_{\rm g}$\fi}
\newcommand{  \rS	}{\ifmmode r_{\rm S}  \else $r_{\rm S}$\fi}

\newcommand{  \mseed    }{\ifmmode M_{\rm seed} \else $M_{\rm seed}$\fi}
\newcommand{  \mbul     }{\ifmmode M_{\rm Bulge} \else $M_{\rm Bulge}$\fi} 
\newcommand{  \mstar    }{\ifmmode M_{*} \else $M_{*}$\fi} 
\newcommand{  \mgal     }{\ifmmode M_{*} \else $M_{*}$\fi} 
\newcommand{  \mhost    }{\ifmmode M_{\rm Host} \else $M_{\rm Host}$\fi}
\newcommand{  \mmsmall  }{\ifmmode M_{\rm BH}/M_{*} \else $M_{\rm BH}/M_{*}$\fi}
\newcommand{  \mmlarge  }{\ifmmode M_{*}/M_{\rm BH} \else $M_{*}/M_{\rm BH}$\fi}

\newcommand{  \mmwp     }{\ifmmode \left(M_{*}/M_{\rm BH}\right) \else $\left(M_{*}/M_{\rm BH}\right)$\fi}
\newcommand{  \ml       }{\ifmmode M_{*}/L_{*} \else $M_{*}/L_{*}$\fi}
\newcommand{  \mlwp     }{\ifmmode \left(M_{*}/L\right) \else $\left(M_{*}/L\right)$\fi}
\newcommand{  \mlk      }{\ifmmode \left(M_{*}/L_{K}\right) \else $\left(M_{*}/L_{K}\right)$\fi}
\newcommand{  \sigs     }{\ifmmode \sigma_{*} \else $\sigma_{*}$\fi}
\newcommand{  \Reff     }{\ifmmode R_{\rm e} \else $R_{\rm e}$\fi}
\newcommand{  \nser     }{\ifmmode n_{\rm s} \else $n_{\rm s}$\fi}
\newcommand{  \LFIR     }{\ifmmode L_{\rm FIR} \else $L_{\rm FIR}$\fi}
\newcommand{  \Lfir     }{\ifmmode L_{\rm FIR} \else $L_{\rm FIR}$\fi}

\newcommand{  \mdyn     }{\ifmmode M_{\rm dyn} \else $M_{\rm dyn}$\fi} 
\newcommand{  \mgas     }{\ifmmode M_{\rm gas} \else $M_{\rm gas}$\fi} 
\newcommand{  \mh       }{\ifmmode M_{\rm h} \else $M_{\rm h}$\fi}
\newcommand{  \mhalo    }{\ifmmode M_{\rm halo} \else $M_{\rm halo}$\fi}

\newcommand{  \Lcii     }{\ifmmode L_{\cii} \else $L_{\cii}$\fi}

\newcommand{\bj}{\ifmmode b_{\rm J} \else $b_{\rm J}$\fi}

\newcommand{\iab}{\ifmmode i_{\rm AB} \else $i_{\rm AB}$\fi}

\newcommand{\jab}{\ifmmode J_{\rm AB} \else $J_{\rm AB}$\fi}
\newcommand{\hab}{\ifmmode H_{\rm AB} \else $H_{\rm AB}$\fi}
\newcommand{\kab}{\ifmmode K_{\rm AB} \else $K_{\rm AB}$\fi}

\newcommand{\jveg}{\ifmmode J_{\rm Vega} \else $J_{\rm Vega}$\fi}
\newcommand{\hveg}{\ifmmode H_{\rm Vega} \else $H_{\rm Vega}$\fi}
\newcommand{\kveg}{\ifmmode K_{\rm Vega} \else $K_{\rm Vega}$\fi}

\newcommand{  \Chisq    }{\ifmmode \chi^{2} \else $\chi^{2}$}
\newcommand{  \nelec    }{\ifmmode n_{e} \else $n_{e}$\fi}     % electron density
\newcommand{  \nh       }{\ifmmode n_{H} \else $n_{H}$\fi}     % hydrogen density
\newcommand{  \Ncol     }{\ifmmode N_{col} \else $N_{col}$\fi} % column density
\newcommand{  \NH       }{\ifmmode N_{H} \else $N_{H}$\fi}     % column density

\newcommand{\MgasCO}{\ifmmode M_{\rm gas, \, CO} \else $M_{\rm gas, \, CO}$\fi}
\newcommand{\Mgasdust}{\ifmmode M_{\rm gas, \, dust} \else $M_{\rm gas, \, dust}$\fi}
\newcommand{\lMgasCO}{\relax\ifmmode \log \left( M_{\rm gas, \, CO} \right) \else $ \log \left( M_{\rm gas, \, CO}  \right)$\fi}
\newcommand{\lMgasdust}{\ifmmode  \log \left( M_{\rm gas, \, dust} \right) \else $ \log \left( M_{\rm gas, \, dust} \right)$\fi}
\newcommand{\Mdust}{\ifmmode M_{\rm dust} \else $M_{\rm dust}$\fi}
\newcommand{\lMdust}{\ifmmode  \log \left( M_{\rm dust} \right) \else $ \log \left( M_{\rm dust} \right)$\fi}

\newcommand{\Hmol }{\ifmmode {\rm H_2} \else ${\rm H_2}$\fi}

\def\deg{\hbox{$^\circ$}}

\def\micron{\hbox{$\mu$m}}
\def\ion#1#2{#1$\;${\small\rm\@Roman{#2}}\relax}

\LetLtxMacro\oldcitep\citep % command to wrap each \citep in an \mbox to prevent the error that occurs when a link spans two pages
\RenewDocumentCommand{\citep}{O{} O{} m}{\oldcitep[#1][#2]{#3}}
\NewDocumentCommand{\citex}{O{} O{} m}{\oldcitep{#3}}

\LetLtxMacro\oldcitet\citet % command to wrap each \citet in an \mbox to prevent the error that occurs when a link spans two pages
\RenewDocumentCommand{\citet}{O{} O{} m}{\oldcitet[#1][#2]{#3}}

\newcommand\blfootnote[1]{%
  \begingroup
  \renewcommand\thefootnote{}\footnote{#1}%
  \addtocounter{footnote}{-1}%
  \endgroup
}
\title[Outflow scaling relations]{Incidence, scaling relations and physical conditions of ionised gas outflows in MaNGA}
\author[Avery et al.]{%mn@ras.org.uk (KTS)}
Charlotte R. Avery,$^{1,{\color{blue} \star}}$
Stijn Wuyts,$^{1}$
Natascha M. F\"{o}rster Schreiber,$^{2}$
Carolin Villforth,$^{1}$ \newauthor
Caroline Bertemes,$^{1,3}$
Wenjun Chang,$^{4,5,6}$
Stephen L. Hamer,$^{1}$
Jun Toshikawa,$^{1,7}$
Junkai Zhang$^{1}$
\\
$^{1}$ Department of Physics, University of Bath, Claverton Down, Bath, BA2 7AY, UK\\
$^{2}$ Max-Planck-Institut f\"{u}r Extraterrestrische Physik, Giessenbachstr. 1, 85748 Garching, Germany\\
$^{3}$ Zentrum f\"{u}r Astronomie der Universit\"{a}t Heidelberg Astronomisches Rechen-Institut, M\"{o}nchhofstr 12-14, 69120 Heidelberg, Germany\\
$^{4}$ CAS Key Laboratory for Research in Galaxies and Cosmology, Department of Astronomy, University of Science and Technology of China, \\ Hefei, Anhui, 230026, China\\
$^{5}$ School of Astronomy and Space Sciences, University of Science and Technology of China, Hefei, 230026, China\\
$^{6}$ Department of Physics \& Astronomy, University of California, Riverside, CA, 92521, USA\\
$^{7}$ Institute for Cosmic Ray Research, The University of Tokyo, Kashiwa, Chiba 277-8582, Japan
}

\date{Accepted 9 March 2021.}

\pubyear{2020}

\begin{document}
\label{firstpage}
\pagerange{\pageref{firstpage}--\pageref{lastpage}}
\maketitle
\raggedbottom  %use this option if we don't ever want double-spacing between paragraphs (as TeX sometimes introduces it to align the text vertically)

% Abstract of the paper
\begin{abstract}
In this work, we investigate the strength and impact of ionised gas outflows within $z \sim 0.04$ MaNGA galaxies. We find evidence for outflows in 322 galaxies ($12\%$ of the analysed line-emitting sample), 185 of which show evidence for AGN activity. Most outflows are centrally concentrated with a spatial extent that scales sublinearly with $R_{\rm e}$. The incidence of outflows is enhanced at higher masses, central surface densities and deeper gravitational potentials, as well as at higher SFR and AGN luminosity. We quantify strong correlations between mass outflow rates and the mechanical drivers of the outflow of the form $\dot{M}_{\rm out} \propto \rm SFR^{0.97}$ and $\dot{M}_{\rm out} \propto L_{\rm AGN}^{0.55}$. We derive a master scaling relation describing the mass outflow rate of ionised gas as a function of $M_{\star}$, SFR, $R_{\rm e}$ and $L_{\rm AGN}$. Most of the observed winds are anticipated to act as galactic fountains, with the fraction of galaxies with escaping winds increasing with decreasing potential well depth. We further investigate the physical properties of the outflowing gas finding evidence for enhanced attenuation in the outflow, possibly due to metal-enriched winds, and higher excitation compared to the gas in the galactic disk. Given that the majority of previous studies have focused on more extreme systems with higher SFRs and/or more luminous AGN, our study provides a unique view of the non-gravitational gaseous motions within `typical' galaxies in the low-redshift Universe, where low-luminosity AGN and star formation contribute jointly to the observed outflow phenomenology. \\

\noindent \textbf{Key words:} galaxies: active – galaxies: evolution – galaxies: ISM – galaxies: starburst  \\
\end{abstract}

%%% -------------------------------------------------------------------------------------------------- %%%

\section{Introduction}
\label{sec:Intro}

\blfootnote{ {\color{blue} $\star$} {E-mail: c.r.avery@bath.ac.uk}}

The efficiency of galaxy formation is low given the amount of available baryons in the Universe. This is evident from studies matching the abundances of galaxies and their dark matter halos which reveal peak stellar-to-halo mass ratios of only $\sim 20\%$ of the cosmic baryon fraction at halo masses of $\mathrm{log(M_{h}/M_{\odot}) \sim 12}$. Below and above this mass the efficiency drops steeply \citep[e.g.,]{Moster2013, Behroozi2013}.  At low stellar masses, this inefficiency has been widely attributed to feedback induced by the energetic processes associated with massive star evolution, most notably supernova explosions \citep{Dekel1986, Efstathiou2000}, although stellar winds and radiation effects are also found to be important \citep{Murray2005, Hopkins2014, Tollet2019}.  At high masses, feedback from accreting central supermassive black holes is thought to be required to drive gas out of the deep potential wells and/or to sustain heating of the halo (see, e.g., the review by \citealp{Fabian2012} and references therein).  The energy and momentum injected by star formation and nuclear accretion into the surrounding interstellar medium (ISM) gives rise to galactic-scale winds.  These can lead to fountains that re-accrete or they can expel gas from the gravitational potential altogether. \par Although major advances in numerical modelling have allowed studying such outflow phenomena within the context of cosmological hydrodynamical simulations \citep{Nelson2019, Mitchell2020}, the detailed physics describing their driving and coupling to the multi-phase ISM requires sub-grid processes that cannot yet be implemented from first principles.  Higher resolution simulations \citep[e.g.,]{Hopkins2014, Tanner2017} provide complementary insights, but lack the cosmological context or statistics to evaluate the impact of large-scale winds across the galaxy population.  Observational guidance on how outflow properties and outflow strength scale with star formation and/or AGN activity, or with properties of the host galaxy more generally, thus remain invaluable to inform galaxy formation models. \par
Quantifying the impact of winds on galaxy evolution based on observational data has been an active area of research for over a decade. 
At low redshift, studies of star-forming galaxies (SFGs) have found the outflow velocity to correlate with stellar mass ($M_{\star}$), SFR and circular velocity in the cool neutral \citep{Rupke2005}, low-ionisation \citep{Chisholm2015, Heckman2016} and warm ionised gas phases \citep[][see also recent reviews by \citealp{Rupke2018} and \citealp{Veilleux2020} for overviews of results on these, and other wind components such as those traced by molecular gas and dust]{Arribas2014, Cicone2016}. By and large, these pioneering efforts have focussed on more `extreme' objects, such as starbursting outliers above the main sequence of SFGs or luminous AGNs \citep[e.g.,][]{Rupke2011, Harrison2018}, or relied on stacking of large numbers of (single-fibre) galaxy spectra. Further examples of how extreme a feedback can be induced
by luminous AGNs are provided by, e.g., \citet{Farrah2012}, \citet{Cicone2012} and \citet{Borguet2013}. 
\par
Outflow studies have also been pushed to higher redshifts, around $1 \lesssim z \lesssim 3$ when the cosmic star-formation history peaked \citep{Madau2014} and both star formation and AGN activity were elevated by over an order of magnitude compared to typical nearby galaxies. Here, outflows are found to be ubiquitous among SFGs, especially where star formation or star formation surface densities are enhanced \citep{Shapley2003, Weiner2009, Rubin2010, Genzel2011, Kornei2012, Newman2012, Rubin2014, Davies2019, ForsterSchreiber2019, Swinbank2019}.  In active galaxies, correlations between AGN-driven winds and stellar mass, central concentration (\citealp{Genzel2014, ForsterSchreiber2019}) and AGN luminosity have further been recorded (\citealp{Harrison2016, Leung2019}; also at low redshift in \citealp{Cicone2014} and \citealp{Lutz2020}).  For recent reviews on outflow phenomenology at high redshift, we refer the reader to \citet{ForsterSchreiber2020} and \citet{Veilleux2020}. \par
For galaxies in the nearby Universe, a new generation of highly multiplexed fibre bundle spectrographs, most notably SAMI \citep{Croom2012} and MaNGA \citep{Bundy2015}, are now enabling a more systematic assessment of wind scaling relations with a robust understanding of the relation to the underlying galaxy population.  The samples covered by the SAMI and MaNGA surveys span over two orders of magnitude in stellar mass and amass several thousands of objects \citep{Green2018, Wake2017}.  Other than sheer number statistics, these integral-field unit (IFU) surveys introduce several key advantages to study the physics of outflows by merit of their combined spatially and spectrally resolved information.  These include a better ability to measure outflow properties by measuring the spatial extent of outflows \citep{Roberts-Borsani2020}, and by removing the large-scale rotational velocity field of gas in the disk, which may otherwise contribute to broadening of line profiles and thus inhibit the detection of weak outflow signatures in galaxy-integrated spectroscopy. Furthermore, IFU capabilities allow the localisation of the launching sites of outflows \citep{RodriguezDelPino2019} and identification of the presence of low-luminosity AGN activity. This can be achieved by extracting nuclear line diagnostics that may otherwise be diluted by line emission excited by star formation in galaxy-integrated spectra. The relative contributions of star formation and AGN to the driving of galactic-scale outflows and their impact on galaxy evolution remains heavily debated.  The ability to detect weak AGN activity and identify launching sites is important to understand the relative wind driving contributions. \par
Among other applications, these unique strengths have been exploited to assess the low coupling efficiencies (measured as the kinetic power of the outflow divided by AGN luminosity) of ionised gas outflows driven by low-luminosity AGN \citep{Wylezalek2020}, the geometry of galactic-scale winds \citep{Bizyaev2019}, their ionisation state \citep{Ho2014}, and even the presence of star formation occurring in the outflowing gas \citep{Gallagher2019}.
\par 
Here, we exploit the exquisite MaNGA data set to systematically investigate the incidence and scaling relations of ionised gas outflows among typical local galaxies via the broad velocity components seen in their strong optical emission lines.  We critically make use of the resolved information to remove the large-scale velocity field, to measure the extent of galactic winds, and to identify low-luminosity AGN which are only dominant in light over star formation (SF) in the very central region of galaxies.  We consider the combined effects of both SF and AGN, and investigate outflow strength as well as physical conditions inferred from broad-component line ratio information.  We provide fitting formulae describing the relation between outflow properties and the internal properties of their host galaxies. \par 
The paper is organised as follows.  After laying out the methodology and sample definition in Section\ \ref{method_sample.sec}, we present results on the incidence of outflows across the MaNGA sample in Section\ \ref{incidence.sec}, quantify outflow scaling relations with internal galaxy properties in Section\ \ref{scaling_relations.sec}, and address the outflow geometry and physical conditions in Sections\ \ref{geometry.sec} and\ \ref{physical_conditions.sec}, respectively.  We discuss the results obtained in Section\ \ref{discussion.sec} and summarize our key findings in Section\ \ref{summary.sec}. \\
Throughout the paper, we adopt a \citet{Chabrier2003} IMF and a flat $\Lambda$CDM cosmology with $\Omega_{\Lambda} = 0.7$, $\Omega_m = 0.3$ and $H_0 =70\ \rm{km}\ \rm{s}^{-1}\ \rm{Mpc}^{-1}$.

%%% -------------------------------------------------------------------------------------------------- %%%

%%% Method & sample
\section{Method and sample selection}
\label{method_sample.sec}

\subsection{Parent sample}
\label{parent_sample.sec}

We use the data cubes from the SDSS-IV MaNGA galaxy survey \citep{Bundy2015, Blanton2017} which is part of the fifteenth data release of SDSS \citep{Aguado2019}. The IFU capabilities of MaNGA provide spectra over the 2D field of view of hexagonally bundled arrangements of 19-127 fibres (corresponding to $27''-36''$ in diameter) which are fed into the BOSS spectrographs \citep{Drory2015}.  The MaNGA fibre bundles cover galaxies out to a radius of $1.5-2.5R_{\rm e}$ for the primary and secondary samples \citep{Wake2017}, respectively.  MaNGA galaxy data have a typical spatial resolution of $2\farcs 5$ FWHM, corresponding to 1.8 kpc at the median redshift $z \sim 0.04$.  This work makes use of the objects in the primary, secondary and ancillary target samples. We make use of the outputs from the MaNGA Data Analysis Pipeline (DAP; \citealp{Westfall2019}), which provides 2D maps of the stellar and gaseous kinematics, and emission-line properties extracted from the data cubes reduced by the data reduction pipeline \citep{Law2016}. 
\par 
The spectral resolution is wavelength dependent with a typical value of R$\sim 2000$ (increasing to R $\sim2300$ at $8500$\AA). The large wavelength coverage of 3,600\AA - 10,000\AA \ provides data on the nebular emission lines H$\mathrm{\beta}$, [OIII]$\lambda4959,5007$,  [NII]$\lambda6548,6583$, H$\mathrm{\alpha}$ and [SII]$\lambda6717,6731$ which can be used to investigate gas excitation mechanisms using the BPT diagnostic diagrams \citep{Baldwin1981}. The BPT diagnostic uses specific line ratios to determine whether gas is predominantly photoionised by star-forming regions or some other form of excitation, such as the central AGN, shocks or evolved stellar populations. The [SII] doublet further allows us to place constraints on the local electron density \citep{Osterbrock1989}.  
\par
We cross-match the MaNGA DR15 sample to the MPA-JHU database \citep{Kauffmann2003, Brinchmann2004, Salim2007}, from which we extract total galaxy stellar masses and star formation rates and their associated error estimates.  There are a total of 4239 MaNGA objects successfully analysed by the MaNGA DAP with corresponding MPA-JHU data.

\subsection{Methodology}
\label{method.sec}

We take advantage of the IFU capabilities of MaNGA to distinguish outflowing gas and disk gas.  Outflows are typically identified by their kinematic signature, in particular through telltale high velocity wings on emission lines (e.g., \citealp{Veilleux2005}).

\subsubsection{Spectral stacking}
\label{stacking.sec}

High signal-to-noise ratio (S/N) spectral data are required to separate outflows from the (typically dominant) galaxy disk. This is achieved by the following careful stacking analysis. \par
We stack spaxels within elliptical apertures with major axes equal to $0.5R_{\rm e}$, $R_{\rm e}$, and $1.5R_{\rm e}$ for individual objects.\footnote{The motivation to consider differently sized apertures, rather than a single aperture size in units of $R_{\rm e}$, stems from the fact that outflows may in principle span a range in spatial extents depending on nature and location of their drivers, and serves to optimise the robustness of outflow detection, which depends on a balance between S/N of the data and minimising the addition of spaxels at large radii that may dilute the relative broad-component signal in some of the stacks.} The large-scale velocity gradient, which would otherwise contribute to the broadening of line profiles during stacking, is removed from each data cube using the emission line velocity fields provided by the DAP.  The resulting velocity shifted spaxels are interpolated over a common velocity grid. Poor quality pixels and spaxels with poorly defined velocity measurements were excluded from the stacking process using the mask flags provided.  The remaining spaxels were combined to create an average spectrum for each of the three aperture sizes, while keeping note of the normalisation factor to later determine total line fluxes.  Errors were combined following Equation 9 of \cite{Law2016} in order to account for covariance between spatially adjacent pixels. \par
To reliably retrieve underlying broad components to nebular emission lines, the following methodology steps are taken on the 2744 stacked spectra with S/N > 10 in the H$\mathrm{\beta}$, [OIII]$\lambda5007$,  [NII]$\lambda6583$ and H$\mathrm{\alpha}$ emission lines. We will refer to these 2744 objects as the `analysed MaNGA sample' throughout this paper. The majority of MaNGA DR15 objects weeded out by this S/N criterion on line emission belong to the class of quiescent galaxies. We find that our stacking procedure is successful in producing high quality data, with typical S/N $\sim 65, 38, 208, 94, $ in the H$\beta$, [OIII]$\lambda5007$, H$\alpha$ and [NII]$\lambda6583$ lines respectively for objects within the analysed sample. \par 
The stellar continuum is subtracted from each individual stacked spectrum using \texttt{ppxf} \citep{Cappellari2017} with the following set-up. Two Gaussian components were used to fit the emission lines, each with two kinematic moments which were tied between spectral lines. The MILES library of stellar spectra was used to fit the stellar continuum with 4 moments in the line-of-sight velocity distribution. The line ratios [OIII]$\lambda 5007/4959$ and [NII]$\lambda 6583/6548$ were fixed to their theoretical values, Balmer line ratios were tied to their intrinsic values whilst the gas reddening was left as a free parameter in the fit and [SII] doublets were limited to their physically allowed range.  In this step, the act of simultaneously fitting the gas emission and stellar continuum was to simply achieve a good continuum fit to the stacked spectra.  From the \texttt{ppxf} results, we adopt the best-fit stellar continuum model to generate continuum-subtracted emission line spectra.  The resulting gaseous emission line profiles are next fed into our custom-made line profile analysis scripts. \par
The error on the resulting gas spectrum is given by the sum in quadrature of the error on the stacked spectrum and the error on the continuum fit.  The continuum fit error is estimated by masking the emission lines and taking the RMS of the residuals of the \texttt{ppxf} fit in the continuum regions.

\subsubsection{Line profile fitting}
\label{profile_fit.sec}

% FIG line profile
\begin{figure*}
\centering
\includegraphics[width=0.9\linewidth]{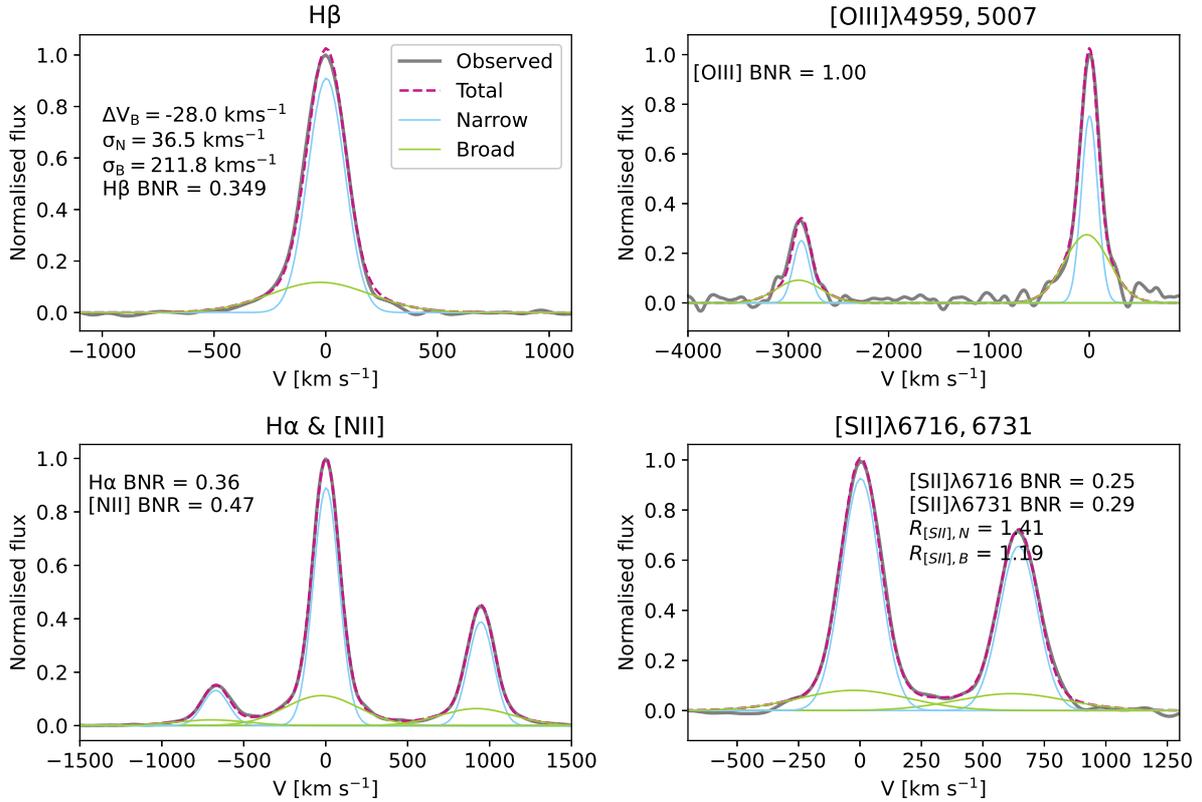}
\caption{Double-Gaussian decomposition of the strong optical emission lines for one of the galaxies in our outflow sample.  Constraints on which parameters are fixed and/or tied are detailed in Section \ref{profile_fit.sec}.  The observed gas spectrum is constructed by stacking spaxels within the outflow radius as described in Section \ref{Rout_def.sec}.}
\label{fig:lineprofile}
\end{figure*}

We extract the following four line-emitting regions: the H$\mathrm{\beta}$ line, the [OIII]$\mathrm{\lambda4959,5007}$ doublet, the [NII]$\lambda6548,6583$/H$\mathrm{\alpha}$ complex, and the [SII]$\mathrm{\lambda6716,6731}$ doublet from the continuum subtracted stack using a $\sim 2000$-$4000 \ \rm{km} \ \rm{s}^{-1}$ window for each emitting region as appropriate.  We normalise each spectral window to its strongest line while keeping note of the normalisation factor to allow for further analysis of line ratios and absolute flux values.  We fit a first-order polynomial to the continuum about the emission features to account for any poorly subtracted stellar continuum in the \texttt{ppxf} subtraction. \par
We use \texttt{mpfit} (\citealp{mpfit}, updated for Python by Sergey Koposov) to perform a single Gaussian and double Gaussian decomposition of the nebular emission lines.  We find that in a number of cases the stacked spectra are significantly better fit when, in addition to a narrow Gaussian component (N) with velocity dispersion $\mathrm{\sigma_N}$, a broad Gaussian component (B) is added with a velocity offset (from the N component) $\mathrm{\Delta V_{\rm B}}$ and a velocity dispersion $\mathrm{\sigma_{\rm B}}$. All kinematic moments are tied between the different emission lines during the fitting. We require $\mathrm{\sigma_{\rm B} > \sigma_N + 50\ \rm{km}\ \rm{s}^{-1}}$ to ensure a significantly broader B component. $\mathrm{[OIII]\lambda5007/4959}$ and [NII]$\mathrm{\lambda6583/6548}$ line ratios are fixed to 3; their values set by atomic physics, and the $\mathrm{[SII]\lambda6731/6716}$ flux ratio is limited to [0.2, 2], encompassing the physically allowed range.  An example line profile decomposition is illustrated in Figure\ \ref{fig:lineprofile}.
\par
Intrinsic gas kinematics are recovered from the stacked spectra by fitting a Gaussian model of width $\sigma_{\rm observed} = (\sigma^2 +\sigma_{\rm LSF}^2)^{1/2}$; i.e. the sum in quadrature of the intrinsic $\sigma$ and the instrumental broadening $\sigma_{\rm LSF}$.  The wavelength-dependent spectral line-spread function (LSF) is provided for each spaxel in the cube as part of the MaNGA data reduction pipeline.  Here, $\sigma_{\rm LSF}$ refers to the total stack $\sigma_{\rm LSF}$ which is obtained by combining the individual spaxel LSF measurements following Equation 3 of \cite{Westfall2019}.  Measurement errors in the LSF are on the order of a few percent \citep{Westfall2019}.

\subsubsection{Outflow sample definition}
\label{outflow_sample.sec}

After processing the $0.5R_{\rm e}$, $R_{\rm e}$, and $1.5R_{\rm e}$ stacks of all 2744 objects in the MaNGA analysed sample through the line profile fitting scripts outlined in Section \ref{profile_fit.sec}, we investigate for the presence of outflows. We take the following conservative selection criteria on the fitted line profiles as robust evidence for outflows.

\begin{enumerate}
    \item The double Gaussian decomposition must provide an improved fit compared to the single Gaussian fit according to the Bayesian Information Criterion (BIC) statistic, where we take $\mathrm{\Delta BIC > 10}$ as evidence for an additional broad component required in the fitting. This approach is the same as taken by \cite{Swinbank2019}.
    \item $\mathrm{\sigma_N+50 \ \rm{km}\ \rm{s}^{-1} < \sigma_{\rm B} < 500\ \rm{km}\ \rm{s}^{-1}}$, where the upper limit is imposed to avoid the broad component fitting any stellar continuum residuals that may not be subtracted properly.
    \item $\mathrm{|\Delta V_{\rm B}| < 500\ \rm{km}\ \rm{s}^{-1}}$ to avoid fitting of noise.
    \item The amplitude of the H$\mathrm{\alpha}$ broad component must be greater than 3 times the RMS of the continuum around the H$\mathrm{\alpha}$ line.
    \item The broad-to-narrow flux ratios (BNR) in all BPT lines must be greater than zero and the BNR $> 0.05$ in H$\mathrm{\alpha}$ and [OIII]$\lambda$5007.
\end{enumerate}
For a galaxy to enter our `outflow sample', the above criteria need to be met within at least one of the apertures considered.  The number of objects from our original sample of 2744 line-emitting galaxies remaining after application of the cumulative criteria (i) to (v) is 1315, 872, 835, 603 and eventually 376. We point out to the reader that in Section\ \ref{DIG.sec} we further reduce this `outflow sample' to 322 objects in order to mitigate the contribution of diffuse ionised gas (DIG) to the broad component emission.

\subsubsection{SF \& AGN subsamples}
\label{SF_AGN_def.sec}
\begin{figure}
\centering
\includegraphics[width=\linewidth]{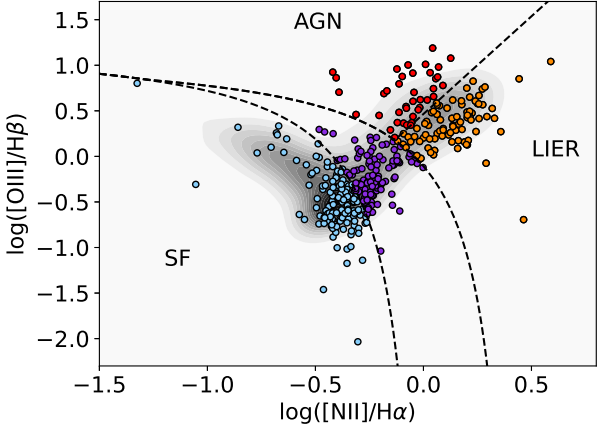}
\caption{BPT positions of the outflow sample determined within the central $0.25R_{\rm e}$ apertures. The nuclear BPT line ratios of all line-emitting MaNGA objects are shown in greyshades.}
\label{fig:025Re_BPT}
\end{figure}

The outflow sample is subdivided into galaxies with evidence for nuclear activity (AGN) and those without (SF) based on the \citet{Kauffmann2003} separation curve on the [NII]-BPT diagnostic diagram.  Here we leverage the spatially resolved nature of the MaNGA dataset by evaluating the narrow-component line ratios within the central $0.25R_{\rm e}$ stack, hence minimising the diluting effect of line emitting gas excited by HII regions which are located further away from the galaxy centre. For the few objects where the signal-to-noise ratio was too low to allow for a BPT diagnostic within $0.25R_{\rm e}$ a larger aperture size was adopted.  We present the BPT positions of the nuclear spectra in Figure \ref{fig:025Re_BPT}.  We highlight to the reader that a significant fraction of our AGN outflow sample falls within the composite and LIER regions of the BPT diagram. We expect the nuclear regions of objects that fall in the composite region to have a mixture of star formation, AGN and shock ionisation whereas LIERs are more difficult to interpret. Nevertheless, we assign these objects to the ‘AGN’ subsample given their emission line ratios are higher than the theoretical limit for ionisation from star-forming regions indicating some contribution from an alternative excitation mechanism to star formation.  The SF and AGN outflow subsamples count 139 and 237 objects, respectively.  The two types are labelled separately in all of our analysis, and distinctions in their outflow phenomenology are commented on where relevant. The subsets of our AGN subsample (i.e. composites, LIERs and ‘true’ AGN) are indicated in figures presenting relationships between outflow and host galaxy properties. 

\subsubsection{AGN luminosity}
\label{LAGN_def.sec}
In order to understand relationships between outflow properties and their physical drivers, we estimate the AGN luminosity ($L_{\rm AGN}$) for galaxies within the AGN subsample as follows. We use the MaNGA DAP line flux maps, produced through single Gaussian fitting, to identify BPT-AGN spaxels \citep{Kauffmann2003}, taking only spaxels with S/N $>5$ in all BPT lines to ensure reliable BPT line ratios. Using these BPT-AGN spaxels, we calculate the total luminosity of AGN-photoionised [OIII] emitting gas in each object. We then correct the [OIII] luminosity for extinction using Eq. 1 of \cite{Lamastra2009} before applying an interpolated version of the \cite{Lamastra2009} bolometric correction to convert $L_{\rm [OIII]}$ to a measure of the bolometric $L_{\rm AGN}$.  Eddington ratios were computed by dividing $L_{\rm AGN}$ by the black hole mass as inferred from the $M_{\rm BH} - \sigma$ relation by \citet{Kormendy2013}.

We note that one issue with this method is that we assume all spaxels with line ratios too strong for pure HII photoionisation are considered BPT-AGN leading to a potential overestimation of the luminosity of AGN-photoionised [OIII] emitting gas. We therefore also derive an estimation of $L_{\rm AGN}$ using mixing models described in \cite{Davies2014}. We adapt this method assuming that for each object the spaxels with lowest and highest [OIII]/H$\beta$ have an associated AGN fraction of $0\%$ and $100\%$ respectively, again taking only spaxels with S/N $>5$ in all BPT lines. We then calculate the total [OIII] luminosity using an AGN fraction weighting. For objects where the [OIII]/H$\beta$ spaxel extremes do not fall below the SF-composite boundary or above the composite-AGN boundary, we extrapolate the mixing sequence such that the $0\%$ or $100\%$ fall at the relevant boundary. Despite the numerous assumptions used in both models, we find that the derived $L_{\rm AGN}$ values are in good agreement with each other with no systematic offset and a scatter of 0.5 dex. Furthermore, we re-run our analysis using the $L_{\rm AGN}$ derived using the \cite{Davies2014} mixing model and find no significant changes in the finding of this paper. For simplicity and reproducibility, we use the former derivation of $L_{\rm AGN}$ in the work presented in this paper.  We further note that \cite{Lamastra2009} did also not apply a mixing sequence approach to the [OIII] luminosity measurements that served as basis for their bolometric correction recipe.

\subsubsection{Spatial Extent of the Outflows}
\label{Rout_def.sec}

% FIG Rout
\begin{figure}
\centering
\includegraphics[width=\linewidth]{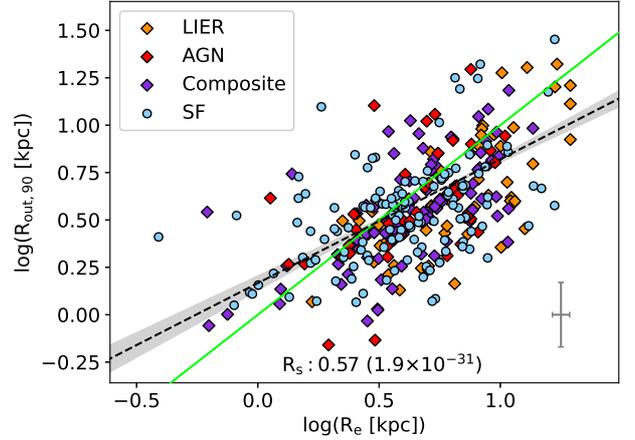}
\caption{Spatial extent of the ionised gas outflows, defined by the radius containing 90\% of the broad H$\alpha$ flux ($R_{\rm out, 90}$), plotted as a function of the effective radius of galaxies in the outflow sample.  Diamonds denote galaxies hosting an active nucleus (AGN), whereas objects for which only star formation has been identified as a possible driver of outflow activity are shown as circles (SF). Subdivisions of the AGN outflow sample (LIERs/`true AGN'/Composites) are marked using different colours in line with Figure\ \ref{fig:025Re_BPT}. The median error bar of the data points is shown in the bottom right.  The black dashed line and grey-shaded region mark the best-fit linear regression to the outflow sample as a whole and the associated central 68th percentile on the posterior distribution, respectively.  The Spearman's rank correlation coefficient and associated p-value are indicated. For reference, the one-to-one line is shown in green.}
\label{fig:Rout}
\end{figure}

To determine the spatial extent of the outflow in each object, the data cubes of galaxies in the outflow sample were binned in elliptical annuli \footnote{Petrosian ellipticities and position
angles are provided as part of the
MaNGA DAP products.} of width $0.25R_{\rm e}$ from the galactic centre out to 1.5$R_{\rm e}$, with a final outer bin containing spaxels beyond 1.5$R_{\rm e}$. The spaxels within each annular bin were stacked, and the resulting stack continuum subtracted, according to the method outlined in Section \ref{stacking.sec}. The continuum subtracted annular bins were then fit with double Gaussian components (Section \ref{profile_fit.sec}), and non-zero broad component fluxes were assigned to the annuli provided criteria (i) - (iii) from Section\ \ref{outflow_sample.sec} were met.\footnote{Given the lower S/N of stacked spectra constructed from annuli, relative to the 0.5/1/1.5 $R_{\rm e}$ elliptical apertures, we drop the more conservative criteria on broad component detection to construct a `more complete' curve of growth of $\mathrm{F_B(H\alpha)}$.}

From this, we construct the radial curve of growth of the broad-component H$\alpha$ flux $\mathrm{F_B(H\alpha)}$ for each object by interpolating between annuli. We quantify $R_{\rm out,90}$ as the radius enclosing 90\% of the galaxy-integrated $\mathrm{F_B(H\alpha)}$. 
Figure\ \ref{fig:Rout} shows that for the majority of objects the bulk of broad-component emission is contained within the galaxy's effective radius $R_{\rm e}$, quantified on the optical broad-band NASA-Sloan Atlas (NSA; Blanton M. http://www.nsatlas.org) image.  A positive correlation between $R_{\rm out,90}$ and $R_{\rm e}$ is evident, with a sub-unity slope such that outflow signatures within larger galaxies are more centrally concentrated. The best-fit linear regression to the outflow sample in Figure\ \ref{fig:Rout} is adopted to define the `outflow radius' $R_{\rm out}$ as a measure of the spatial extent of the outflow within an object of effective radius $R_{\rm e}$:
\begin{equation}
\log(R_{\rm out}) = (0.65 \pm 0.11) \ \log(R_{\rm e}) + (0.17 \pm 0.08).
\label{eq:Rout}
\end{equation}
The formal uncertainties on the best-fit coefficients of the linear relation of Eq. \ref{eq:Rout} (obtained using {\tt linmix}; \citealp{Kelly2007}) may be regarded as lower limits given potential surface brightness limitations at large radii and resolution limitations at small radii in constructing the broad-component curve of growths. We note that a substantial scatter is present around the relation, estimated by {\tt linmix} to be intrinsically at the level of $\sim 0.2$ dex once accounting for measurement uncertainties.  \par 

In order to include as much of the outflow emission as possible whilst minimising the addition of noise within the stack, we re-stack the individual objects making up our outflow sample within elliptical apertures of radius $R_{\rm out}$.  We repeat the methods outlined in Sections \ref{stacking.sec} and \ref{profile_fit.sec}, and check that our multi-component fitting are reliably tracing the outflow features within each $R_{\rm out}$ stack using the criteria outlined in Section \ref{outflow_sample.sec}.  For a minority of objects, the broad Gaussian component extracted within this aperture did not satisfy all the criteria outlined in Section \ref{outflow_sample.sec}.  For these objects we adjust $R_{\rm out}$ to the closest aperture size for which a broad component is required in the fit such that all criteria in Section \ref{outflow_sample.sec} were satisfied (considering apertures of either 0.25, 0.5, 0.75, 1, 1.25 or 1.5 $R_{\rm e}$, or the galaxy-integrated spectrum).  We base the inferences of outflow strength and properties presented in the remainder of this paper on the resulting $R_{\rm out}$ stacks. \par 

\subsubsection{Removing DIG from the Outflow Sample}
\label{DIG.sec}
We are wary of Diffuse Ionised Gas (DIG) emission components within galaxies which may be kinematically offset from the large-scale galaxy disk component (e.g., \citealp{Bizyaev2017}), and could plausibly contribute to broad component emission or even be mistakenly identified as outflowing gas. Given that the outflow properties estimated in this work are based on spectra constructed from spaxels within the $R_{\rm out}$ aperture, we remove any objects from our outflow sample where DIG dominates the emission within $R_{\rm out}$.  We identify such objects using the  H$\alpha$ equivalent width (EW) MAP output from the MaNGA DAP, where we recognise DIG dominated objects as having H$\alpha$ EW $< 3$ \citep{Lacerda2017} in more than half the spaxels within $R_{\rm out}$. These amount to 54 objects, 40 of which fall in the LIER region in Figure \ref{fig:025Re_BPT}. This cut reduces our outflow sample to 322, of which 137 and 185 fall in the SF and AGN subsamples, respectively. \par

\subsubsection{Definition of outflow properties}
\label{outflow_def.sec}

To quantify the physical characteristics of outflows across the galaxy population, we consider the parameters defined in this Section. The following definitions assume that the narrow emission line component is associated with gas in the disk, predominantly coming from star-forming HII region gas, whereas the broad component traces gas in the outflow exhibiting non-gravitational motions.

We follow \cite{Genzel2011} and \cite{Davies2019} in defining the outflow velocity as:
\begin{equation}
v_{\rm out} = |\Delta V_{\rm B} - 2\sigma_{\rm B}|.
\label{eq:vout}
\end{equation}

$v_{\rm out}$ represents one of the ingredients to compute the mass outflow rate $\dot{M}_{\rm out}$, which is an important parameter quantifying the amount of gas driven away from its launch site by winds. We follow the approach of \cite{Newman2012} and quantify the mass outflow rate as:
\begin{equation} 
\dot{M}_{\rm out}\ [M_{\odot}\ \rm{yr}^{-1}] = 1.586\times 10^{-26}\ \frac{1.36\ m_{\rm H}}{\gamma_{\rm H\alpha}\ n_{\rm e}}\ \frac{v_{\rm out}}{R_{\rm out}}\ L_{\rm H\alpha,B}
\label{eq:Mdot_out}
\end{equation}

where $ \gamma_{\rm H\alpha} = 3.56 \times 10^{-25} \ \mathrm{erg \ cm^3 \ s^{-1}} $ is the H$\alpha$ emissivity at $T = 10^4\ \rm{K}$, and $n_{\rm e}$ is the electron density of the outflowing gas in $\rm cm^{-3}$ assuming the same temperature. $R_{\rm out}$ is the maximum radial extent of the outflow in km and is estimated as described in Section\ \ref{Rout_def.sec}.  $L_{\rm H\alpha,B}$ is the H$\alpha$ broad component luminosity in $\mathrm{erg \ s^{-1}}$.  The first term converts the units of $\dot{M}_{\rm out}$ from $\rm{g}\ \rm{s}^{-1}$ to $M_{\odot}\ \rm{yr}^{-1}$.  This equation assumes that the outflow velocity and mass outflow rate are constant with radius.  As discussed further in Section\ \ref{ne.sec}, we adopt a constant value of $n_{\rm e} = 192 \ \rm{cm}^{-3}$ for the broad-component gas as inferred from double-Gaussian decompositions of the [SII] line doublet.  The parameters $v_{\rm out}$, $R_{\rm out}$ and $L_{\rm H\alpha,B}$ are measured for each galaxy individually, where $v_{\rm out}$ and $L_{\rm H\alpha,B}$ are extracted from the Gaussian decomposition of the $R_{\rm out}$ stacked spectra. \par 

The mass loading factor $\eta$ is defined as the outflow rate normalised by the star formation rate:
\begin{equation}
\eta \equiv \frac{\dot{M}_{\rm out}}{\rm SFR}.
\end{equation}
An estimate of the SFR can be obtained from the narrow Gaussian component of H$\alpha$ tracing the ongoing star formation activity in the galaxy disk:
\begin{equation}
\mathrm{SFR}\ [\mathrm{M} _{\odot}\ \rm{yr}^{-1}] = 2.1\times 10^{-41} \ \it{L}_{\rm H\alpha,N}\ [\rm{erg}\ \rm{s}^{-1}],
\end{equation}
which through combination with Eq.\ \ref{eq:Mdot_out} boils down to the following dependency on the H$\alpha$ BNR:
\begin{equation}
\eta = 2.1\times 10^{41}\ \frac{1.36m_{\rm H}}{\gamma_{\rm H\alpha}\ n_{\rm e}}\ \frac{v_{\rm out}}{R_{\rm out}}\ \rm BNR_{\rm H\alpha}.
\label{eq:eta}
\end{equation}
This parameter quantifies the relative rate at which gas is expelled versus consumed by star formation, and in the case of star-formation driven outflows can be thought of as a diagnostic of outflow efficiency.  Here we add the caveat that $\dot{M}_{\rm out}$ in our analysis captures specifically the warm ionised gas component, and that the observed winds do not necessarily escape from the gravitational potential of the galaxy altogether.  Both aspects complicate the comparison to galaxy formation models, and we return to these points in Section\ \ref{discussion.sec}.

We further note that, whereas by default we assume any dust correction factors to be divided out in Eq.\ \ref{eq:eta}, a dust correction factor based on single-component Gaussian fits of the Balmer decrement ($H\alpha/H\beta$) is applied in computing $L_{\rm H\alpha,B}$ in Eq.\ \ref{eq:Mdot_out}.  Possible adjustments to this approach and its impact on the results obtained are discussed in Section\ \ref{attenuation.sec}.

\section{Results}
\label{results.sec}

With the results of the emission line decomposition and inferred physical properties of outflows based thereupon in hand, we now transition to presenting our key findings.  We do this in a three-pronged approach: considering first the incidence of outflows among the entire sample of line-emitting MaNGA galaxies analysed (Section\ \ref{incidence.sec}), focussing next on the outflow sample itself to address how outflow properties scale with internal galaxy properties (Section\ \ref{scaling_relations.sec}), and finally evaluating more closely the geometry and physical conditions of the broad-component gas (Sections\ \ref{geometry.sec} and \ref{physical_conditions.sec}).  Further considerations of outflow detectability, driving mechanisms and ultimate fate are discussed in Section\ \ref{discussion.sec}.

\subsection{Incidence of outflows among the MaNGA sample}
\label{incidence.sec}

\subsubsection{Most MaNGA outflows are centrally concentrated}
\label{rad_profile.sec}

Figure \ref{fig:Rout} demonstrates that the spatial extent of broad-component line emission, interpreted to be relatively high-velocity ionised gas associated with a large-scale galactic wind, shows a positive trend with galaxy size, yet one with a slope that is shallower than unity.  Moreover, for a galaxy of typical size, the bulk of the broad-component line emission, 90\% as quantified by the parameter $R_{\rm out,90}$, emerges from within the galaxy's effective radius (see also Eq\ \ref{eq:Rout}). Spatially within MaNGA galaxies, outflows are centrally concentrated with $\sim 67\%$ of outflow galaxies showing the bulk of their broad-component wind signature encompassed within $R_{\rm e}$.  \cite{Roberts-Borsani2020} find neutral gas outflows in MaNGA galaxies also to be centrally concentrated within $\sim1R_{\rm e}$.  This finding is further in line, at least at a qualitative level, with galaxy formation models which invoke feedback to expel low angular momentum gas from galaxy centres in order to produce galactic disks of realistic size \citep[e.g.,]{Dutton2009}.

\subsubsection{Which galaxies show outflow signatures?}
\label{incidence_galprop.sec}

% FIG Incidence
\begin{figure*}
\centering
\includegraphics[width=\textwidth]{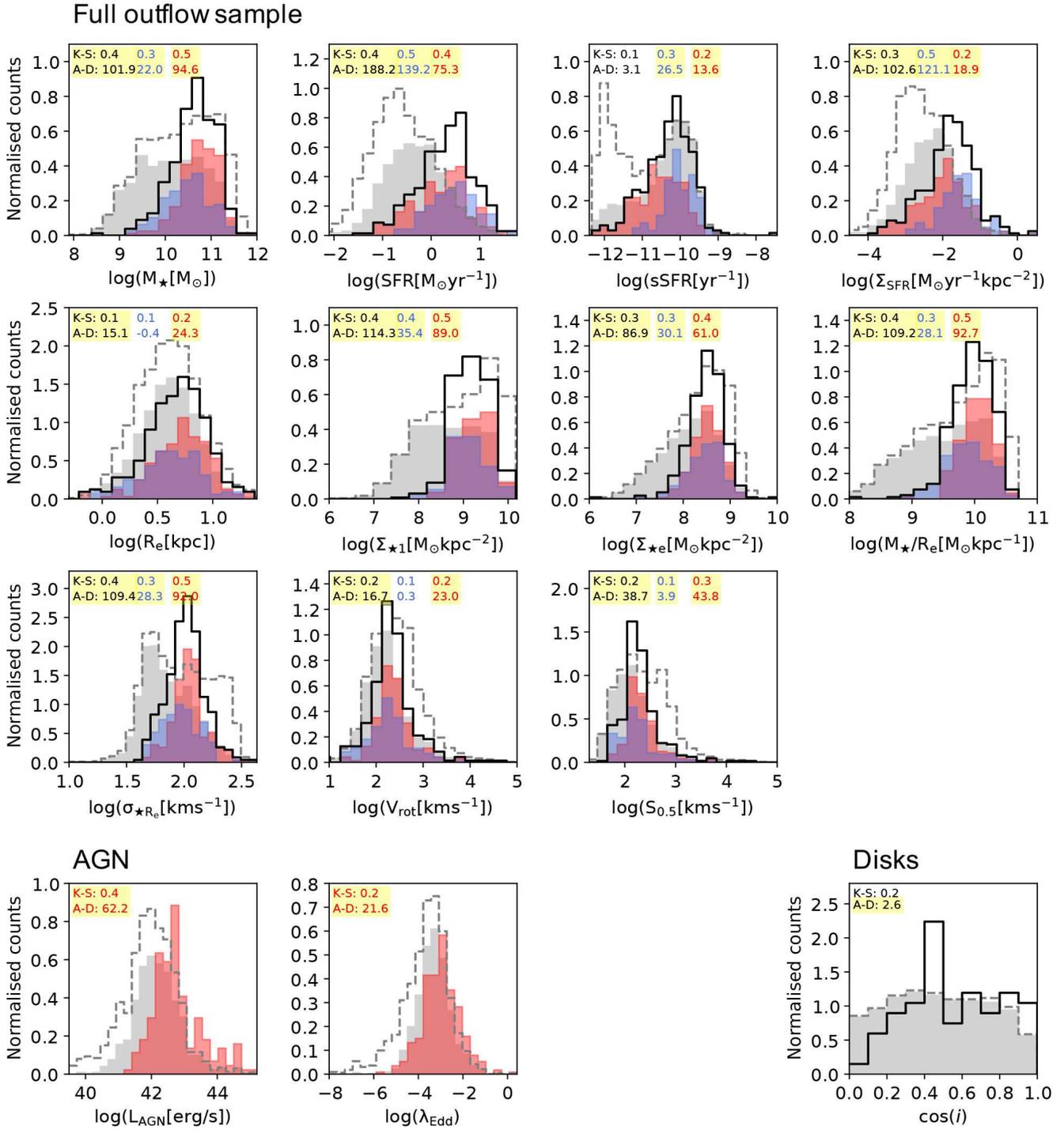}
\caption{Distribution of internal galaxy properties for galaxies exhibiting signatures of outflow activity ({\it black open histograms}) and the subpopulations with ({\it red}) and without ({\it blue}) AGN.  The underlying full MaNGA sample is shown as the grey-dashed distribution, and the MaNGA analysed (i.e., line-emitting) sample for which line profile fitting was carried out is indicated as the grey-shaded distribution. Inset numbers denote the Kolmogorov-Smirnov and Anderson-Darling test values comparing outflow galaxies to the underlying sample.  Highlighted values indicate tests with p-values < 0.05.  More detail on the internal galaxy properties considered is provided in the text.  AGN luminosities ($L_{\rm AGN}$) and Eddington ratios ($\lambda_{\rm Edd}$) are shown only for those galaxies with identified AGN.  Likewise, the comparison of inclinations ($\cos(i)$) is restricted to a secure morphological disk sample for which axial ratios provide a sound measure of inclination.  Note that the black, red and blue distributions are normalised to the black open distribution, and that both grey distributions are normalised to the grey-shaded histogram.}
\label{fig:incidence}
\end{figure*}

As outlined in Section\ \ref{method_sample.sec}, MaNGA DR15 counts 4239 galaxies with counterparts in the MPA-JHU database, of which 2744 objects feature line emission at a level significant enough that they entered our `analysed MaNGA sample' to which single- and double-Gaussian line profile fitting was applied.  Of these, just 322 objects feature significant evidence for broad-component line emission that can be interpreted as a signature of large-scale winds being driven from the galaxy (see criteria in Section\ \ref{outflow_sample.sec}) whilst ruling out a significant DIG contribution (Section \ref{DIG.sec}).  At 12\% of the analysed sample of line-emitting MaNGA galaxies (and 8\% of the full underlying MaNGA sample), outflows are thus only detectable within a small portion of the nearby galaxy sample considered here, even with the high-S/N resolved information offered by MaNGA.  For reference, outflow fractions reported for higher redshift samples are typically higher by factors of several, with exact numbers varying from $\sim20 - 30\%$ to $\gg 50\%$ depending on tracer and sample definition, despite the overall lower S/N and resolution of typical high-z observations complicating the detection of faint outflow signatures.  Outflow incidence is directly related to outflow strength (as stronger outflows
are more easily identified), and the higher incidence at higher redshift is in line with the stronger star-formation
and AGN activity around the epoch of cosmic noon ($1 \lesssim z \lesssim 3$).

In Figure \ref{fig:incidence}, we show for the analysed MaNGA sample and for the sample of galaxies with significant evidence for outflow activity, the logarithmic distributions of a set of key internal properties of the host galaxies.  In blue and red histograms, we further highlight the distribution of galaxy properties for the SF and AGN subsets of our fiducial outflow sample separately.  Specifically, we consider their stellar mass ($M_{\star}$), star formation rate (SFR), specific star formation rate ($\rm{sSFR} \equiv  \rm{SFR}/ \it{M_{\star}}$), star formation rate surface density ($\Sigma_{\rm SFR} \equiv \mathrm{SFR}/2\pi R_{\rm e}^{2}$), size ($R_{\rm e}$), stellar surface density within the central 1 kpc ($\Sigma_{\rm \star 1}$) or within $1 R_{\rm e}$ ($\Sigma_{\rm \star e} \equiv M_{\star} / 2\pi R_{\rm e}^2$), the stellar velocity dispersion within $1R_{\rm e}$ ($\sigma_{\rm \star R_{\rm e}}$), the rotational velocity taken from the H$\alpha$ velocity field and corrected for inclination on the basis of the galaxy's axial ratio ($V_{\rm rot}$), and a kinematic measure which adds the gaseous velocity dispersion in quadrature, which we dub $S_{0.5} = \sqrt{ 0.5 V_{\rm rot}^2 + \sigma_{\rm H\alpha, R_{\rm e}}^2 } $.\footnote{The $S_{0.5}$ parameter is calculated using the DAP entries \texttt{HA\_GVEL\_HI\_CLIP} and \texttt{HA\_GVEL\_LO\_CLIP} (with additional inclination correction) and \texttt{HA\_GSIGMA\_1RE}, which is different from the local disk velocity dispersion.  Although adopting its name from\ \cite{Weiner2006} because of its similarity in functional form, the quantity may not be directly comparable to the values reported by these authors due to different definitions of the velocity dispersions used.}
\par 
To test for the significance of differences between galaxy properties in the analysed MaNGA sample compared to the outflow sample and SF/AGN subsamples, we perform a two-sample Kolmogorov–Smirnov (K–S) test and an Anderson–Darling (A–D) test on each pair of histograms.  Relative to the K-S test, the A-D test assigns more weight to the tails of the distribution.  Values of both statistics are shown in each panel, and highlighted in yellow if the corresponding p-value is less than 0.05 indicating that the null hypothesis that the two samples are drawn from the same distribution is firmly rejected.
\par 
In all galaxy properties, except for sSFR, we find that the objects with evidence for outflow activity exhibit significantly different host galaxy properties when compared to the underlying MaNGA sample. \par 

Of notable interest is the enhanced incidence of outflows at the massive (high $M_{\star}$) end, and in the regime of high (central) stellar surface densities ($\Sigma_{\star 1}$, $\Sigma_{\rm \star e}$) and/or deep gravitational potentials ($M_{\star}/R_{\rm e}$, $\sigma_{\rm \star R_{\rm e}}$). All these trends of incidence are more pronounced for AGN outflows.  The increased incidence of AGN outflows at high mass and high central mass concentrations echoes findings at higher redshifts \citep{ForsterSchreiber2019}. \par

We find that outflows are detected relatively uniformly across a wide range in sSFR, and the offset in the sSFR distribution between the AGN and SF outflow subsamples is primarily reflecting the distribution of the underlying SFG and AGN-hosting populations across the SFR-M$_{\star}$ diagram. This is evident by comparing the SFR-M$_{\star}$ distribution of AGN and SF outflows in MaNGA (Section\ \ref{MstarSFR_scaling.sec}) to Figure 1 in \cite{Leslie2016}.\par

Unsurprisingly, the incidence of SF outflows is strongly tied to signatures of elevated star formation activity (SFR, sSFR and $\Sigma_{\rm SFR}$).  This connection is anticipated (see, e.g., \citealp{Ho2016}), as by lack of detectable nuclear activity the energy and momentum injection associated with the late stages of massive stellar evolution counts as the most plausible physical driver of large-scale winds.  Analysing the Na D $\lambda5889,\ 5895\AA$ neutral gas absorption profiles of a set of 405 inactive star-forming galaxies in MaNGA, \citet{Roberts-Borsani2020} likewise find an enhanced outflow incidence (and strength) at higher $\Sigma_{\rm SFR}$ (and $\Sigma_{\star}$).  We do note that SF outflows are observed at galaxy-averaged star formation surface densities well below $\Sigma_{\rm SFR} \sim 0.1 \ \mathrm{M_{\odot} \ yr^{-1}}$, proposed by \cite{Heckman2002} as an empirical threshold for so-called `superwinds' and interpreted by \cite{Ostriker2011} as the point where the energy from supernovae and galactic winds can overcome the gravity of the galaxy disk.  This may either be because surface densities local to the star-forming regions which act as the wind launching sites are higher than the total galaxy $\Sigma_{\rm SFR}$ referred to here, or simply because the MaNGA spectra allow probing more sensitively into galaxies of more moderate star formation surface density, with associated weaker levels of feedback (see also \citealp{Ho2016, Roberts-Borsani2019, Davies2019} and \citealp{Roberts-Borsani2020}).  We characterise dependencies on star formation activity among the outflow sample more quantitatively in Section\ \ref{scaling_relations.sec}.  Finally, it is noteworthy that in absolute SFR, the sample of AGN outflows is also skewed to higher values than the underlying population (yet the opposite is seen for specific SFRs).  One possible reason for this is that a correlation between SFR and stellar mass (the so-called main sequence of SFGs) introduces an indirect imprint of the aforementioned trend of AGN outflow incidence with mass.  One should further keep in mind that our AGN outflow sample spans a significant range in AGN luminosity ($L_{\rm AGN}$), down to low $L_{\rm AGN}$ where the energetics contributed by star formation and nuclear activity are on a par.  We therefore refrain from labelling the AGN outflow sample as a whole as AGN-driven outflows, and merely adopt the term AGN outflow to indicate the presence of an active nucleus.  For similar reasons, Section\ \ref{all_scaling.sec} establishes a master scaling relation for SF and AGN outflows jointly. \par
Turning to the bottom row of Figure\ \ref{fig:incidence}, the first two panels present for galaxies with evidence for AGN activity (Section\ \ref{SF_AGN_def.sec}) the logarithmic distributions of the AGN luminosity $L_{\rm AGN}$ and the Eddington ratio $\lambda_{\rm Edd}$. 
\par
We find evidence for outflows to be more prevalent among those AGN that have higher Eddington ratios and especially higher AGN luminosities.  A closer look at the distribution of AGN luminosities and accretion rates, as well as the relation between outflow rate and each of these AGN characteristics within the outflow sample is presented in Section \ref{AGN_scaling.sec}.

% FIG Correlation grid
\begin{figure*}
\centering
\includegraphics[width=\textwidth]{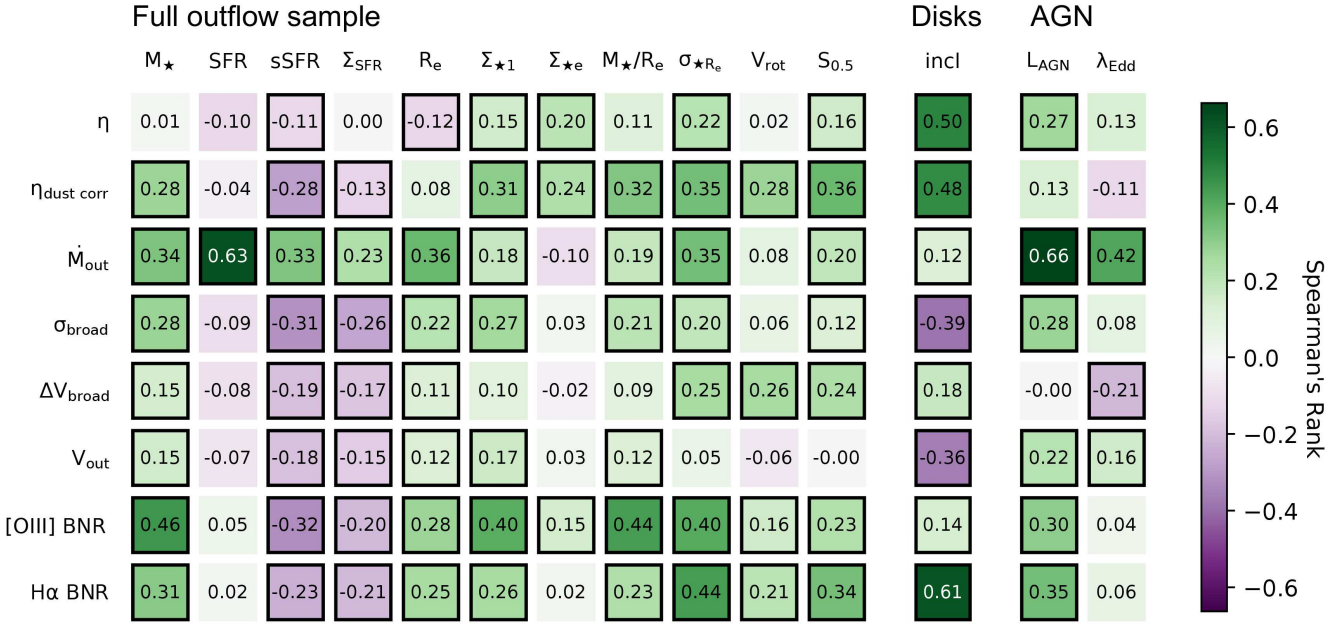}
\caption{Spearman's rank correlation coefficients quantifying the strength of correlations between outflow and host galaxy properties.  Black outlined boxes indicate a statistically significant correlation, i.e., p-value $< 0.05$.  The dependence on inclination is only investigated for the subsample of morphological disks, and the dependence on $L_{\rm AGN}$ and $\lambda_{\rm Edd}$ only for the outflow galaxies with AGN.  Strong correlations are identified, most notably between the observed outflow rate ($\dot{M}_{\rm out}$) and the intensity of its potential drivers: the star formation rate and AGN luminosity.}
\label{fig:correlation_grid}
\end{figure*}

Lastly, our analysis reveals only a minor hint of an inclination dependence of outflow signatures.  Here, the comparison is compiled for a subsample of star-forming galaxies (log(sSFR) > -11) with disk morphologies\footnote{Disk morphologies are defined as S\'{e}rsic indices < 2 and visual morphologies which are not classified as “odd looking” according to the Galaxy Zoo 2 classification by \citet{GalaxyZoo2}.} drawn from the outflow and underlying MaNGA samples, such that an estimate of inclination can reliably be derived from the axial ratio:
\begin{equation}
\cos(i) = \sqrt{\frac{(b/a)^2 - \rm{thickness}^2}{1 - \rm{thickness}^2}},
\label{eq:incl}
\end{equation}
where $b/a$ represents the galaxy's semi-minor to semi-major axis ratio.  A characteristic disk thickness (i.e., ratio of scale height over scale length) of 0.15 is assumed, appropriate for thin nearby disk galaxies.  We return to the absence of strong inclination dependencies in Section\ \ref{geometry.sec} when discussing the physical conditions of winds with an eye on outflow geometry.

\subsection{Outflow scaling relations}
\label{scaling_relations.sec}

The presence of outflow activity should not be considered as a binary flag, since a continuum of outflow strength and physical conditions exists, with a detectability threshold cutting through it.  To this end, we focus in the remainder of this Section on the outflow sample of 322 objects featuring broad-component line emission, and address their broad-component characteristics in light of the internal physical properties of the galaxies themselves.

\subsubsection{How outflow strength scales with internal galaxy properties}
\label{R_corr.sec}

In Figure\ \ref{fig:correlation_grid} we present the Spearman's rank correlation coefficients quantifying the strength of correlations between outflow properties and host galaxy properties.  Statistically significant correlations with p-values < 0.05 are denoted by a black-edged square.\footnote{The vast majority (90\%) of correlations with $p < 0.05$ remain significant if adopting a more stringent threshold for significance of $p < 0.01$.} 
Table \ref{tab:outflow_scaling} lists the corresponding best-fit parameters of a linear regression to the relationships between host galaxy properties and the mass-loading factor, mass outflow rate and outflow velocity, respectively.  Fits are performed using {\tt linmix} \citep{Kelly2007} which takes into account errors on both x and y values. Appendix\ \ref{appendix_SFAGN.sec} summarises the results from a linear regression applied to the SF and AGN outflow samples separately. \par

A first observation from Figure\ \ref{fig:correlation_grid} is that strong trends are present between mass outflow rate and a number of galaxy properties, most notably and importantly, with SFR and $L_{\rm AGN}$, i.e., the strength of the physical wind drivers.  With the functional form for $\dot{M}_{\rm out}$ given by Eq.\ \ref{eq:Mdot_out}, we note that these relations stem predominantly from a strong tie between the strength of the physical driver and the observed broad-component luminosity $L_{\rm H\alpha,B}$, more so than from a relation to the observed outflow velocity (which is insignificant in the case of SFR). \par 
When normalising the mass outflow rate by the SFR, i.e., considering the fiducial mass loading factor $\eta$ as computed using Eq.\ \ref{eq:eta}, the strong trend with SFR vanishes and even reverses sign, and the trend with $L_{\rm AGN}$ weakens.  For AGN-hosting galaxies it should further be noted that a portion of the (centrally emitted) narrow-component H$\alpha$ flux within the $R_{\rm out}$ aperture will not trace star formation but be excited by the AGN, hence rendering the values of $\eta$ as a lower limit to the actual mass loading.  Other effects that could lead to $\eta$ values that are underestimates of the true mass loading include outflow contributions in other phases than the warm ionised gas, and the possibility that dust attenuation factors do not divide out in Eq.\ \ref{eq:eta}.  The latter is hard to pin down observationally, but a motivation to account for additional attenuation to the broad-component gas, using the broad-component Balmer decrement $(H\alpha/H\beta)_B$, is discussed in Section\ \ref{attenuation.sec} and implemented in the alternative mass loading estimate labelled $\eta_{\rm dust\ corr}$.

A second observation from Figure\ \ref{fig:correlation_grid} is that most outflow-related observables or physical parameters based thereupon show significant positive correlations with galaxy stellar mass, or equally the central stellar velocity dispersion.

% FIG SFR-Mstar
\begin{figure*}
\centering
\begin{subfigure}[t]{\textwidth}
\includegraphics[width=0.48\textwidth]{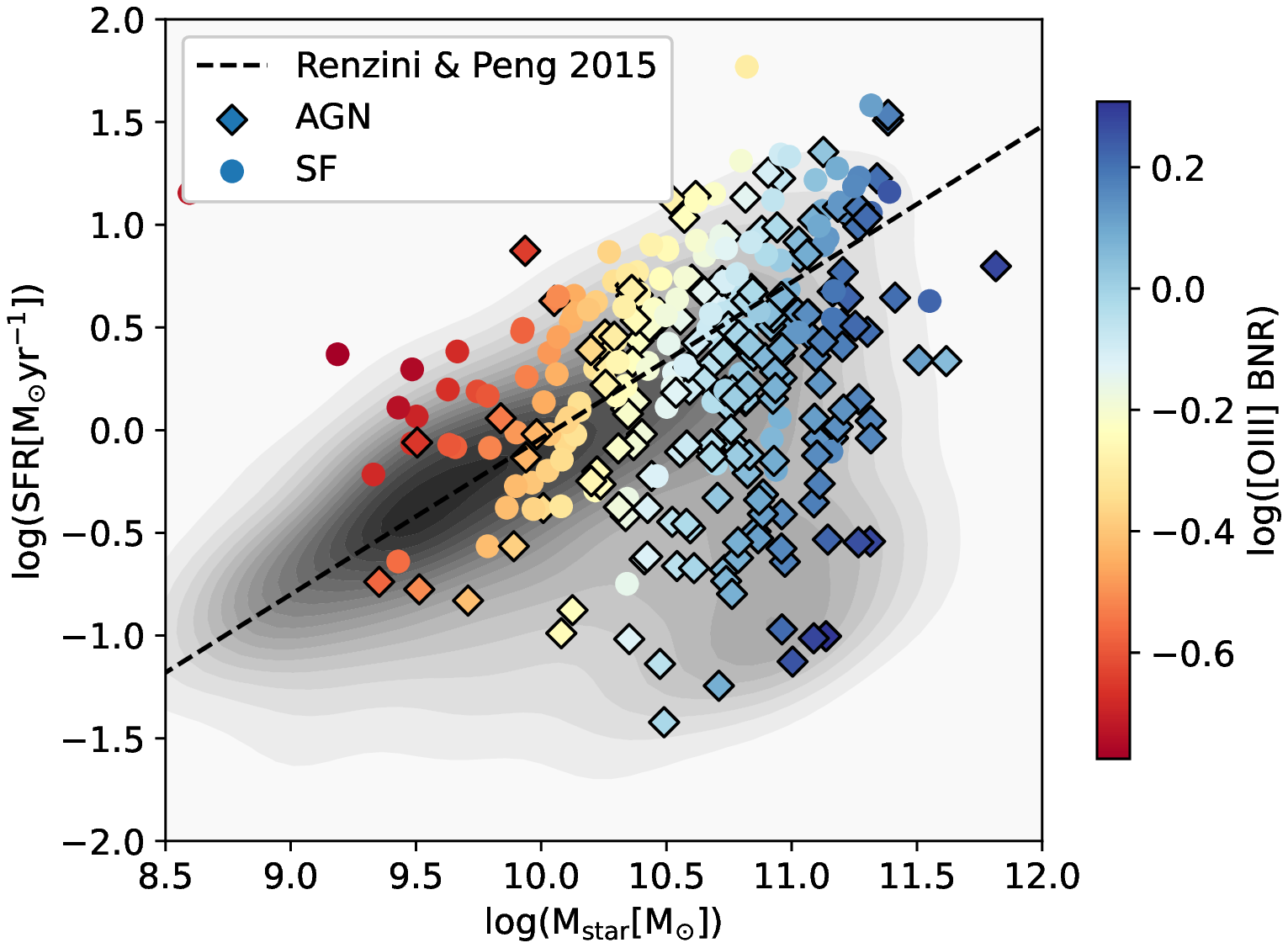}
\includegraphics[width=0.48\textwidth]{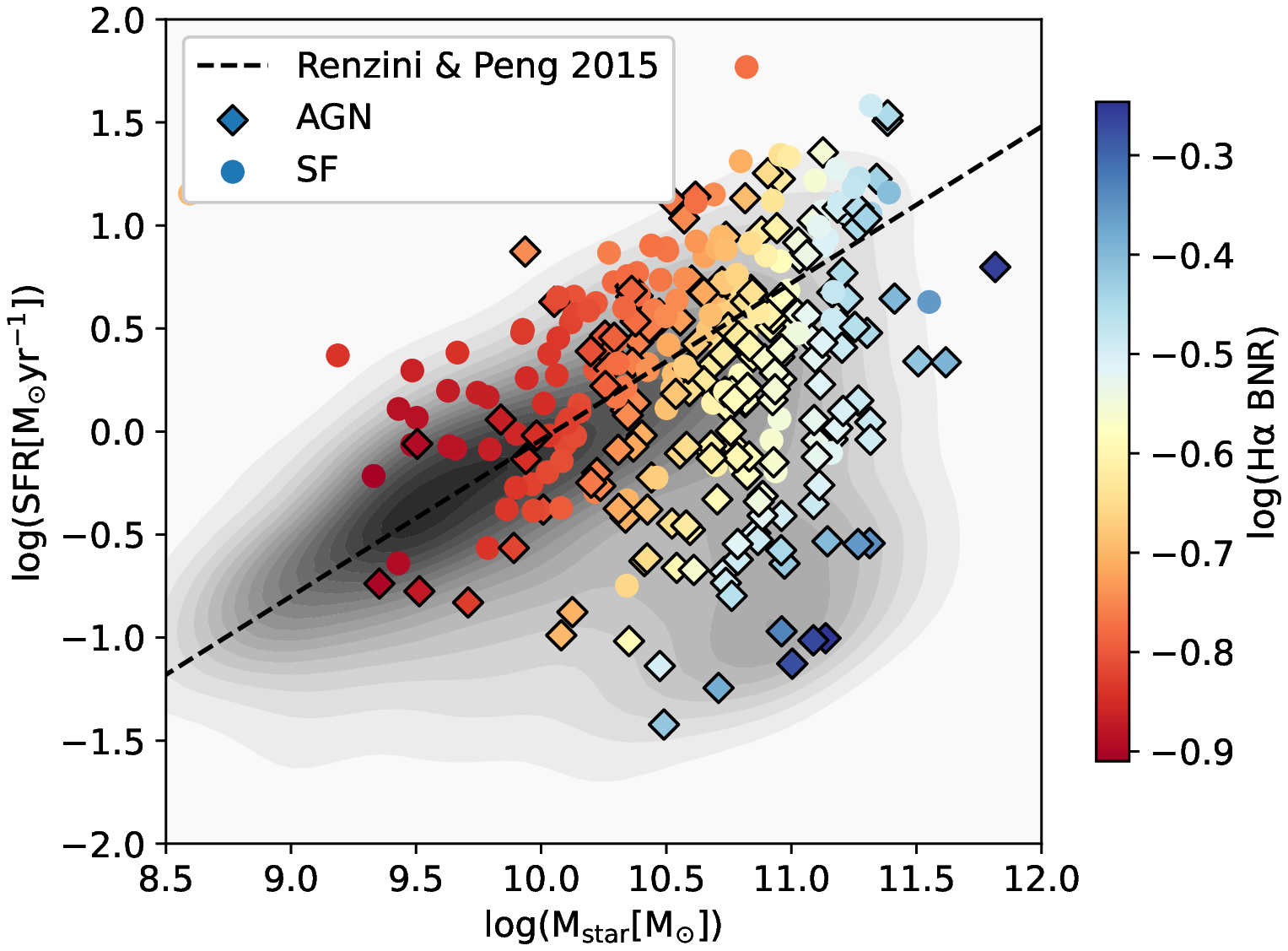}
\subcaption{Variation of [OIII] (left) and H$\alpha$ (right) BNR across the SFR-M$_{\star}$ plane assuming the same dust correction for disk/outflow.}
\end{subfigure}

\begin{subfigure}[t]{\textwidth}
\includegraphics[width=0.48\textwidth]{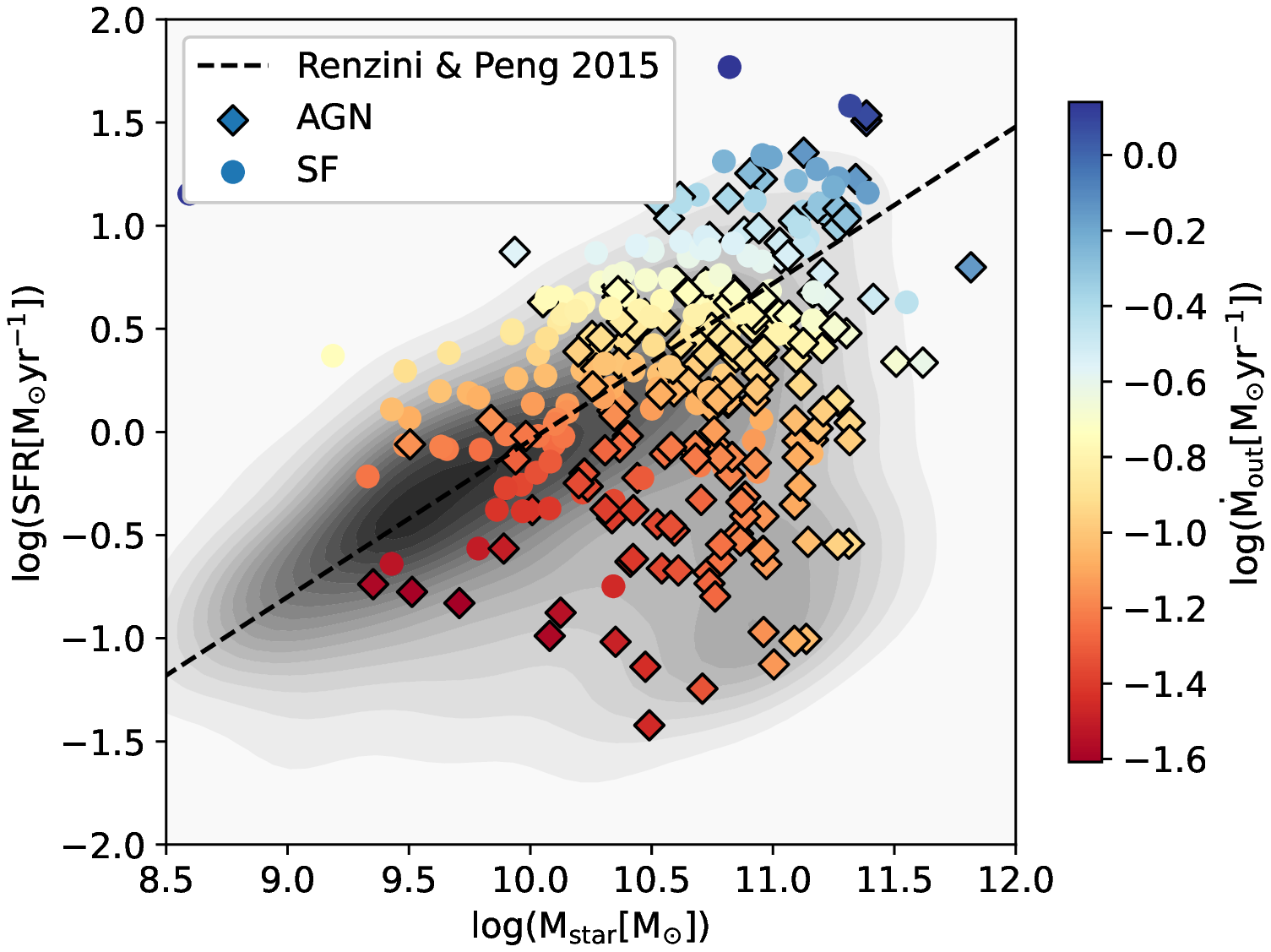}
\includegraphics[width=0.48\textwidth]{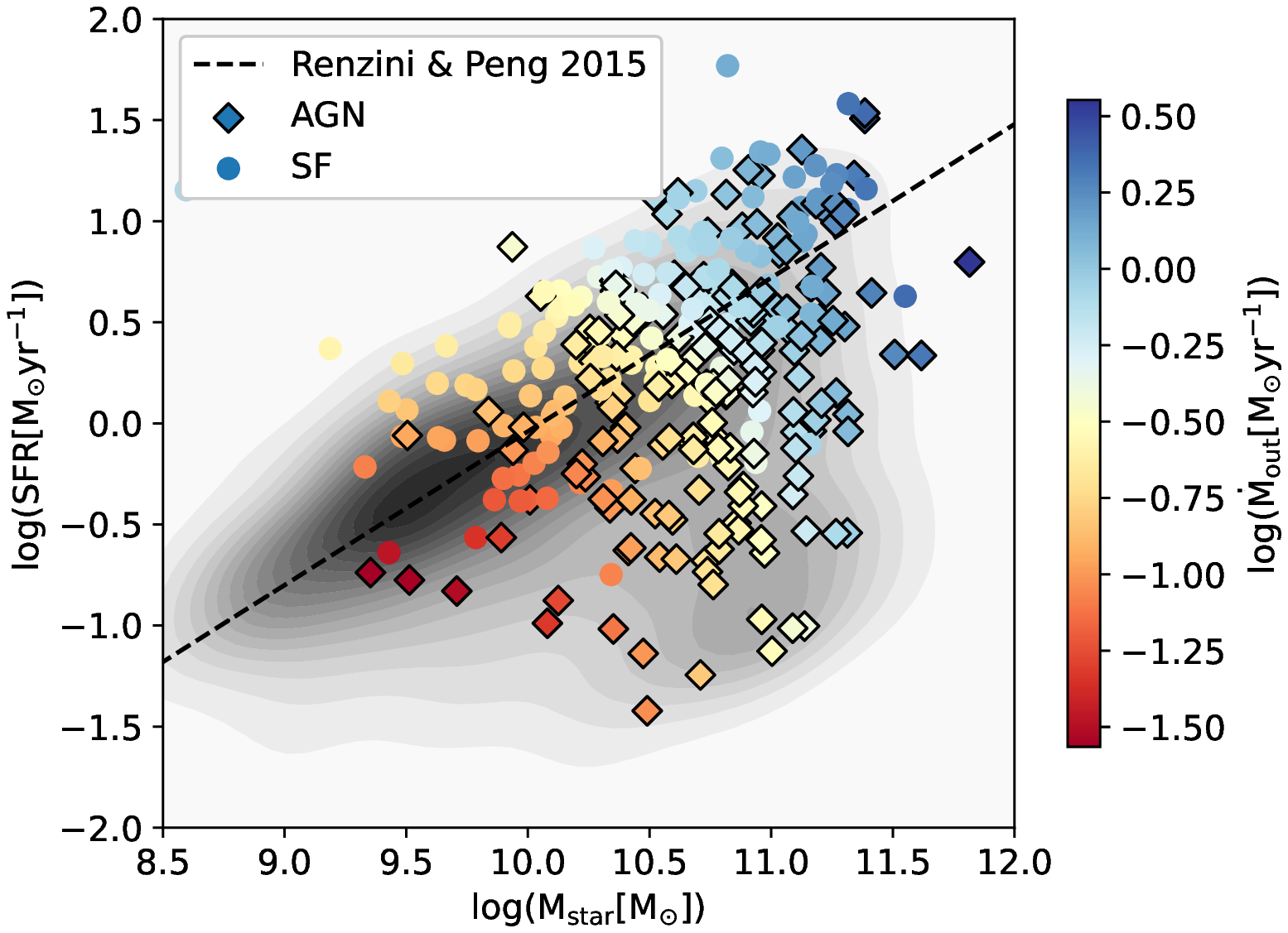}
\subcaption{Variation of mass outflow rate across the SFR-M$_{\star}$ plane assuming the same (left) and separate (right) dust correction for disk/outflow.}
\end{subfigure}

\begin{subfigure}[t]{\textwidth}
\includegraphics[width=0.48\textwidth]{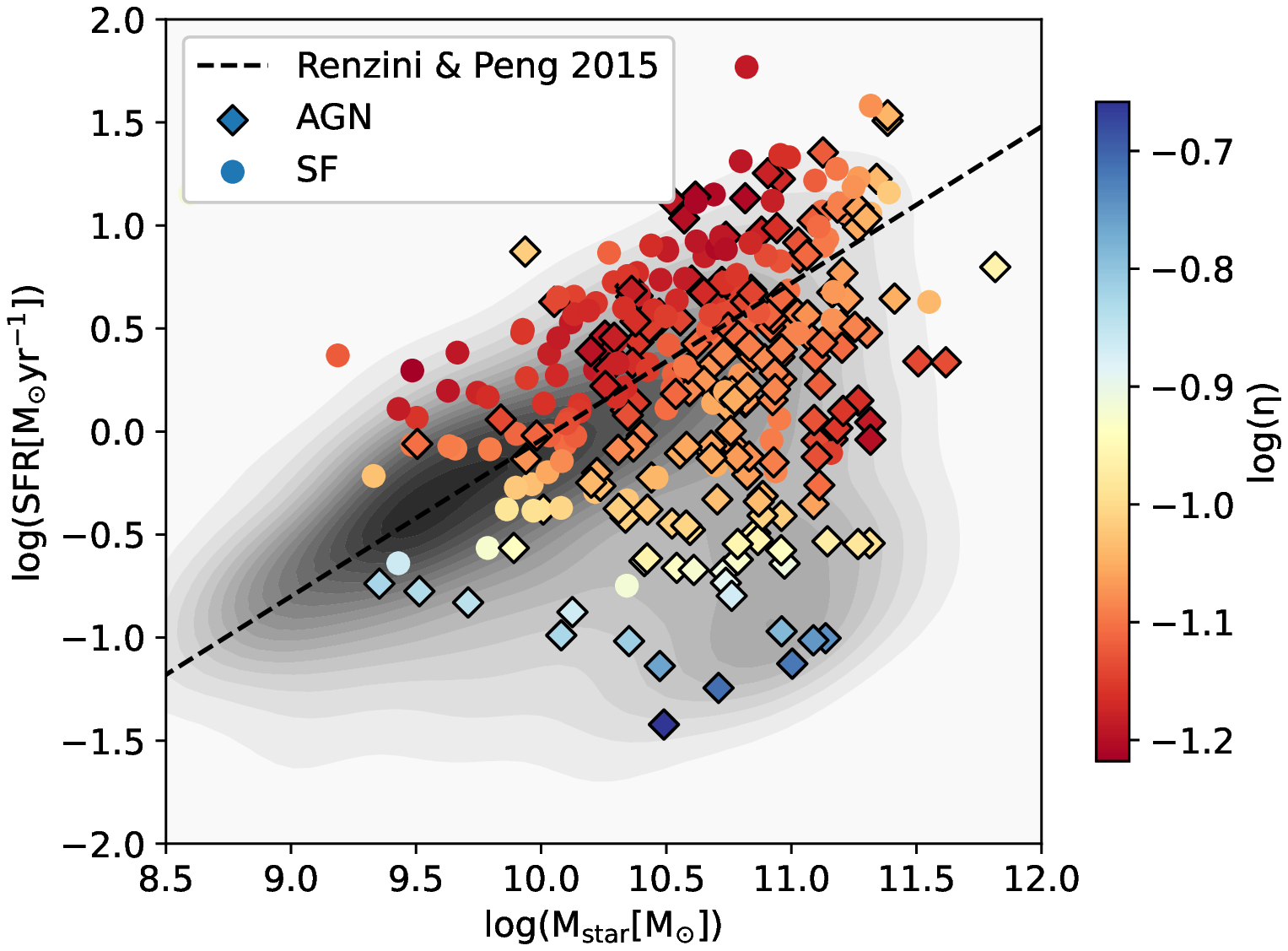}
\includegraphics[width=0.48\textwidth]{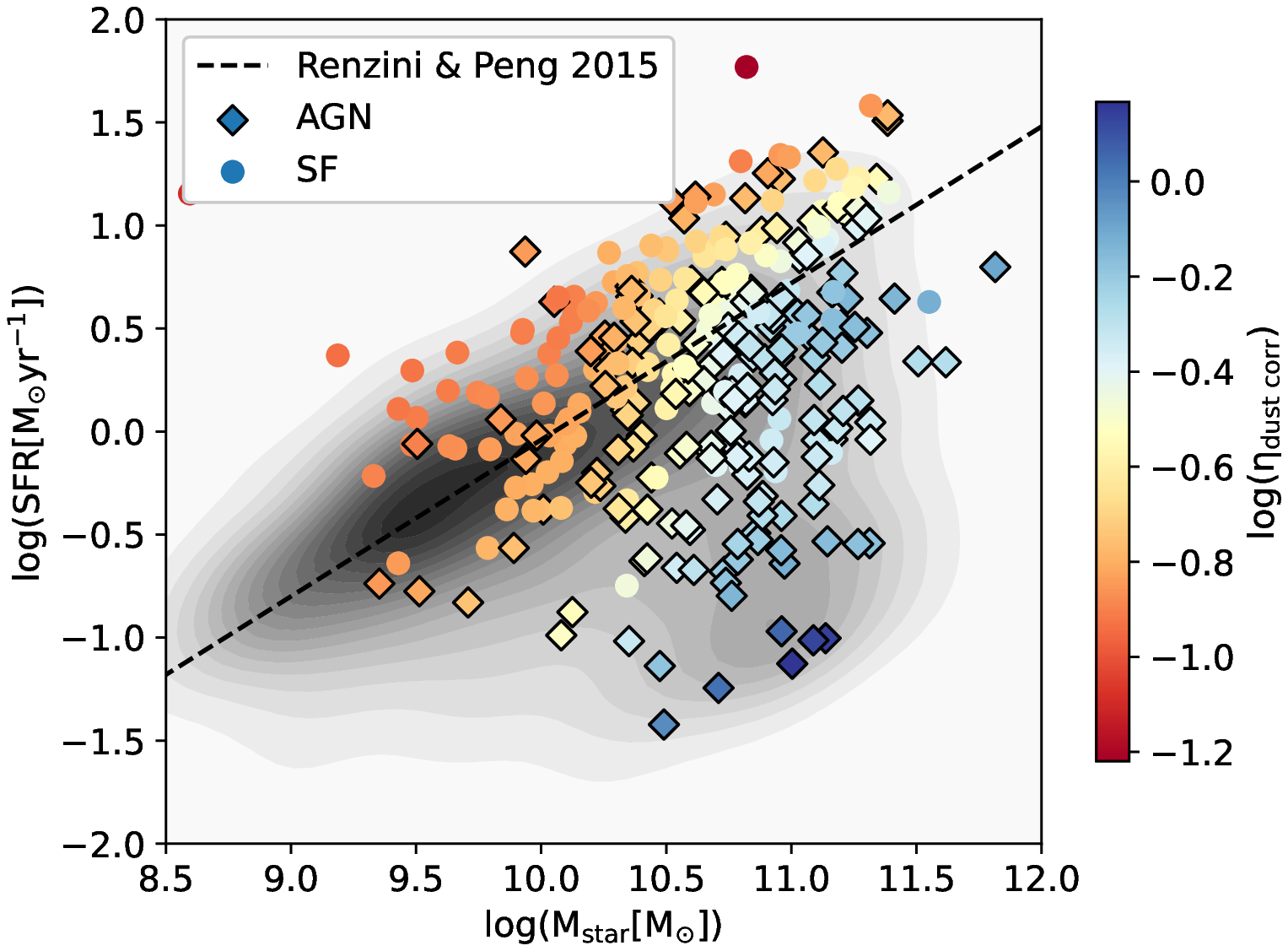}
\subcaption{Variation of mass loading across the SFR-M$_{\star}$ plane assuming the same (left) and separate (right) dust correction for disk/outflow.}
\end{subfigure}

\caption{Variation in outflow properties across the SFR - $M_{\star}$ diagram.  Galaxies from the outflow sample are colour-coded using an LOESS-smoothing on the outflow properties indicated in subcaptions. Where we drop the default assumption that disk and the outflowing gas are attenuated by the same amount ({\it middle right \& bottom right panels}), we compute the mass outflow rate and mass loading using a dust correction to the broad H$\alpha$ component based on the Balmer decrement of the broad component. This is as opposed to the default approach of adopting the Balmer decrement quantified from a single-Gaussian fit to the H$\alpha$ and H$\beta$ line profiles ({\it middle left \& bottom left panels}). Outflow galaxies with AGN are marked with a diamond symbol and extend further below the star-forming main sequence (dashed line; \citealt{Renzini2015}).  The underlying MaNGA sample for which a line profile analysis was carried out is depicted in grey shades.}
\label{fig:SFRMstar}
\end{figure*}

We find that these observations hold when considering just the SF or AGN subsets separately as shown in Figures \ref{fig:correlation_grid_SF} and \ref{fig:correlation_grid_AGN} (Appendix \ref{appendix_SFAGN.sec}).  In particular, the strong correlation between SFR and $\dot{M}_{\rm out}$ holds also for galaxies with AGN, which emphasises the idea that the energetic processes associated with star-formation and nuclear activity can both contribute significantly to the galactic outflow (see also \citealp{Talia2017}). \par
In the next sections, we examine a few of the relevant single-parameter dependencies more closely, before formulating a master scaling relation in Section\ \ref{all_scaling.sec} which captures the joint dependence of outflow strength on a few of the most critical host properties.

\subsubsection{Scaling with stellar mass and SFR}
\label{MstarSFR_scaling.sec}

Stellar mass is arguably one of the most fundamental parameters defining a galaxy.  Importantly, it also represents the most robustly measurable product of stellar population modelling, available for many reference samples (with total $M_{\star}$ in our case taken from the MPA-JHU database) and to outflow studies spanning wide ranges in lookback time.  We therefore consider it first, despite not being the best predictor of outflow strength revealed by our analysis.  As illustrated in Figures\ \ref{fig:incidence} and\ \ref{fig:SFRMstar}, the dynamic range in stellar mass sampled by our outflow sample covers over two orders of magnitude, with SF objects probing further down below $\log(M_{\star})<10$ whereas AGN objects are more prevalent at $\log(M_{\star}) \gtrsim 10.5$.

The strongest correlation of outflow properties with $M_{\star}$ is found for the [OIII] BNR.  Whereas this observable does not enter directly into any of the equations detailed in Section\ \ref{outflow_def.sec}, the relative strength of the broad-component feature to [OIII] relative to that seen in the Balmer lines helps characterise the outflowing gas component as the kinematic moments of the different optical emission lines are tied in the fitting procedure.  Assessed over the full outflow sample, we find $\rm BNR_{\rm [OIII]}$ to be on average $\sim 3$ times higher than $\rm BNR_{\rm H\alpha}$, with object-to-object variations varying from $\rm BNR_{\rm [OIII]}/BNR_{\rm H\alpha}$ near unity to a factor $\sim 10$.

\begin{figure}
\centering
\includegraphics[width=0.98\linewidth]{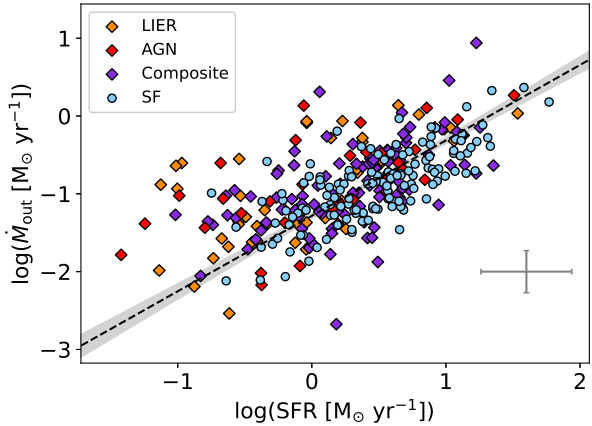}
\caption{Mass outflow rate versus star formation rate where our SF and AGN outflow subsamples are shown using circular and diamond shaped markers, respectively ({\it Plot style as in Figure\ \ref{fig:Rout}}).}
\label{fig:SFR}
\end{figure}

Linear regression of the $\dot{M}_{\rm out} - M_{\star}$ relation returns a positive trend with power-law slope of $0.37 \pm 0.06$ (Table\ \ref{tab:outflow_scaling}), but with substantial scatter ($\sim 0.51$ dex).  Evaluating the variation in mass outflow rate across the SFR - $M_{\star}$ plane, Figure\ \ref{fig:SFRMstar} furthermore reveals that the mass dependence of $\dot{M}_{\rm out}$ largely reflects an indirect imprint of a much stronger relation between $\dot{M}_{\rm out}$ and SFR, where the latter is broadly related to stellar mass for galaxies in the outflow sample.  That is, at fixed SFR little evidence for variations in outflow rate with galaxy mass is found.  All outflow properties displayed in Figure \ref{fig:SFRMstar} are adaptively smoothed using the two-dimensional locally weighted regression (LOESS) method \citep{Cleveland1988, Cappellari2013}.  The projection of the SFR - $M_{\star}$ diagram along stellar mass, showing mass outflow rate as a function of SFR is presented in Figure\ \ref{fig:SFR}, featuring a power-law slope of $0.97 \pm 0.07$ and scatter of $\sim 0.48$ dex.  Slopes of linear relations fit to the SF and AGN subsamples individually are both consistent with unity (when accounting for the slope error at the $2\sigma$ level), with the AGN extending to lower SFR values and in the regime of low star-formation activity counting more outliers above the linear relation, reflected in a larger scatter ({0.33 dex for SF outflows versus 0.60 dex for AGN outflows}; see tables \ref{tab:outflow_scaling_SF} and \ref{tab:outflow_scaling_AGN}).   
\cite{Arribas2014} find a similar yet slightly steeper slope of $\dot{M}_{\rm out} \propto \rm SFR^{1.11}$ for a sample of low-redshift luminous and ultra-luminous infrared galaxies without AGN activity.  Using H$\alpha$-based SFR estimates they find a somewhat higher zero point of the relation, possibly associated with their (U)LIRGs being significantly more compact at fixed SFR, and the adopted $R_{\rm out}$ (0.7 kpc) being smaller.  We further note that a higher $n_{\rm e,B}$ (315 cm$^{-3}$) is found for their compact (U)LIRGs. Additionally, as illustrated in their paper, the H$\alpha$-based SFRs they adopt do not recover the full amount of star formation revealed bolometrically via far-infrared measurements, an effect commonly expected from saturation of dust attenuation probes in the regime of large column densities \citep[see, e.g.,]{Wuyts2011}.  When adopting the SFR($L_{\rm IR}$) they provide instead, their sample shifts to lie closely along our inferred $\dot{M}_{\rm out}$ - SFR relation, extending the dynamic range to the higher SFR regime.  A comparison to the more luminous samples presented by \citet{Arribas2014} and \citet{Fluetsch2019} is presented in Appendix\ \ref{appendix_literature.sec}. \par 

In terms of mass loading of the winds, the LOESS regression on our fiducial estimates of $\eta$ reveal systematic variations in the 5 - 30\% range across the SFR - $M_{\star}$ plane with moderate mass galaxies occupying the upper half of the star-forming main sequence featuring the lowest mass-loading factors, and AGN outflows located below the main sequence the highest ones.\footnote{We remind the reader that the mass-loading factor $\eta$ is derived entirely based on the MaNGA spectrum extracted within an aperture of size $R_{\rm out}$ whereas the galaxy-integrated SFR throughout this paper is adopted from the MPA-JHU database.  The latter include a multi-band photometry-based aperture correction to account for star formation happening outside the SDSS fibre, and for galaxies with AGN signatures in the SDSS spectra adopt the D4000 break rather than Balmer emission lines as SFR diagnostic.}  The dependence of $\eta$ on SFR is negative, with a power-law slope of $-0.12 \pm 0.08$.  Of particular interest to theory \citep[e.g.,]{Lilly2013, Dave2017} is the dependence of $\eta$ on $M_{\star}$ in star formation feedback.  We find no significant correlation, but discuss in Section \ref{fate.sec} why the mass loading factor quantified here may not be directly related to the effective mass loading which enters these gas regulator models.

Finally, we note that when dropping the default assumption that gas belonging to the disk and wind components are attenuated by the same amount, a significantly different pattern of and range in mass loading factors as well an overall increase in the mass outflow rate is observed. As discussed in Section\ \ref{attenuation.sec}, broad-component gas in some of the galaxies may suffer enhanced levels of attenuation, raising the estimated $\eta_{\rm dust\ corr}$ accordingly, although the effect is challenging to quantify on an individual object basis. This difficulty is primarily due to the uncertainly in quantifying the broad component Balmer ratio for individual objects, where the weakness of the H$\beta$ line can make the broad component difficult to detect.

% FIG SFR

% FIG AGN
\begin{figure}
%\begin{subfigure}[ht!]{\linewidth}
\centering
\includegraphics[width=0.98\linewidth]{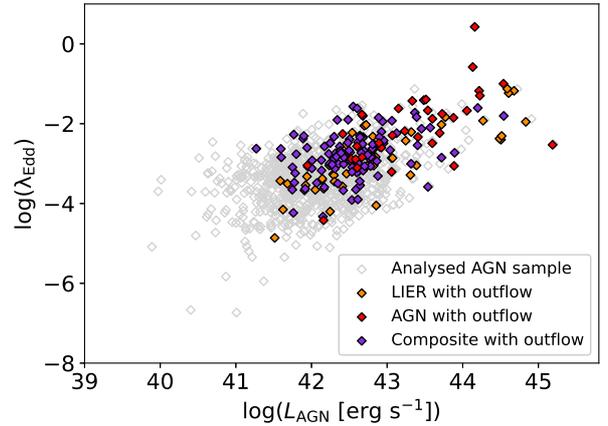}
\includegraphics[width=0.98\linewidth]{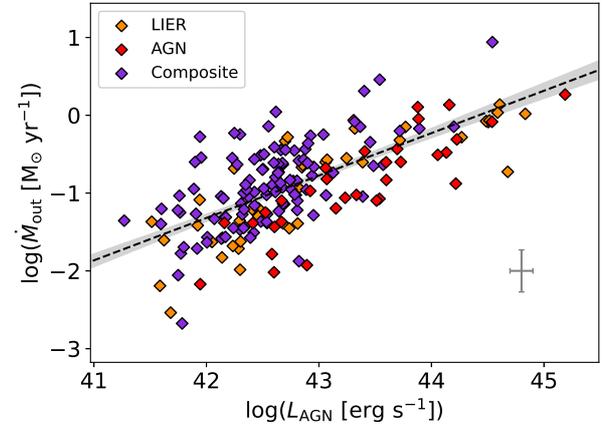}
\includegraphics[width=0.98\linewidth]{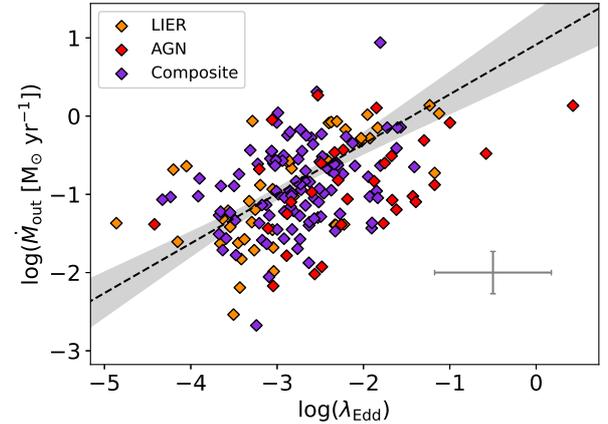}
\caption{{\it Top:} Incidence of outflows ({\it coloured diamonds}) among MaNGA AGNs ({\it all diamonds}) in the $\lambda_{\rm Edd}$ - $L_{\rm AGN}$ plane.  Significant evidence for outflows is more prevalent at higher AGN luminosities.  {\it Middle and bottom:} Mass outflow rates contrasted to the [OIII]-based AGN luminosity and Eddington ratio ({\it plot style as in Figure\ \ref{fig:Rout}}).}
\label{fig:AGN}
%\end{subfigure}
\end{figure}

\subsubsection{Scaling with nuclear activity}
\label{AGN_scaling.sec}

We measure the (normalised) strength of nuclear activity and the energy injection rate associated with it via the Eddington ratio and AGN luminosity, respectively.  The top panel in Figure \ref{fig:AGN} illustrates the incidence of outflows among active galaxies in a diagram of $\lambda_{\rm Edd}$ versus $L_{\rm AGN}$.  As hinted already in Figure \ref{fig:incidence}, we find that outflows are more prevalent at higher AGN luminosities, and, more subtly, at higher $\lambda_{\rm Edd}$.  

In the middle and bottom panels of Figure \ref{fig:AGN}, we show mass outflow rates contrasted to the [OIII]-based AGN luminosity and Eddington ratio.  Linear regression of the $\dot{M}_{\rm out} - L_{\rm AGN}$ relation gives a positive trend with a power-law slope of $0.55 \pm 0.04$, with an intrinsic scatter of $\sim 0.43$ dex (Table \ref{tab:outflow_scaling}). The relation with $\lambda_{\rm Edd}$ has a power-law index of $0.64 \pm 0.14$, and features a larger scatter ($\sim 0.58$ dex). We again emphasise that our sample extends down to low luminosity AGN, some of which would not be identifiable as such from their galaxy-integrated spectrum, and varying contributions of other, star formation related drivers may contribute to the observed scatter and subunity trendline. We see a hint of this effect in Figure \ref{fig:AGN}, where we find that composites feature enhanced outflow rates compared to AGN/LIERs at a given AGN luminosity.  This is most plausibly interpreted by their outflow driving having relatively more significant contributions from star formation feedback motivating the master scaling approach in Section \ref{all_scaling.sec} which aims to encompass the different driving mechanisms simultaneously. When considering the AGN and LIER population solely, the best-fit linear relation between outflow rate and AGN luminosity becomes steeper with a slope of $0.62 \pm 0.04$ and a tighter scatter of $\sim 0.37$ dex.  \par

Measurements by \citet{Fluetsch2019} of mass outflow rates in the molecular phase show a slightly steeper dependence on $L_{\rm AGN}$, of slope 0.68, which can possibly be attributed to enhanced relative contributions of the molecular phase to the outflowing gas at high $L_{\rm AGN}$, as reported by the same authors.  In line with our results, \citet{Fluetsch2019} find substantially more scatter in the relation with $\lambda_{\rm Edd}$ (compared to $L_{\rm AGN}$), despite the fact that this quantity is fundamental in the energy-driven and radiation pressure-driven scenarios often invoked for AGN outflows \citep{King2015, Ishibashi2018}.  Uncertainties in the estimates of black hole mass, based on the $M_{\rm BH} - \sigma$ relation in our case and entering in the denominator of $\lambda_{\rm Edd}$ via the Eddington luminosity, may contribute to the observed scatter.

We also estimate the momentum and energy ratio of the ionised AGN winds, defined as $\dot{p}_{\rm out}/(L_{\rm AGN}/c)$ and $\dot{E}_{\rm out}/L_{\rm AGN}$, respectively, where the outflow momentum flux $\dot{p}_{\rm out} = \dot{M}_{\rm out} v_{\rm out}$ and $L_{\rm AGN}/c$ represents the AGN radiative momentum rate. The kinetic power of the outflow is given by $\dot{E}_{\rm out} \equiv \frac{1}{2} \dot{M}_{\rm out} v_{\rm out}^{2}$.  We find significant correlations between the outflow momentum flux and $L_{\rm AGN}/c$ and between kinetic power and $L_{\rm AGN}$.  Typical inferred momentum ratios of the ionised gas winds are of order unity, but objects in our AGN outflow sample are found to span the full $\sim 0.1 - 20$ range.  Energy ratios computed based on the ionised gas phase alone typically fall below $\sim 10^{-1.5}$, and reach two to three orders of magnitude further down from there, consistent with the low kinetic coupling efficiencies reported by \citet{Wylezalek2020}.

\begin{table*} %[] 
\centering 
\resizebox{\linewidth}{!}{%
\begin{tabular}{l|rrrrrrrrrrr} 
\hline \hline 
Galaxy property & $A\eta$ & $B\eta$ & $\Delta\eta$ & $A{\dot{M}_{\rm out}}$ & $B{\dot{M}_{\rm out}}$ & $\Delta{\dot{M}_{\rm out}}$ & $A{v_{\rm out}}$ & $B{v_{\rm out}}$ & $\Delta{v_{\rm out}}$ & $x_0$ \\
\hline 
\multicolumn{11}{c}{Full outflow sample} \\ 
\hline 
log($M_{\star} \ [\rm M_{\odot}]$)& $0.01 \pm 0.06$ & $-1.04 \pm 0.03$ & $0.48$ & $0.37 \pm 0.06$ & $-0.75 \pm 0.04$ & $0.51$ & $0.05 \pm 0.02$ & $2.62 \pm 0.01$ & $0.14$ & $11.0$ \\
log(SFR [$\rm M_{\odot} \ yr^{-1}$])& $-0.12 \pm 0.08$ & $-1.00 \pm 0.04$ & $0.48$ & $0.97 \pm 0.07$ & $-1.28 \pm 0.04$ & $0.48$ & $-0.02 \pm 0.03$ & $2.61 \pm 0.01$ & $0.14$ & $0.0$ \\
log(sSFR \ [yr$^{-1}]$)& $-0.14 \pm 0.09$ & $-1.06 \pm 0.03$ & $0.48$ & $0.51 \pm 0.10$ & $-0.82 \pm 0.03$ & $0.53$ & $-0.09 \pm 0.03$ & $2.58 \pm 0.01$ & $0.14$ & $-10.0$ \\
log($\Sigma_{\rm SFR} \ [\rm M_{\odot} \ yr^{-1} \ kpc^{-2}]$)& $0.03 \pm 0.07$ & $-1.05 \pm 0.03$ & $0.48$ & $0.26 \pm 0.08$ & $-0.97 \pm 0.04$ & $0.53$ & $-0.05 \pm 0.02$ & $2.61 \pm 0.01$ & $0.14$ & $-2.0$ \\
log($R_{\rm e}$ [kpc])& $-0.20 \pm 0.10$ & $-1.11 \pm 0.04$ & $0.47$ & $0.80 \pm 0.11$ & $-0.63 \pm 0.04$ & $0.50$ & $0.06 \pm 0.03$ & $2.62 \pm 0.01$ & $0.14$ & $1.0$ \\
log($\Sigma_{\star 1} \ [\rm M_{\odot} \ kpc^{-2}]$)& $0.20 \pm 0.07$ & $-1.07 \pm 0.03$ & $0.47$ & $0.26 \pm 0.08$ & $-0.94 \pm 0.03$ & $0.53$ & $0.07 \pm 0.02$ & $2.59 \pm 0.01$ & $0.14$ & $9.0$ \\
log($\Sigma_{\rm \star e} \ [\rm M_{\odot} \ kpc^{-2}]$)& $0.26 \pm 0.08$ & $-0.90 \pm 0.05$ & $0.47$ & $-0.21 \pm 0.09$ & $-1.00 \pm 0.06$ & $0.53$ & $0.02 \pm 0.02$ & $2.61 \pm 0.01$ & $0.14$ & $9.0$ \\
log($M_{\star} / R_{\rm e} \ [\rm M_{\odot} \ kpc^{-1}])$& $0.20 \pm 0.09$ & $-1.03 \pm 0.03$ & $0.47$ & $0.29 \pm 0.11$ & $-0.88 \pm 0.03$ & $0.53$ & $0.07 \pm 0.03$ & $2.60 \pm 0.01$ & $0.14$ & $10.0$ \\
log($\sigma_{\rm \star R_e} \ [\rm km \ s^{-1}]$)& $1.44 \pm 0.30$ & $-1.13 \pm 0.03$ & $0.48$ & $2.38 \pm 0.31$ & $-1.05 \pm 0.04$ & $0.53$ & $0.30 \pm 0.09$ & $2.58 \pm 0.01$ & $0.15$ & $2.0$ \\
log($V_{\rm rot} \  [\rm km \ s^{-1}])$& $-0.03 \pm 0.06$ & $-1.03 \pm 0.04$ & $0.48$ & $0.04 \pm 0.07$ & $-0.91 \pm 0.04$ & $0.54$ & $-0.00 \pm 0.02$ & $2.60 \pm 0.01$ & $0.14$ & $2.0$ \\
log($S_{0.5} \ [\rm km \ s^{-1}]$)& $0.04 \pm 0.08$ & $-1.05 \pm 0.04$ & $0.48$ & $0.13 \pm 0.08$ & $-0.94 \pm 0.04$ & $0.54$ & $0.02 \pm 0.02$ & $2.59 \pm 0.01$ & $0.14$ & $2.0$ \\
\hline
\multicolumn{11}{c}{Disks} \\
\hline
${\rm cos(i \ [\deg])}$ & $-0.97 \pm 0.22$ & $-1.54 \pm 0.11$ & $0.38$ & $-0.05 \pm 0.25$ & $-0.92 \pm 0.13$ & $0.46$ & $0.27 \pm 0.08$ & $2.71 \pm 0.04$ & $0.15$ & $2.0$ \\
\hline
\multicolumn{11}{c}{AGN} \\
\hline
log($L_{\rm AGN} \ [\rm erg \ s^{-1}]$)& $0.19 \pm 0.05$ & $-0.88 \pm 0.04$ & $0.47$ & $0.55 \pm 0.04$ & $-0.78 \pm 0.03$ & $0.43$ & $0.05 \pm 0.01$ & $2.64 \pm 0.01$ & $0.14$ & $43.0$ \\
log($\lambda_{\rm Edd}$)& $0.13 \pm 0.12$ & $-0.95 \pm 0.04$ & $0.48$ & $0.64 \pm 0.14$ & $-0.99 \pm 0.05$ & $0.58$ & $0.04 \pm 0.04$ & $2.62 \pm 0.01$ & $0.15$ & $-3.0$ \\
\hline \hline  
\end{tabular}  
}  
\caption{Scaling relations between ionised gas outflow properties and internal galaxy properties including mass, star formation activity, structure, orientation (for star-forming disks only) and AGN luminosity and Eddington ratio (for active galaxies only). With the exception of cosine inclination, linear regression of the form $\log({\rm outflow\ property}) = A [\log({\rm galaxy\ property}) - x_0] + B$ is carried out in $\log-\log$ space with errors representing the central 68th percentile of the posterior distribution on slope and intercept. Standard deviations of residuals from the best-fit linear relation are denoted as $\Delta$.  Fit parameters are determined using \texttt{linmix} \citep{Kelly2007}.}
\label{tab:outflow_scaling}
\end{table*}

\subsubsection{An all-encompassing outflow scaling relation}
\label{all_scaling.sec}

Figure \ref{fig:correlation_grid} presents a number of strong correlations between outflow properties, in particular mass outflow rate, and host galaxy properties.  Given that many internal galaxy properties correlate with one another, an observed scaling with one parameter does however not necessarily reveal any form of causation.  In this section we therefore aim to construct an `all-encompassing outflow scaling relation' where the mass outflow rate of the ionised gas in MaNGA galaxies is described with the least possible scatter by expressing its dependence on a number of galaxy properties jointly.  

Here we explore three forms of such a master outflow scaling relation of increasing complexity from Equation \ref{eq:master_3par} to Equation \ref{eq:master_scaling_delMS}.  

\begin{equation}
   \log(\dot{M}_{\rm out}) = a\ + b \ \log\left(\dfrac{M_{\star}}{5\times 10^{10}\ \mathrm{M_{\odot}}}\right) + c \ \log\left(\dfrac{\rm SFR}{\rm SFR_{\rm MS}}\right)  
\label{eq:master_3par} 
\end{equation}

\begin{align}
   \log(\dot{M}_{\rm out}) = &\ a\ + \nonumber \\
   &\ b \ \log\left(\dfrac{M_{\star}}{5\times 10^{10}\ \mathrm{M_{\odot}}}\right) +  \nonumber \\
   &\ c \ \log\left(\dfrac{\rm SFR}{\rm SFR_{\rm MS}}\right) + \nonumber \\
   &\ d \ \log \left(\dfrac{R_{\rm e}}{5\ \rm{kpc}}\right) 
\label{eq:master_4par} 
\end{align}

\begin{align}
   \log(\dot{M}_{\rm out}) = &\ b \ \log\left(\dfrac{M_{\star}}{5\times 10^{10}\ \mathrm{M_{\odot}}}\right) + \nonumber \\
   &\ d \ \log \left(\dfrac{R_{\rm e}}{5\ \rm{kpc}}\right) + \nonumber \\
   &\ \log\left[10^a \left(\dfrac{\rm SFR}{\rm SFR_{\rm MS}}\right)^c + 10^{e}\left(\dfrac{L_{\rm AGN}}{1\times 10^{43}\ \rm{erg}\ \rm{s}^{-1}}\right)^f \right]
\label{eq:master_scaling_delMS} 
\end{align}

Here, $\rm SFR_{\rm MS}$ refers to the star formation rate of a galaxy residing on the star-forming main sequence relation as parameterised by \citet{Renzini2015}:
\begin{equation}
\log(\mathrm{SFR_{\rm MS}}) = 0.76 \log(M_{\star}/\mathrm{M_{\odot}}) - 7.64
\end{equation}
While facilitating a more straightforward reading of the mass dependence of outflow rates for typical (i.e., main sequence) SFGs, the use of $\rm SFR_{\rm MS}$ as a normalisation factor does imply that mass implicitly enters in two of the terms in Eqs.\ \ref{eq:master_3par}-\ref{eq:master_scaling_delMS}.  We therefore also explore an equivalent to Eq.\ \ref{eq:master_scaling_delMS} in which the role of mass and star formation are fully decoupled:

\begin{align}
   \log(\dot{M}_{\rm out}) = &\ b \ \log\left(\dfrac{M_{\star}}{5\times 10^{10}\ \mathrm{M_{\odot}}}\right) + \nonumber \\
   &\ d \ \log \left(\dfrac{R_{\rm e}}{5\ \rm{kpc}}\right) + \nonumber \\
   &\ \log\left[10^a \left(\dfrac{\rm SFR}{3 \ \mathrm{M_{\odot} \ yr^{-1}}}\right)^c + 10^{e}\left(\dfrac{L_{\rm AGN}}{1\times 10^{43}\ \rm{erg}\ \rm{s}^{-1}}\right)^f \right]
\label{eq:master_scaling_SFR} 
\end{align}

Furthermore, we test for a formulation of $\dot{M}_{\rm out}$ depending on the drivers alone:

\begin{align}
   \log(\dot{M}_{\rm out}) = &\ \log\left[10^a \left(\dfrac{\rm SFR}{3 \ \mathrm{M_{\odot} \ yr^{-1}}}\right)^c + 10^{e}\left(\dfrac{L_{\rm AGN}}{1\times 10^{43}\ \rm{erg}\ \rm{s}^{-1}}\right)^f \right]
\label{eq:master_SFRL_AGN} 
\end{align}

The above equations incorporate between two and four physical parameters describing the host galaxy: $M_{\star}$, the offset from the star-forming main sequence $\rm SFR/SFR_{MS}$ (or alternatively the absolute SFR), $R_{\rm e}$, and $L_{\rm AGN}$.  We express $\dot{M}_{\rm out}$ as having a power-law dependence on each physical parameter with power-law indices $b$, $c$, $d$ and $f$ respectively, leaving two parameters ($a$ and $e$) to set the normalisation for the SF and AGN drivers.  Having the SF and AGN driving terms appear in additive rather than multiplicative form makes sense physically, and prevents divergence for objects with SF outflows where $L_{\rm AGN}$ is taken to be zero.  Similar considerations were followed by \citet{Fluetsch2019}, although we point out these authors do not allow for different power-law indices quantifying the dependence on SFR and AGN luminosity.

The motivation for Eq.\ \ref{eq:master_3par} is not so much that it is the pairing of two input parameters that yields the smallest dispersion (that would involve the two drivers, star formation and AGN activity, as captured in Eq.\ \ref{eq:master_SFRL_AGN}), but rather that it may be of use as a reference to other studies that have less rich datasets to work with, for example those at high redshift where measurements of $L_{\rm AGN}$ or $R_{\rm e}$ are more challenging due to limited depth or spatial resolution.

\begin{figure}
\centering
\includegraphics[width=\linewidth]{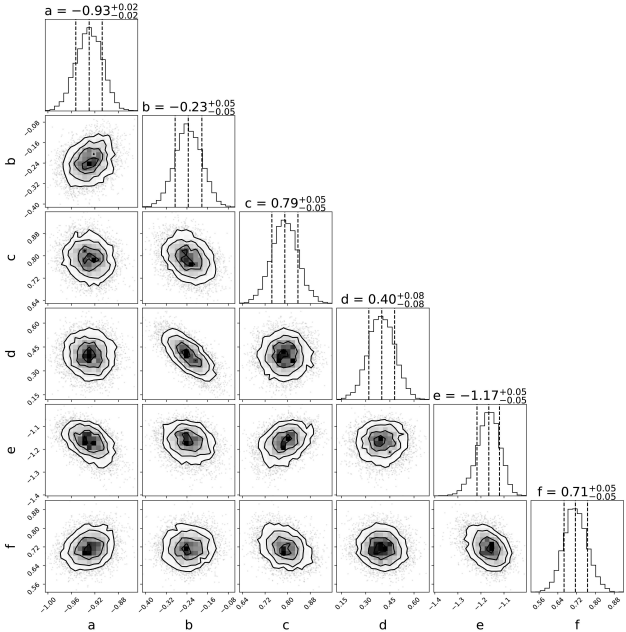}
\caption{Corner plot showing the {\tt emcee} sampling of the posterior probability distribution for the parameters in the outflow scaling relation of Eq.\ \ref{eq:master_scaling_SFR}, describing a mapping from internal galaxy properties to the observed mass outflow rate.}
\label{fig:corner}
\end{figure}

\begin{figure*}
\centering
\includegraphics[width=0.9\textwidth]{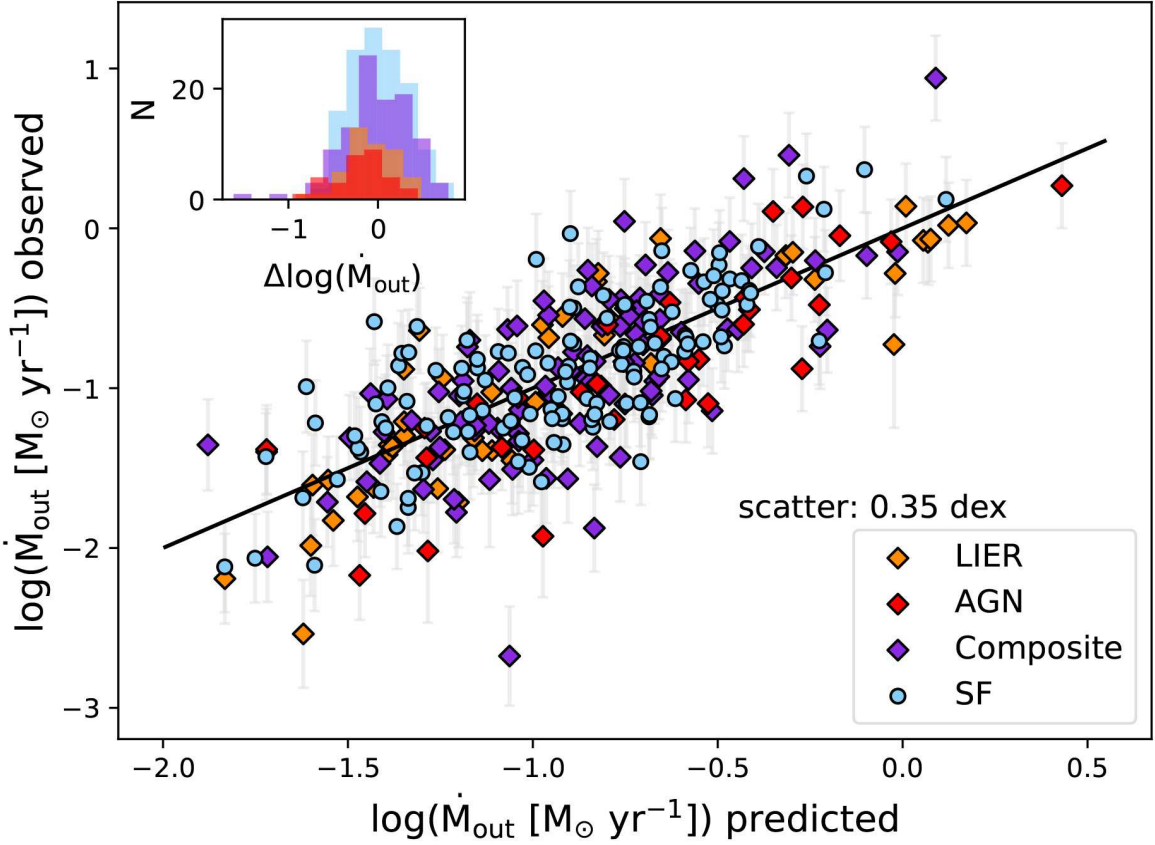}
\caption{Observed mass outflow rates for galaxies in the MaNGA outflow sample, contrasted to the outflow rate predicted based on the best-fit outflow scaling relation of Eq.\ \ref{eq:master_scaling_SFR}, taking galaxy stellar mass ($M_{\star}$), size ($R_{\rm e}$), star formation rate (SFR) and AGN luminosity ($L_{\rm AGN}$) as inputs.  SF and AGN galaxies are denoted by circles and diamonds, respectively.  The black line indicates the one-to-one relation and a histogram of residuals, with a scatter of 0.35 dex, is shown in the inset panel.}
\label{fig:master}
\end{figure*}

We use the \texttt{emcee} Monte Carlo Markov Chain implementation to derive the best-fit parameters in Eqs. \ref{eq:master_3par} - \ref{eq:master_SFRL_AGN} and we compare these functional forms as descriptors of the observed outflow rate by measuring the scatter between the predicted and the observed $\dot{M}_{\rm out}$. The results are presented in Table \ref{tab:master_scaling}.  We find that the dispersion of residuals around the best-fit relation reduces as the additional dependence on $R_{\rm e}$ is accounted for.  The scatter is ultimately reduced further to $\sim 0.35$ dex by including an AGN term, and remains equally small when formulating the scaling relation using absolute SFR or main sequence offset in the `outflow driver' term.  Figure \ref{fig:corner} shows the posterior distributions from \texttt{emcee} for the six fit parameters of Eq.\ \ref{eq:master_scaling_SFR}.  Figure \ref{fig:master} presents the corresponding $\dot{M}_{\rm out}$ predicted using Eq. \ref{eq:master_scaling_SFR} versus the empirically observed mass outflow rate.  
%We show the corresponding posterior distributions from \texttt{emcee} for the six fit parameters in Figure \ref{fig:corner}.  
By removing the dependence on galaxy mass and size from Eq. \ref{eq:master_scaling_SFR}, we find only a small increase in the scatter showing that $\dot{M}_{\rm out}$ is heavily influenced by the strength of its physical drivers.

Introducing additional galaxy properties such as redshift or inclination only yielded a marginal further reduction in scatter (0.34 dex) and reduced chi-square statistic (1.5).

% FIG corner

% FIG master

We thus conclude that the mass outflow rates inferred from broad components to the ionised gas line emission can be reproduced to within a factor $\sim 2$ starting from measurements of the galaxies' mass, size, star formation rate and AGN luminosity.  The same relation performs equally well for SF and AGN winds within our total sample of 322 objects whose spectra exhibit outflow features.

% TABLE master

\begin{table*} 
%\rotatebox{90} 
{ 
\begin{tabular}{l c c c c c c c c c c c} 
\hline 
Fit & a & b & c & d & e & f & scatter [dex] & $\chi_{\rm red}^{2}$ \\
\hline 
Eq\ \ref{eq:master_3par}   & -0.855\ (0.019) & 0.572\ (0.034) & 0.789\ (0.044) & - & - & - & 0.405 & 2.145  \\ 
Eq\ \ref{eq:master_4par}   & -0.855\ (0.019) & 0.361\ (0.049) & 0.779\ (0.044) & 0.488\ (0.081) & - & - & 0.394 & 2.037  \\ 
\textbf{Eq\ \ref{eq:master_scaling_delMS}}   & \textbf{-0.933\ (0.023)} & \textbf{0.276\ (0.051)} & \textbf{0.789\ (0.046)} & \textbf{0.371\ (0.081)} & \textbf{-1.215\ (0.047)} & \textbf{0.634\ (0.048)} & \textbf{0.348} & \textbf{1.567}  \\ 
\textbf{Eq\ \ref{eq:master_scaling_SFR}}   & \textbf{-0.930\ (0.022)} & \textbf{-0.235\ (0.052)} & \textbf{0.790\ (0.046)} & \textbf{0.399\ (0.080)} & \textbf{-1.166\ (0.049)} & \textbf{0.715\ (0.050)} & \textbf{0.345} & \textbf{1.540}  \\ 
Eq\ \ref{eq:master_SFRL_AGN}   & -0.915\ (0.020) & - & 0.754\ (0.041) & - & -1.197\ (0.051) & 0.743\ (0.050) & 0.355 & 1.618  \\ 
\hline 
\end{tabular} 
} 
\caption{Power-law indices of galaxy stellar mass (b), star formation activity (c), galaxy size (d), AGN luminosity (f) and normalisation factors (a, e) derived using {\tt emcee} following the functional forms of outflow scaling relations described by Eqs.\ \ref{eq:master_3par}, \ref{eq:master_4par}, \ref{eq:master_scaling_delMS}, \ref{eq:master_scaling_SFR} and \ref{eq:master_SFRL_AGN}. The corresponding scatter and reduced chi-squared values are listed in the two right-most columns.}
\label{tab:master_scaling}
\end{table*}

\subsection{Outflow geometry for star-forming disks}
\label{geometry.sec}

In order to investigate the dependence of outflow properties on galaxy inclination, we extract a Disks subsample comprised of star-forming disk galaxies with log(sSFR) > -11, S\'{e}rsic indices n < 2, and visual morphologies which are not classified as “odd looking” according to the Galaxy Zoo 2 classification \citep{GalaxyZoo2}.  The latter two criteria are to ensure that the projected axial ratio can be reliably used as a proxy for galaxy inclination.  As shown in the bottom-right panel of Figure \ref{fig:incidence}, we find a marginal difference in the $\cos(i)$ distribution of Disk galaxies with outflows compared to the underlying MaNGA Disk population for which line profile fitting was carried out, at least according to the A-D statistic, which is more sensitive to the tails of the distribution.  We do not find significance in the K-S statistic which is primarily based on the peak of the distributions. The Disk subsample criteria cuts the outflow sample to only 67 objects.  Using a less conservative cut on the Disk subsample by dropping the S\'{e}rsic index criterion, and assuming that the Petrosian axial ratio still serves as an adequate inclination indicator for systems that host significant bulge components, we increase the number of outflows in star-forming disk galaxies (i.e., not irregular/disturbed/merger or otherwise odd-looking) to 177 objects.  With this less conservative cut, we find a significant difference between the $\cos(i)$ distributions of the outflow Disks and the parent Disks samples according to both the K-S and A-D statistics.

% FIG Inclination
\begin{figure*}
\centering
\includegraphics[width=0.45\textwidth]{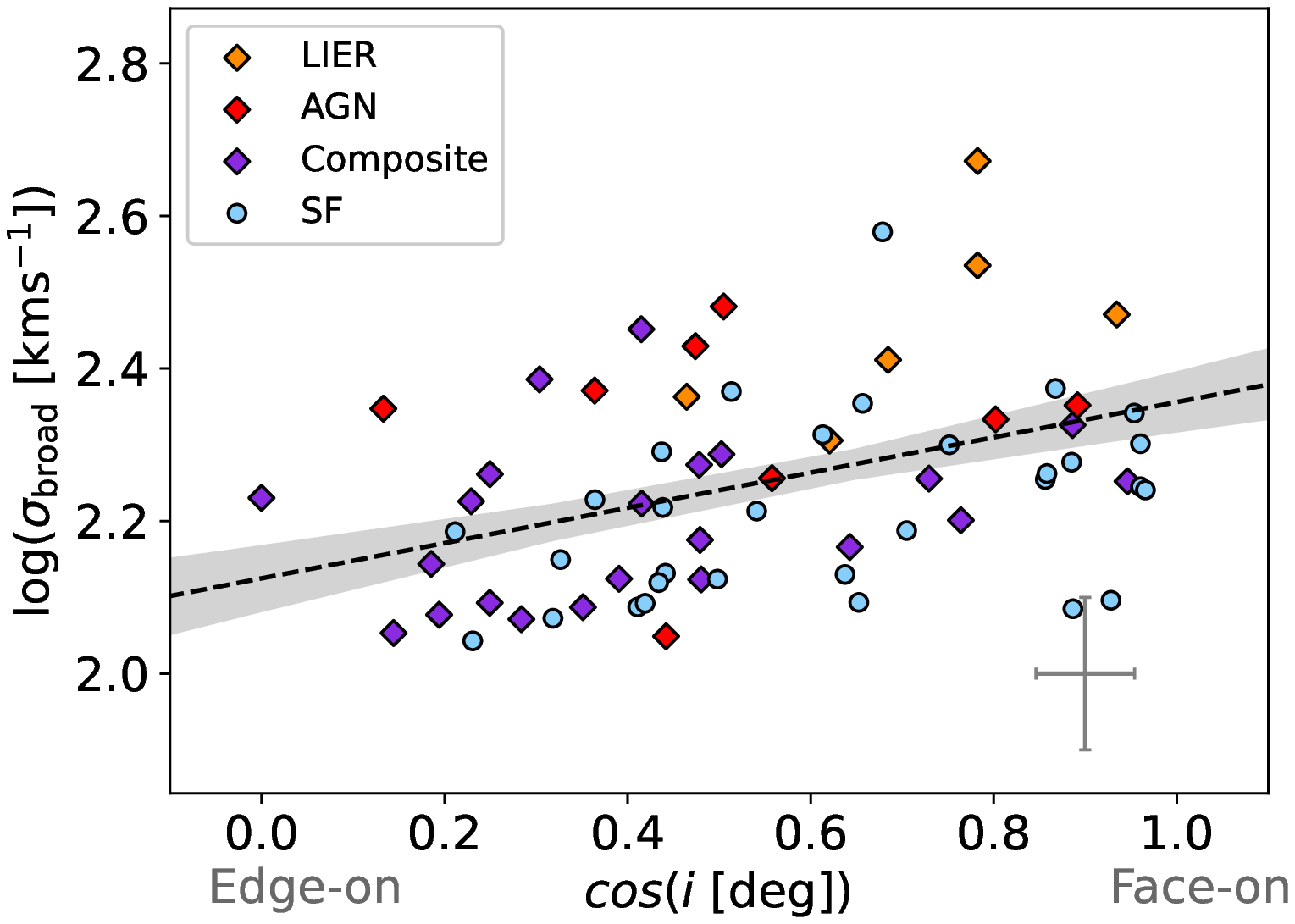}
\includegraphics[width=0.45\textwidth]{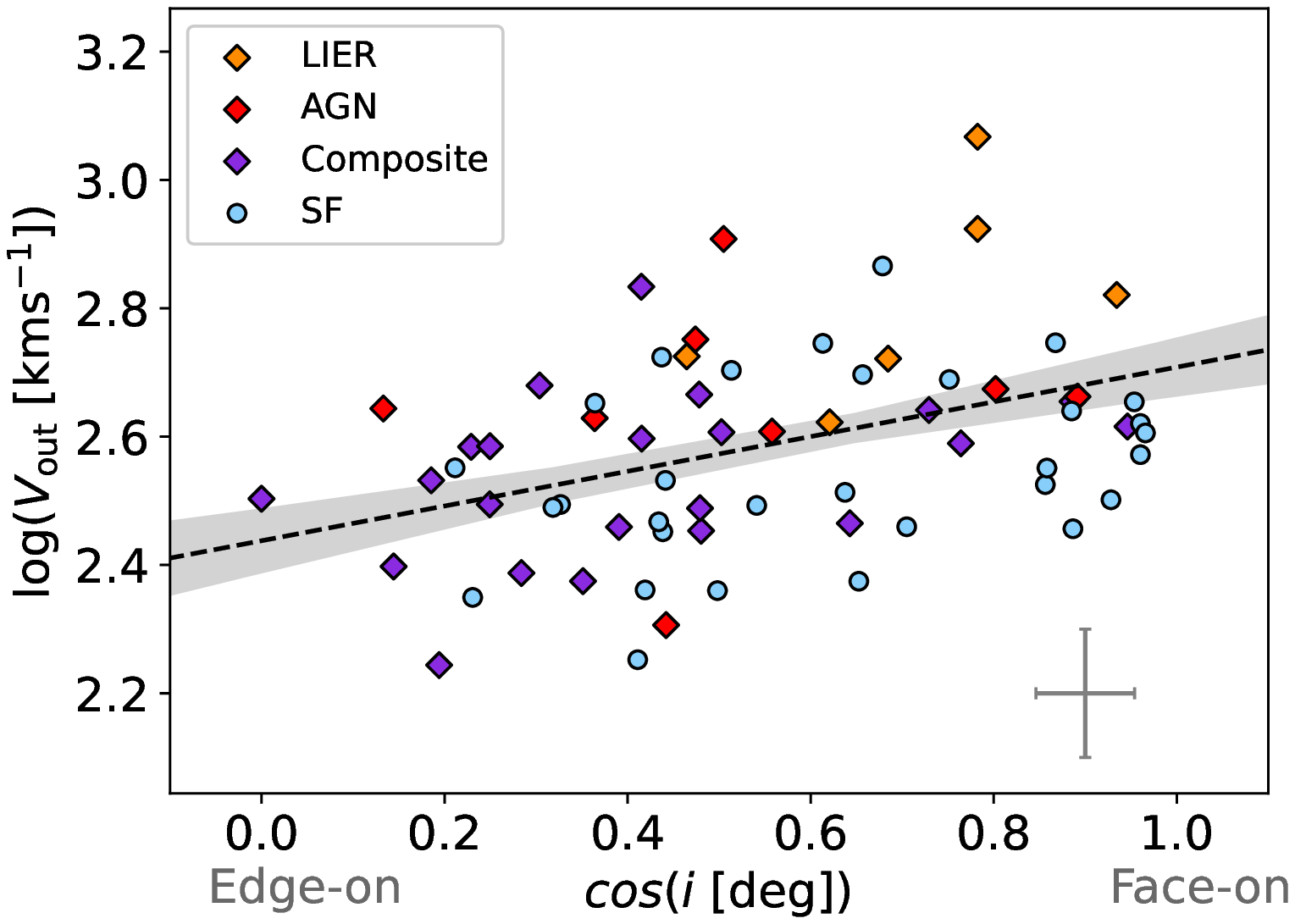}
\includegraphics[width=0.45\textwidth]{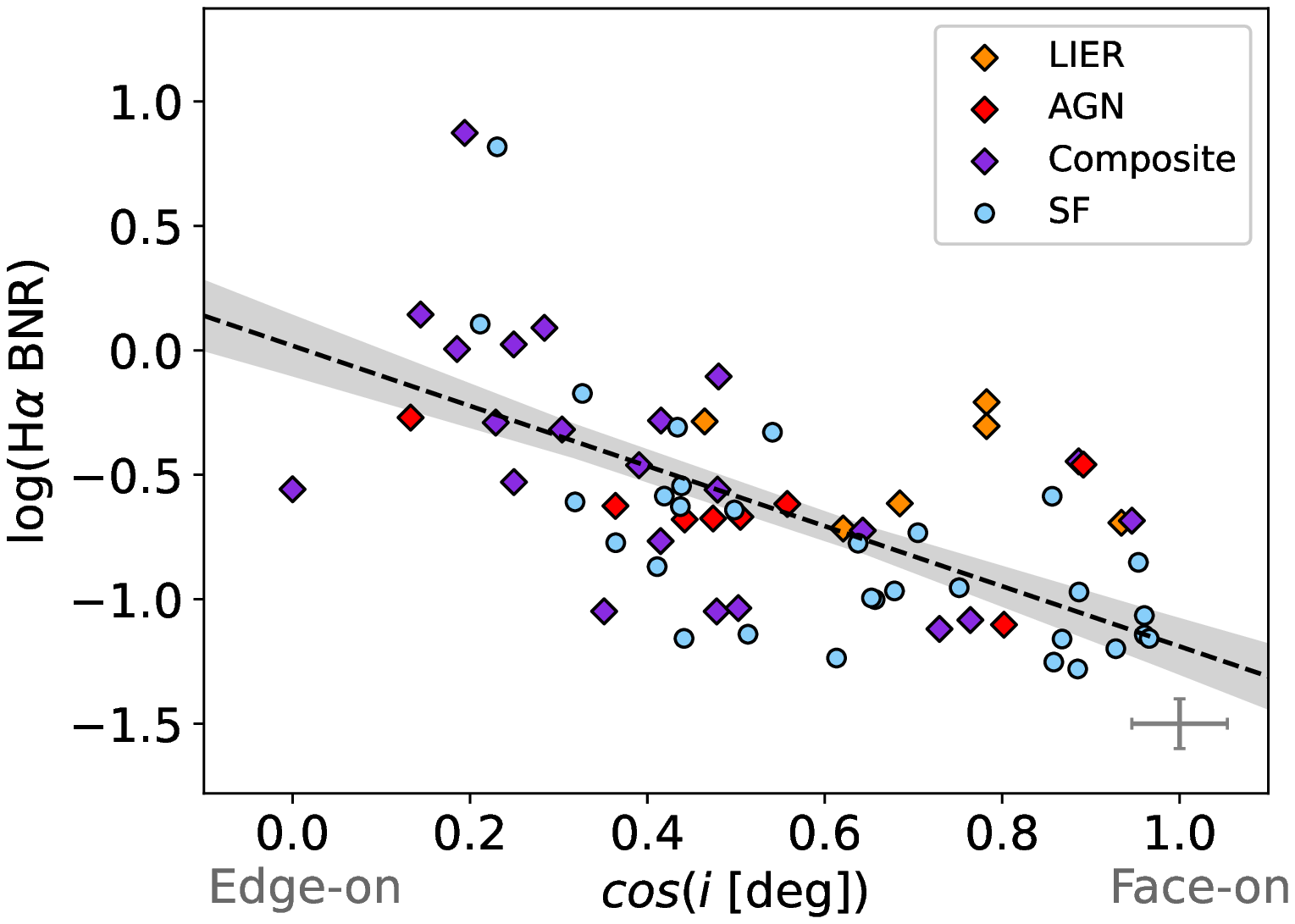}
\includegraphics[width=0.45\textwidth]{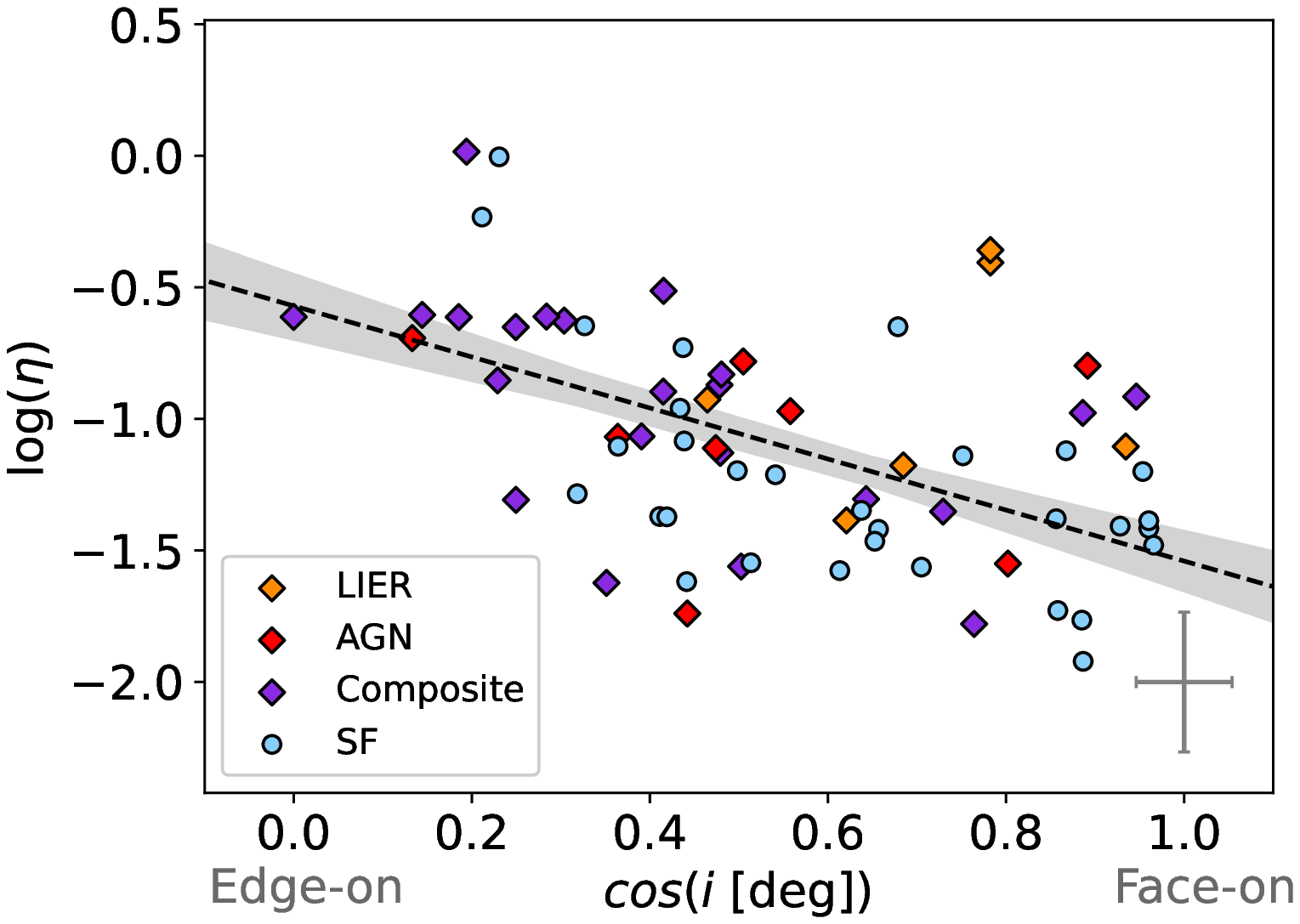}
\caption{{\it Top:} Width of the broad component and outflow velocity versus inclination. {\it Bottom:} H$\alpha$ BNR and mass loading factor $\eta$ versus inclination.}
\label{fig:inclination}
\end{figure*}

The main distinct feature is the lack of outflow objects seen edge on.  To derive a basic inference on what the inclination distribution implies about the outflow geometry, we consider a simple toy model in which a bi-conical outflow is oriented orthogonally to the disk plane.  In this model, a wind of opening angle $\theta$ is only detected if the galaxy is viewed within an angle of $\theta/2$ from face-on.  The toy model is summarised as follows: 
\begin{enumerate}
    \item We assume the distribution of outflow opening angles among galaxies to be characterised by an average opening angle $\bar{\theta}$ and a Gaussian dispersion about the average $\sigma_{\theta}$.
    \item By assuming that the underlying parent sample is viewed from random angles, we infer a mapping between the projected axial ratio and the intrinsic inclination (where we use the $b/a$ values of the Disks sample including $n > 2$ systems). This, unlike Eq. \ref{eq:incl}, allows us to take into account the fact that the surface brightness distribution of galaxies viewed face-on may not be perfectly round, but consistent results are obtained when adopting Eq.\ \ref{eq:incl} instead.
    \item Assuming random viewing angles for galaxies that intrinsically are driving outflow activity (whether detectable or not), we infer the fraction of outflow galaxies at each inclination that will be detectable for a given $\bar{\theta}$ and $\sigma_{\theta}$, and by adopting the mapping derived in (ii) we obtain the predicted normalised cumulative distribution of projected $b/a$ values of galaxies in the outflow sample. 
    \item We derive the values of $\bar{\theta}$ and $\sigma_{\theta}$ for which the predicted normalised cumulative distribution of projected $b/a$ values best matches the observed distribution in our Disks outflow sample, by using a least-squares Levenberg-Markwardt minimisation (\texttt{mpfit}), and independently by minimising the K-S or A-D statistics.
\end{enumerate}

All three ways of optimisation yield consistent results, namely that the opening angle is wide, of order $162 \pm 10 \deg$, corresponding to roughly $84\%$ of all outflow galaxies being detectable thanks to an orientation sufficiently away from an edge-on viewing angle.  Similar results are obtained when using the more conservative Disks sample, albeit suffering from poorer number statistics.  Arguably, the inclination dependence may be consistent with a somewhat smaller opening angle if a broad base is invoked instead of an outflow origin from a small region near the centre as assumed in our toy model. The inclination dependence of outflow detectability for such a geometry would be non-trivial to determine and is beyond the scope of this section. 

As well as finding a significant difference in the inclination distribution of the outflow and underlying MaNGA samples, we find a number of strong correlations between outflow properties and galaxy inclination (Figure \ref{fig:correlation_grid}) which are consistent with the results from the toy model. For outflows with some typical opening angle, at orientations close to edge-on the velocity-extent of the broad component would be reduced, and would be smallest for small opening angles. We can imagine this would be the case for a highly collimated AGN outflow or chimney-like features. This signature is evident in the top panel of Figure \ref{fig:inclination} where we see a positive correlation between $\sigma_{\rm B}$ and cos($i$) which follows through to the relation with $V_{\rm out}$ (Eq. \ref{eq:vout}).

Furthermore, in the bottom panel of Figure \ref{fig:inclination} we find a strong correlation between H$\alpha$ BNR and inclination. We expect two effects come into play here: (i) the fraction of light from the galaxy disk captured by the aperture is expected to be lower at orientations close to edge-on due to intervening dust along the line-of-sight through the disk (thus increasing the BNR), (ii) the fraction of light from the outflow captured by the aperture is expected to be higher at orientations close to edge-on where we are more likely to see both the red- and blue-shifted outflow components. The strong relation between inclination and H$\alpha$ BNR carries forward to the strong correlation with $\eta$. We note that the relation with [OIII] BNR is significantly weaker than with H$\alpha$ BNR although it is observed to some degree (Figure \ref{fig:correlation_grid}).

We considered the impact of choosing an elliptical aperture shape on outflow detection using our methodology, given that elliptical apertures may not capture the extraplanar gas as well as, say, circular apertures if the outflow is expelled along (or close to) the minor axis of the galaxy in an inclined system. As a sanity check we re-ran our experiment consistently adopting circular (rather than elliptical) apertures throughout our analysis, where we used circular apertures of radius 0.5$R_{\rm e}$, $R_{\rm e}$ and 1.5$R_{\rm e}$ to identify galaxy outflows. By using circular apertures, we do not find a significant increase in the number of galaxies with detected outflows, in particular a lack of detection in high inclination systems is common to the two methods differing in aperture shape. In addition, overall, the results from our experiment remain largely the same when using circular apertures.  This includes the inclination dependent trends presented in Figure\ \ref{fig:inclination}.

\subsection{Physical conditions of the outflowing ionised gas}
\label{physical_conditions.sec}

\subsubsection{Excitation pattern}
\label{excitation.sec}

% FIG excitation
\begin{figure*}
\centering
\includegraphics[width=0.48\textwidth]{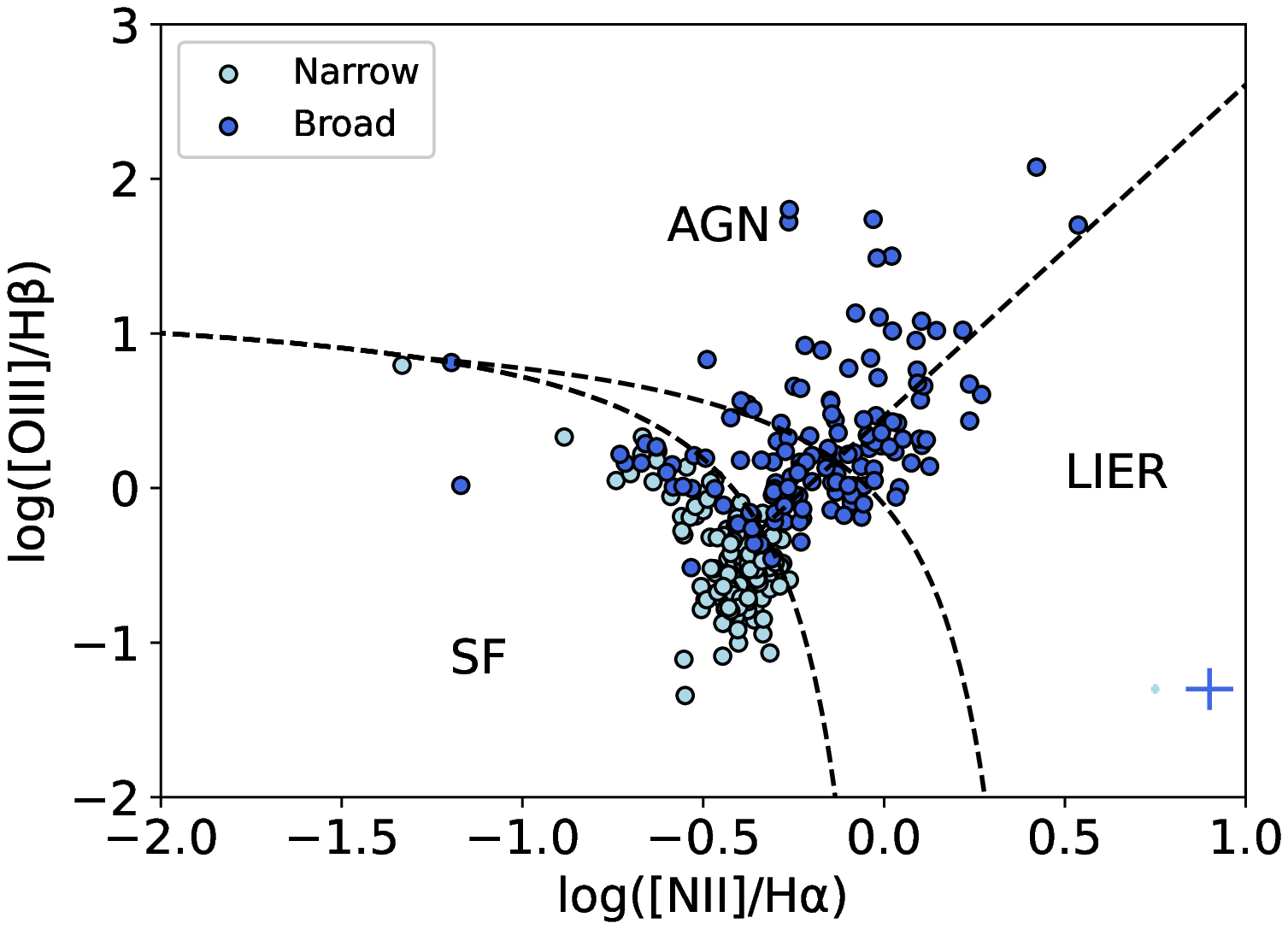}
\includegraphics[width=0.48\textwidth]{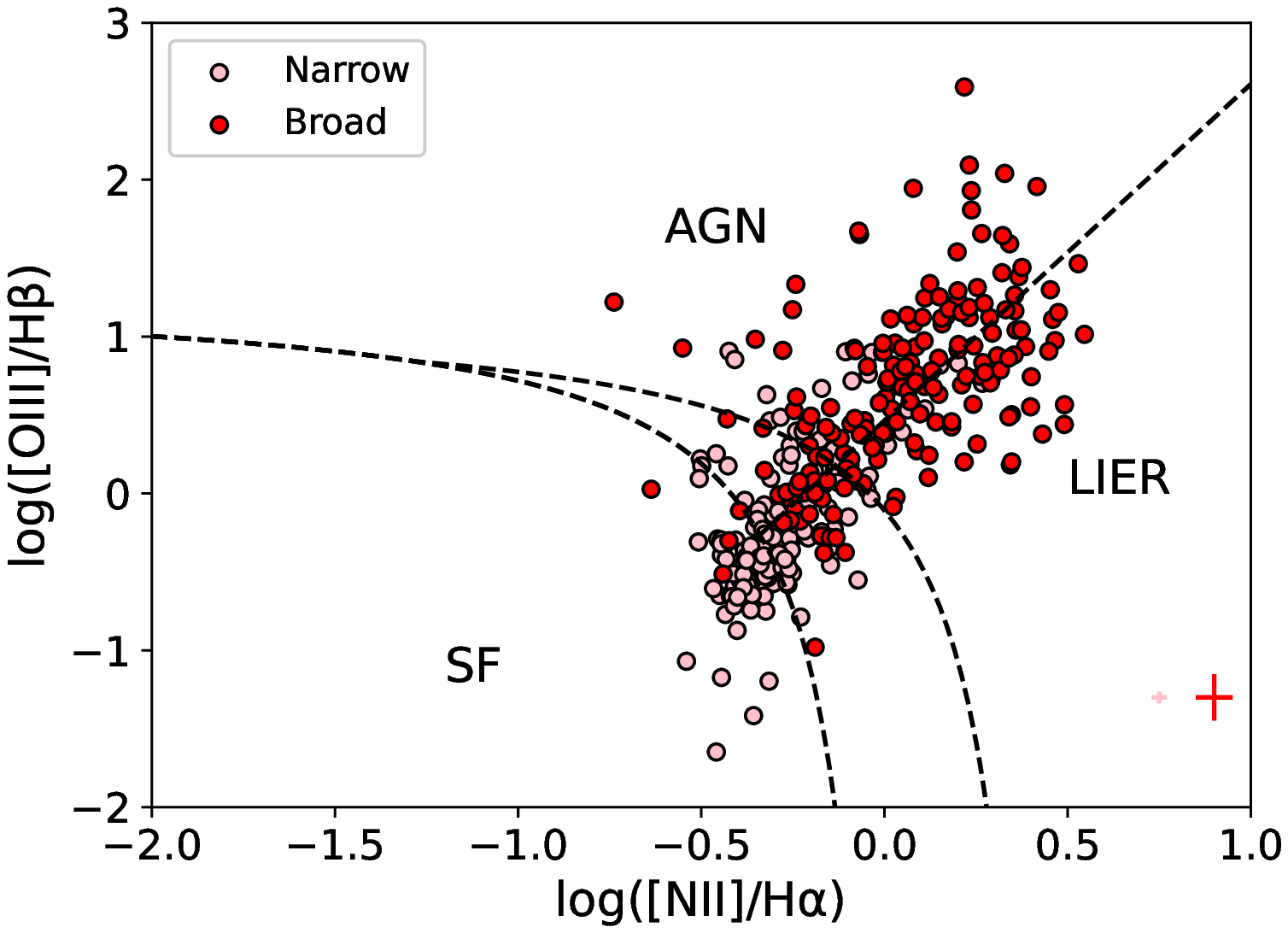}
\includegraphics[width=0.48\textwidth]{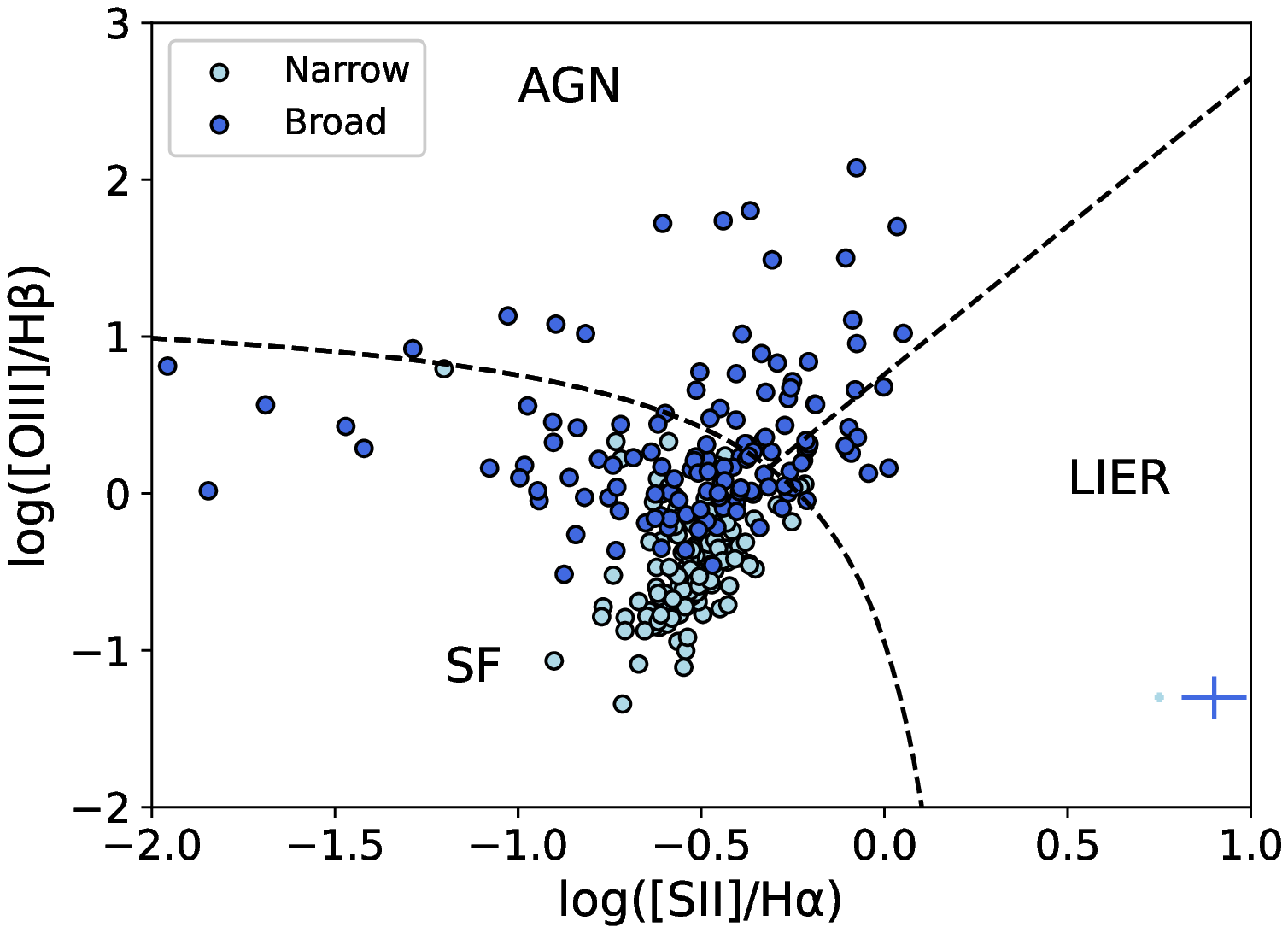}
\includegraphics[width=0.48\textwidth]{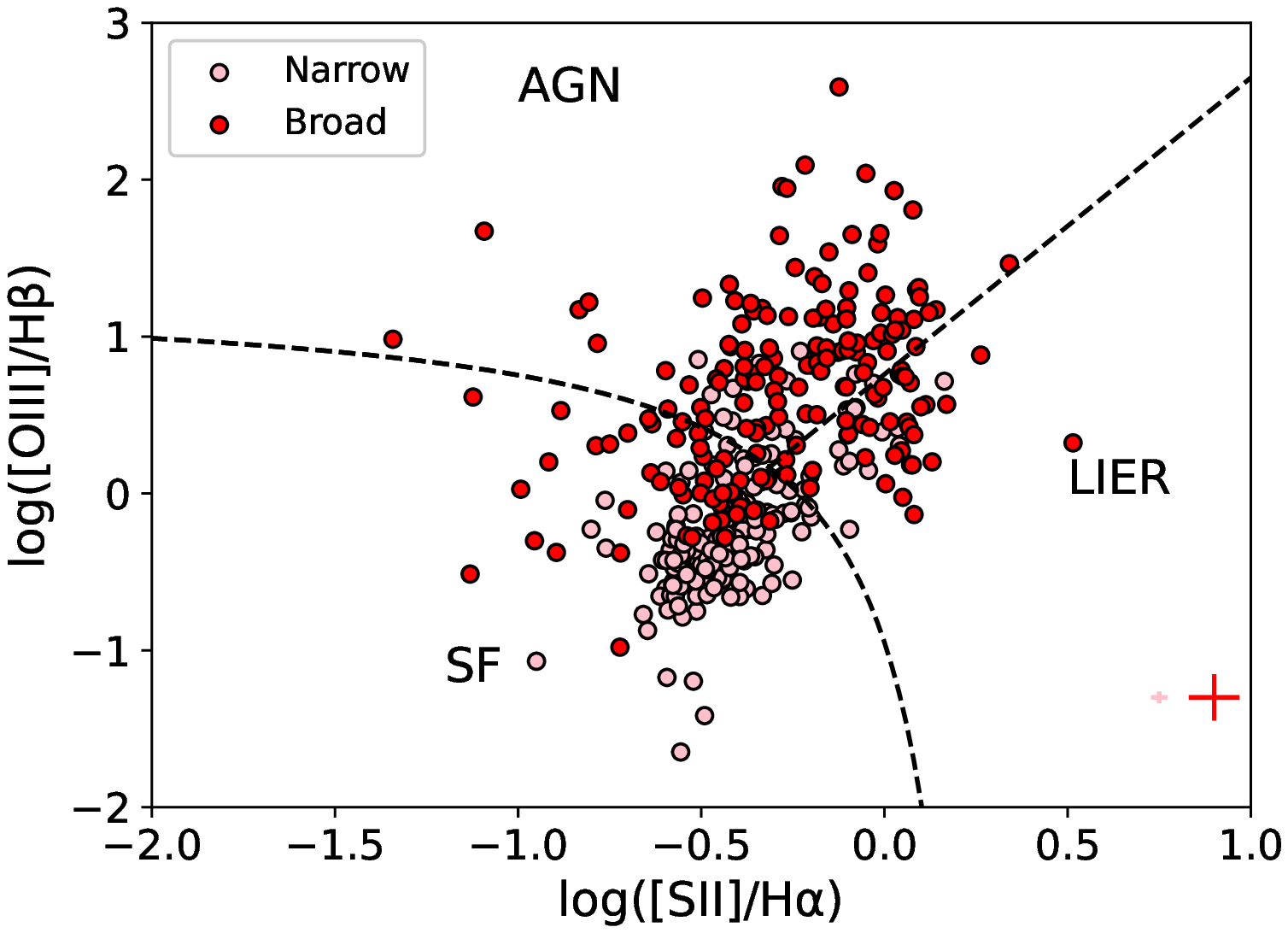}
\caption{[NII]-BPT ({\it top}) and [SII]-BPT ({\it bottom}) diagnostic diagrams displaying the excitation patterns for the broad- and narrow-component gas, as quantified from spectral extractions within an elliptical aperture of size $R_{\rm out}$.  Dashed curves delimiting regions with excitation patterns characteristic for star formation (SF), AGN or low-ionisation emission line regions (LIERs) are taken from \protect\cite{Kauffmann2003} and \protect\cite{Kewley2001} respectively.  Left-hand (right-hand) panels show those objects whose narrow-component line ratios in nuclear spectra, extracted within a $0.25 R_{\rm e}$ aperture, are consistent (inconsistent) with HII-region-like excitation.  In both SF and AGN wind galaxies, the broad-component gas exhibits elevated line ratios compared to the narrow-component gas, consistent with shock excitation within the outflowing gas.}
\label{fig:excitation}
\end{figure*}

The BPT diagnostic diagrams use emission line ratios ([OIII]/H$\beta$, [NII]/H$\alpha$ and [SII]/H$\alpha$) to determine the ionisation mechanisms exciting the gas.  Using our Gaussian decomposition, we determine these ratios for the narrow component tracing the disk gas and the broad component tracing the outflow separately and present the diagnostics for the different components in Figure \ref{fig:excitation}.  We show that the outflowing gas component exhibits harder excitation than the disk gas in both inactive and active galaxies, where the presence of AGN activity is determined by the narrow component line ratios in the $0.25R_{\rm e}$ stacks as described in Section \ref{SF_AGN_def.sec}.  We find cases where the gas excitation is dominated by the harder radiation from the central AGN within $0.25R_{\rm e}$, but by HII regions within $R_{\rm out}$ (right panel of Figure \ref{fig:excitation}).  We also find a few cases where photoionisation by HII regions was the primary ionisation mechanism within $0.25R_{\rm e}$, but narrow components show up in the composite region of the BPT diagram within $R_{\rm out}$ potentially resulting from narrow-line region relics from the time varying nature of AGN \citep[see also]{Wylezalek2020}.  

One plausible explanation for the higher excitation of the broad component gas is an origin from shocks experienced by the outflowing gas, compressing the gas, and causing the elevated line ratios (see, e.g., \citealt{Ho2014}). We see further evidence for this in the velocity widths of the broad components ($\sim 120-310 \ \rm km \ s^{-1}$) found in this work which are consistent with the star-formation shock sequence in \cite{DAgostino2019}, whereas the narrow velocity widths ($\sigma_{\rm N} \sim 40 - 100 \ \rm km \ s^{-1}$) are consistent with the star-formation-AGN mixing sequence in comparison.

\subsubsection{Electron density}
\label{ne.sec}

\begin{figure}
\centering
\includegraphics[width=\linewidth]{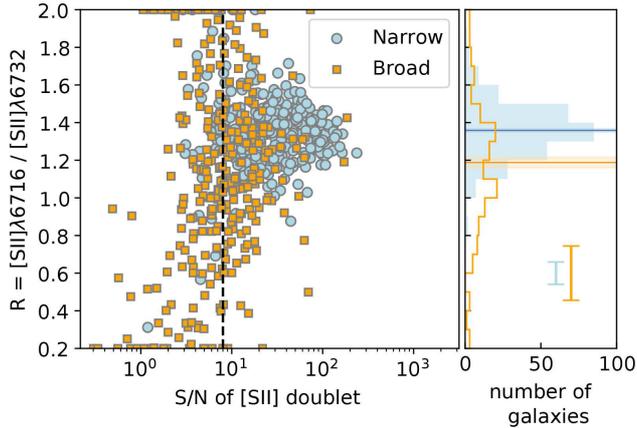}
\caption{[SII] doublet ratio (R) used to determine electron density versus the signal-to-noise ratio of the [SII] doublet. [SII] ratios for narrow and broad components are shown in blue and yellow, respectively. The histogram on the right shows the distribution of R values which have S/N > 8 for narrow and broad components in the outflow sample. Horizontal lines indicate the median of the histograms. Surrounding shading shows the formal error on the median. Error bars show the median error for individual measurements of R. }
\label{fig:ne}
\end{figure}

We estimate the electron density of the disk and wind gas from the [SII]$\lambda6716/6731$ doublet ratio using the relationship in \cite{Osterbrock1989}.  Figure \ref{fig:ne} presents the [SII] doublet ratio $R$ as a function of the S/N of the [SII] doublet for individual objects in the outflow sample.  Due to the small separation of the weak [SII] emission, the doublet becomes heavily blended especially when there is a broad component superimposed at the base of the emission profile making the double Gaussian decomposition challenging.  We find a wide range in the S/N of $R$ down to values $< 3$ and substantial uncertainties on individual measurements, particularly in the broad-component [SII] ratio. We therefore opted not to compute outflow properties based on $n_{\rm e}$ measurements of individual objects, but calculate characteristic values for the electron density of the wind gas (and the disk gas) estimated as the median [SII]$\lambda6716/6731$ ratio of the broad (and narrow) components over objects where S/N of the [SII] doublet > 8 (similar to \citealp{Ho2014}).\footnote{Consistent electron densities are obtained when taking the median statistic over all objects, irrespective of the S/N of $R$.}  This mitigates noise inducing effects on our derived outflow scaling relations that may otherwise come from `poor' [SII] emission line fitting.  We find median values $n_{\rm e} = 192 \pm 61 \ \rm{cm}^{-3}$ for the wind gas and $n_{\rm e} = 48 \pm 10\ \rm{cm}^{-3}$ for the disk gas, the latter consistent with the $R = 1.4$ obtained from a stacking analysis of nearby SAMI galaxies by \citet{Davies2020b}.  Similar electron densities ($\langle n_{\rm e} \rangle \sim 150\ \rm{cm}^{-3}$) were used in deriving the outflow scaling relations presented by \citet{Rupke2017}.  Elevated electron densities for the wind gas compared to the disk gas have been reported previously for both low- and high-redshift outflow samples \citep{Perna2017, Kakkad2018, ForsterSchreiber2019}, albeit quantitatively with substantially larger values, possibly related to a more extreme nature (in terms of outflow strength, star formation and/or nuclear accretion, and compactness) of the targets studied.  Complicating the situation further, not only do $n_{\rm e}$ measurements of individual objects in the literature span a wide range, \citet{Kakkad2018} also show evidence for order-of-magnitude spatial variations in $n_{\rm e}$ {\it within} galaxies, observed for both the disk gas and the outflowing medium, which places a cautionary note to the assumption of uniform density on which our and most analyses rely.  The validity of the [SII] doublet as a density tracer in (luminous) AGN outflows has also been questioned \citep{Davies2020}. 

We conclude that electron density measurements remain a source of significant systematic uncertainty in constraining the strength of feedback processes via ionised gas tracers.  Sensitive integral-field spectrographs with high spatial resolution and sufficient spectral resolution, such as VLT/MUSE+AO for low-z and ELT/HARMONI+AO for high-z studies, are required to make progress in this area.

% FIG ne

\subsubsection{Dust attenuation}
\label{attenuation.sec}

As outlined in Section\ \ref{outflow_def.sec}, the fiducial H$\alpha$ attenuation corrections we applied when computing outflow parameters were based on line ratio diagnostics inferred from single-Gaussian fits.  Specifically, the Balmer decrement, with an intrinsic ratio of $(H\alpha/H\beta)_{\rm int} = 2.86$ for Case B recombination and $T = 10^4\ \rm{K}$ \citep{Osterbrock2006}, serves as the diagnostic of nebular attenuation.  Here, we explore the Balmer decrements obtained via double-Gaussian fitting and consider the relation between observed outflow kinematics and dust richness of the galaxies in our sample.

\begin{figure}
\centering
\includegraphics[width=\linewidth]{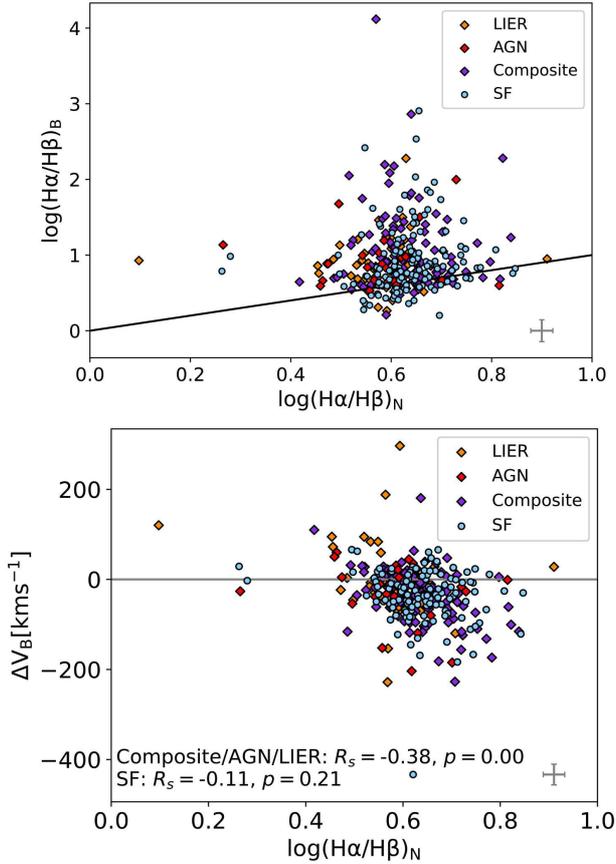}
\caption{{\it Top:} Comparison of Balmer decrement measurements for the broad- and narrow-component gas.  A tail towards larger Balmer decrements for the broad-component gas may imply enhanced attenuation by dust entrailed in the outflowing gas.  {\it Bottom:} On average larger systematic blueshifts of the broad Gaussian component are observed for galaxies with larger narrow-component Balmer decrements, consistent with a scenario in which the back side of a bi-conical outflow is attenuated by enhanced levels of dust within the disk.}
\label{fig:dust}
\end{figure}

The top panel of Figure\ \ref{fig:dust} presents the results from a double-Gaussian decomposition of the H$\beta$ and H$\alpha$ emission lines, contrasting the Balmer decrements of the broad- and narrow-component emission.  For a large number of objects, the level of obscuration inferred for the broad-component gas is consistent within the errors with the better constrained narrow-component attenuation.  However, a tail of the distribution extends to significantly enhanced Balmer decrements for the wind gas (and provided similar Case B conditions and temperatures apply thus elevated attenuation levels).  This feature is seen both among SF and among AGN outflows.  Whereas all objects shown in Figure\ \ref{fig:dust} satisfy the criteria for significant outflow signatures outlined in Section\ \ref{outflow_sample.sec}, one should bear in mind that at the highest $(H\alpha/H\beta)_{\rm B}$, the broad-component contribution to the $H\beta$ line profile becomes increasingly small and a precise inference of attenuation therefore becomes more sensitive to any residual systematics in the preparatory step where the stellar continuum with underlying $H\beta$ absorption is subtracted.  Nevertheless, when taken at face value accounting for the additional attenuation would sensitively increase the inferred outflow rates and mass loading factors, by factors of several, and can even alter patterns across SFR-$M_{\star}$ space as the implied correction is not uniform across the sample (see Figure\ \ref{fig:SFRMstar}).  Specifically, the observed $(H\alpha/H\beta)_{\rm B}$ increases more rapidly with stellar mass than $(H\alpha/H\beta)_{\rm N}$, leading to a stronger mass dependence of $\eta_{\rm dust\ corr} \propto M_{\star}^{0.42}$ compared to $\eta \propto M_{\star}^{0.01}$.  This could potentially hint at outflows in these more massive galaxies expelling preferentially metal-enriched material from the disk ISM.  At the same time, it should be noted that those systems feature stronger underlying H$\beta$ stellar absorption, making them more sensitive to the accuracy of the \texttt{ppxf} spectral decomposition.  Detailed individual object \citep{Perna2019} and sample studies of AGN outflows \citep{VillarMartin2014, RodriguezDelPino2019} have reported evidence for enhanced attenuation of the wind gas relative to the ambient gas before, although unlike what we observe in Figure\ \ref{fig:dust}, \citet{RodriguezDelPino2019} do not find a systematic excess of $(H\alpha/H\beta)_{\rm B}$ for their SF outflows, merely a broad range. 

In the bottom panel of Figure \ref{fig:dust}, the broad-component velocity offset relative to the narrow component is shown as a function of the Balmer decrement measured for the disk gas.  A significant negative correlation between $\Delta V_{\rm B}$ and $(H\alpha/H\beta)_{\rm N}$ is present for galaxies with nuclear activity.  This suggests that dust in the disk could be responsible for the blue-shifted offsets, with thicker dust columns within the disk's ISM blocking more of the receding outflowing gas, thus causing larger blueshifts.  This interpretation seems plausible when also considering the indication of an inclination dependence of the wind gas as discussed in Section \ref{geometry.sec}, implying that outflow detection is weakly favoured in more face-on objects or at least less likely for edge-on systems (see also discussion by \citealp{Veilleux2005}).  This effect may contribute to the enhanced attenuation of the outflowing gas, as shown in the top panel of  Figure \ref{fig:dust}.  However, given that the $\Delta V_{\rm B} - (H\alpha/H\beta)_{\rm N}$ relation is not statistically significant for star-formation driven outflows, the tail towards high Balmer decrements for the outflowing gas (seen among both SF and AGN outflows) may more plausibly be caused by dust entrailed in the outflowing gas itself.  In this context, we note that direct tracers of dust are also increasingly used to augment multi-phase gaseous datasets on galactic outflows (see the review by \citealt{Rupke2018}).

% FIG dust

%%% Discussion

\section{Discussion}
\label{discussion.sec}

In this section, we discuss in more depth four aspects of our outflow analysis and their associated implications: Section \ref{justification.sec} reflects on the interpretation of broad velocity components as galactic-scale winds.  Section \ref{universal_master_scaling.sec} addresses the universality of our derived outflow scaling relation, Section \ref{drivers.sec} considers the driving mechanisms and multi-phase perspective, and in Section \ref{fate.sec} we reflect on the fate and impact of the observed winds.

\subsection{Justifying the outflow interpretation}
\label{justification.sec}

Given the nature of the outflow detection and characterisation used in this work, it is important to justify the interpretation of these broad velocity components as outflowing gas from a quantitative and qualitative standpoint, and show that this rationale holds when considering other effects which could plausibly produce broad component emission.  

From a kinematic standpoint, we find typical $\sigma_B \sim 120 - 310 \ \mathrm{km \ s^{-1}}$, with a median at $183  \ \mathrm{km \ s^{-1}}$ ($v_{\rm out} \sim 240 - 670  \ \mathrm{km \ s^{-1}}$ with a median of $399  \ \mathrm{km \ s^{-1}}$). This is consistent with broad component kinematics measured from studies focusing on star formation driven outflows at higher redshifts (e.g. \citealp{Newman2012}; \citealp{Davies2019}; \citealp{Freeman2019}; \citealp{Swinbank2019}), with measurements of AGN driven outflows at higher redshift extending from similar to larger > 1000 km s$^{-1}$ velocities (e.g. \citealp{Genzel2014}; \citealp{Leung2017}; \citealp{Leung2019}). We also compare our results to the low redshift outflow studies of \citealp{Cicone2016} and \citealp{Concas2019}, where we measure the width of the entire emission line profile using Equation 6 of \citealp{Cicone2016}. We find total line-of-sight velocity dispersions of $\sim 70 - 150 \ (80 - 250) \ \mathrm{km \ s^{-1}}$ for our SF (AGN) subsample. This is of the same order as the outflows found in SFGs and AGN in \citealp{Cicone2016} and \citealp{Concas2019}, respectively. Furthermore, in Figure \ref{fig:dust} we see a tail towards negative $\Delta V_B$, with $70 \%$ of outflow objects featuring systematic blueshifts in their broad velocity component. This is evidence for the preferred detection of blue-shifted emission and is expected when probing outflows in emission, where emission from the receding part of a bi-conical outflow is diminished by the dust within the intervening galaxy disk (at orientations away from edge-on). Other studies such as \citealp{Davies2019} and \citealp{Newman2012} find outflow components which are on average blue-shifted relative to the systematic gas component with offsets consistent with those found in this work. \citealp{Arribas2014} find evidence for stronger blue-shifted offsets in systems that are more extreme in terms of their SFR (ULIRGs versus LIRGs), therefore it may be unsurprising that we find a slightly lower median offset ($\Delta V_B \sim 27 \ \mathrm{km s^{-1}}$) in our sample. Furthermore, \citealp{Arribas2014} find broad component shifts of  $\Delta V \gtrsim -100 \ \mathrm{km s^{-1}}$ for a significant fraction of objects, not too dissimilar from our results. Such consistency with ULIRG studies in this manner is particularly noteworthy given that there is extensive evidence for outflows observed in the form of broad emission lines in well-resolved observations of ULIRG galaxies (e.g., \citealp{Rupke2013} among many others). We note that we find $\Delta V_B \gtrsim 0 \ \mathrm{km \ s^{-1}}$ in a non-negligible fraction ($30\%$) of MaNGA outflows also, where we expect a combination of a favourable galaxy orientation and/or (asymmetric) outflow geometry may be at play, however further investigation would be required to draw such conclusions which is beyond the scope of this paper.

Other clues supporting an outflow interpretation can come from tracers of different gas phases, such as blue-shifted NaID absorption as a probe of neutral gas flows, which we will present in future work.

We considered the possibility of broad components arising as an artifact from beam smearing of the central gas velocity fields due to the limited spatial resolution of the IFU observations. We followed the approach by \citet{Gallagher2019} and compared the velocity dispersion of the broad-component gas to the stellar velocity dispersion quantified within the same aperture of size $R_{\rm out}$, which is affected by the same beam smearing. We find $\sigma_{\rm B} \gg \sigma_{\star}$ for most objects, with $\sim 95(86) \%$ of galaxies featuring $\sigma_{\rm B}$ that exceed $\sigma_{\star}$ at the 1(3)$\sigma$ level (typical $\sigma_{\star} \sim 60 - 160\ \rm{km}\ \rm{s}^{-1}$). Moreover, for star-formation driven winds the two show no significant correlation where we record a correlation coefficient $R_s = 0.0$ and p-value $p = 0.9$. This is unlike what would be expected if broad gas components resulted from beam-smeared gravitational motions of gas within the disk, rather than outflowing motions.

As additional indications of a genuine outflow nature, in Section \ref{physical_conditions.sec} we show that the physical conditions (excitation, $\rm n_e$, dust attenuation) of the broad component emission are distinctly different to the narrow component emission which probes the galaxy disk arguing against broad velocity components arising from beam smearing of disk gas moving at a range of velocities.  Furthermore, we find trends with inclination (Section \ref{geometry.sec}) and between blue-shifts of broad-component velocities and the amount of dust in the disk (Section \ref{physical_conditions.sec}) which are all consistent with the feedback scenario.

An example of an effect which could plausibly contribute to the broad emission is DIG emission. Specifically, where diffuse extraplanar gas is present in the form of co-rotating (rather than outflowing) material, as found in some literature studies (\citealp{Bizyaev2017,Jones2017,Marasco2019}), its lag in rotation with respect to the disk itself may lead to a modestly broadened and even skewed velocity component, but of a width that would normally not be picked up by our imposed minimum criterion on $\sigma_B$. Furthermore, we remove systems from our outflow sample where DIG emission dominates within the outflow radius (Section \ref{DIG.sec}). We find that most of these systems fall in the LIER region of the BPT diagram and at the lower end of the SFR distribution (SFR $\lesssim 0.3 \rm M_{\odot} \ yr^{-1}$). To determine the impact of this sample cut on our results, we re-run our analysis whilst including these DIG dominated objects in our outflow sample. We find that the best-fit trends presented in this paper remain the same within error and the overall conclusions drawn from this work are unchanged. 

Like DIG emission, merging systems could also complicate our analysis by appearing as kinematically disturbed components to the emitting gas. We therefore estimate the number of merging systems in our outflow sample which could potentially disturb the gas kinematics. We identify potential mergers as objects which have a greater than $50\% $ probability of being identified as a merger in the Galaxy Zoo 2 classification scheme. Merger galaxies identified in this way amount to only $2 \%$ of the outflow sample which have associated Galaxy Zoo 2 classifications. 

We note that other than galactic scale winds, the broad components observed in this work may probe the feedback that is localised to star-forming regions in the extended disk (chimneys; see, e.g., \citealt{Ceverino2009}). The assumptions involved in calculating the mass outflow rate and mass loading of the outflow may not be directly applicable to such localised `chimneys'. In particular, these parameters depend on an estimate for the outflow radius which we take as radius containing $90\%$ of the broad component emission. This would be an overestimation for say, localised star formation feedback being driven from outer regions of the disk.  As the $R_{\rm out}$ parameter appears in the denominator of Equation \ref{eq:Mdot_out}, this could consequently lead to an underestimation of the mass outflow rates.

\subsection{A universal master scaling relation?}
\label{universal_master_scaling.sec}

Galaxies with outflow signatures are not drawn randomly from the underlying MaNGA population (Section\ \ref{incidence.sec}).  Among those objects featuring a galactic wind, its strength relates tightly to physical characteristics of the host such as the intensity of star formation and nuclear accretion, as well as its stellar mass and size (Section\ \ref{scaling_relations.sec}).  A dependence on inclination is comparatively weak, albeit present in the sense expected from bipolar outflows with wide opening angles (Section\ \ref{geometry.sec}).

We now tie these findings together and ask the question: If the master scaling relation derived from our outflow sample (Section\ \ref{all_scaling.sec}) were to hold universally, does it explain why we did not detect significant outflow signatures in $\sim 90\%$ of the analysed MaNGA sample?

To address this, we apply Eq.\ \ref{eq:master_scaling_SFR} to all 2744 objects to which line profile fitting was applied, and thus infer their anticipated mass outflow rates $\dot{M}_{\rm out, expected}$.  The median $\dot{M}_{\rm out, expected}$ of the 2368 galaxies without identified outflow signatures is $\sim 0.5$ dex lower than that obtained for the outflow sample itself.  Objects missing from the outflow sample are thus also anticipated to feature, on average, weaker outflow signatures.  
Since the criteria applied in constructing the outflow sample (Section\ \ref{outflow_sample.sec}) do not correspond to a simple threshold in $\dot{M}_{\rm out}$, extra steps are required to evaluate in detail whether the wind gas would be detectable.  To this end, we estimate $R_{\rm out}$ using Eq.\ \ref{eq:Rout} and the wind velocity by randomly sampling a normal distribution with median and width characterised by the $v_{\rm out}$ values found in this work to infer, from $\dot{M}_{\rm out, expected}$, the anticipated luminosity of the H$\alpha$ broad component using Eq.\ \ref{eq:Mdot_out}. Paired with the object's observed narrow-line H$\alpha$ luminosity we estimate the expected H$\alpha$ broad-to-narrow line ratio, and using the broad component width (estimated from the anticipated $v_{\rm out}$) we estimate the amplitude at the line centroid for the broad H$\alpha$ component.  If the estimated broad-component width does not exceed the width of the narrow component by 50 km s$^{-1}$, if the H$\alpha$ broad amplitude at the broad line centroid does not exceed the RMS continuum noise surrounding H$\alpha$ by a factor 3, or $\rm BNR_{\rm H\alpha, expected} < 0.05$, then the object would fail meeting at least one of selection criteria (ii), (iv) or (v) of Section\ \ref{outflow_sample.sec} and thus not make it into the sample.

Of the 2368 analysed objects that did not meet the criteria outlined in Section\ \ref{outflow_sample.sec}, we find that the lack of detection is consistent with expectations from a universal wind scaling relation following the above rationale for $\sim 26 \%$ of the cases. If additionally accounting for viewing angle effects adopting the inferred wide opening angles from Section\ \ref{geometry.sec}, this percentage rises to $\sim 32 \%$.

As for what to make of the remaining $\sim  68 \%$ of objects for which we would have expected to have sufficiently strong H$\alpha$ broad components to be detected above the noise, a number of conjectures may explain why they did not end up in the outflow sample.  For example, given the weakness of the H$\beta$ line, the broad component in H$\beta$ can be very difficult to detect especially when reddening effects from dust extinction in the outflowing gas are significant.  This limitation may result in the object failing criterion (v) in Section \ref{outflow_sample.sec} where we require non-zero flux of the broad component in all BPT lines in order for the object to be considered as an outflow candidate. In this context, we point out the importance of the stellar continuum subtraction for outflow detection since any slight oversubtraction of the stellar light could prevent the broad component being visible in the resulting gas spectrum, especially when the width of the broad component is similar to the width of the stellar absorption features.  

More generally, the BIC statistic may not deem a two-component Gaussian decomposition statistically preferred over a single-Gaussian fit if the outflow is sufficiently weak.  This failing of criterion (i) may be further exacerbated if the kinematic properties of the winds, if present, differ from the selection criteria applied in Section \ref{outflow_sample.sec}.  A broad component that is only marginally broader than the narrow component would be hard to disentangle, whereas at very high velocities the amplitude of the broad component is reduced, rendering detection more difficult in the presence of any residuals from stellar continuum subtraction. \par 
Finally, for a given $\dot{M}_{\rm out}$, the observable signature could be reduced if the electron density of the outflowing medium were lower, although we note that the $n_{\rm e,B}$ we adopt is already on the lower end of values reported in the literature.  

Alternatively, it may of course be that the outflow strength in those objects is genuinely lower than what we inferred on the basis of Eq.\ \ref{eq:master_scaling_SFR}, either because of a hidden dependency on a physical parameter we did not consider in our analysis, or as a consequence of bias by fitting the scaling relation to those galaxies with wind signatures strong enough to make it into the outflow sample.

\subsection{Driving mechanisms and multi-phase perspective}
\label{drivers.sec}
\begin{figure}
\includegraphics[width=\linewidth]{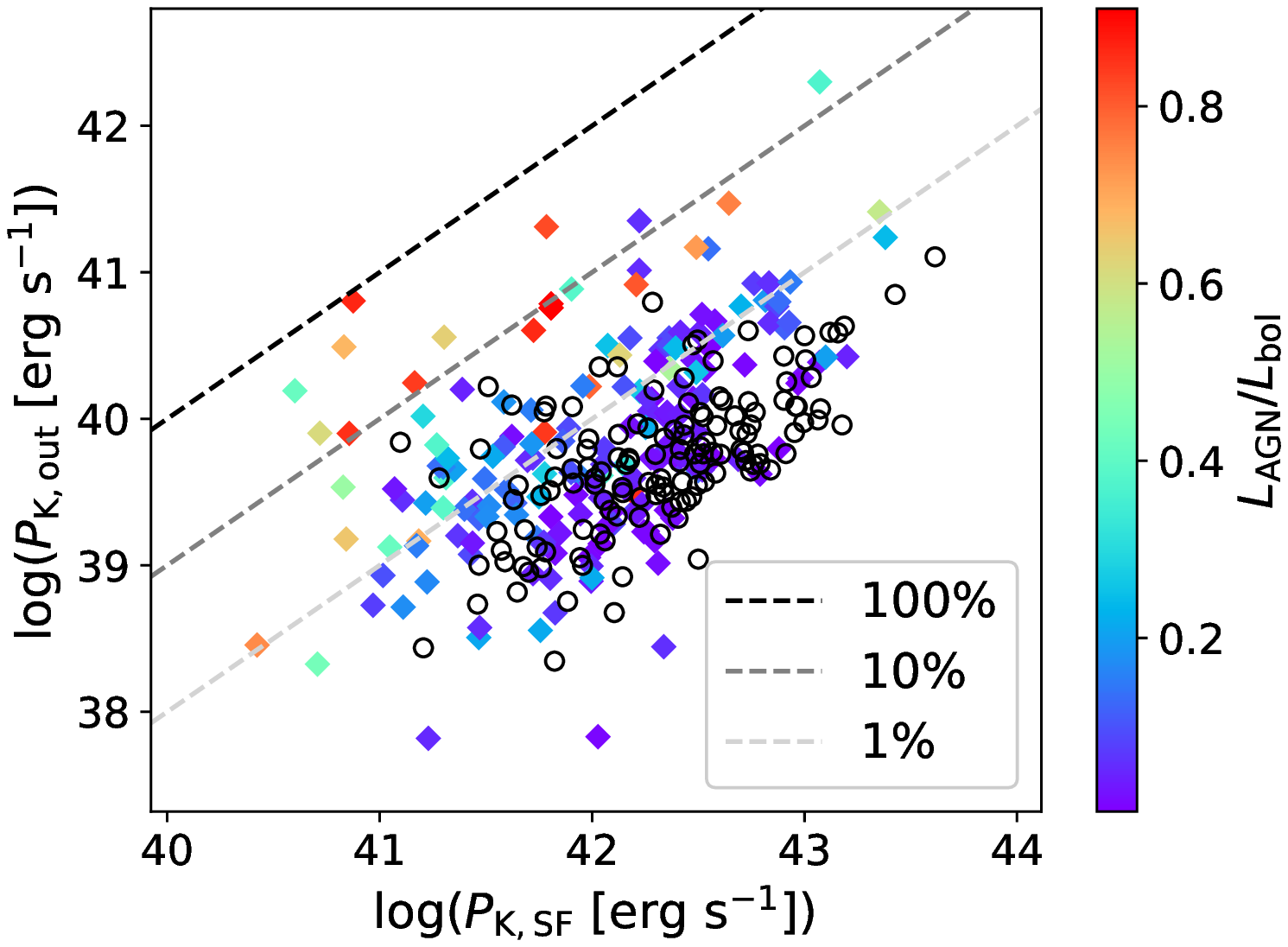}
\includegraphics[width=\linewidth]{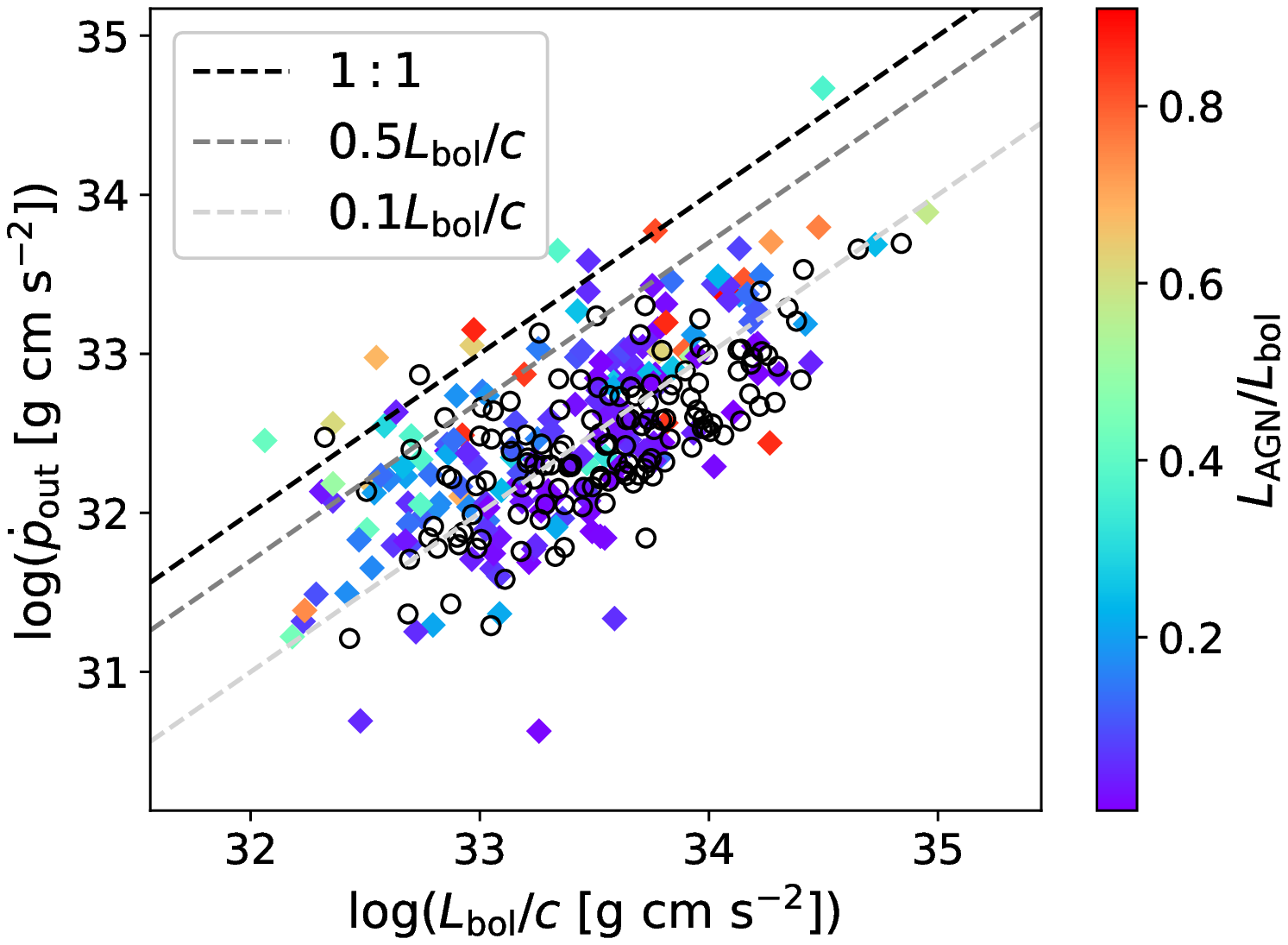}
\caption{{\it Top:} kinetic power of the outflow versus kinetic power generated by supernovae.  The dashed lines indicate coupling efficiencies of 1, 10 and 100 per cent.  {\it Bottom:} relation between outflow momentum rate and the total momentum output from star-formation and AGN.  Galaxies with evidence of AGN activity are shown colour-coded by their AGN luminosity as a fraction of the bolometric luminosity. These plots are inspired by \protect\cite{Fluetsch2019}.}
\label{fig:energy_momentum}
\end{figure}

To further investigate the wind driving mechanisms, we now consider the energy and momentum properties of the observed outflows.  The top panel in Figure \ref{fig:energy_momentum} presents the kinetic power of the outflow, given by $\frac{1}{2} \dot{M}_{\rm out} v_{\rm out}^2 $, compared to the kinetic power which is expected to be generated by supernovae;
$P_{\rm K, SF} \ [\rm erg \ s^{-1}] = 7 \times 10^{41} \rm \ SFR \ [M_{\odot} \ yr^{-1}]$ \citep{Veilleux2005}.  From this we show that the majority of ionised gas outflows in our sample can be accounted for by star formation with coupling efficiencies $\lesssim 1 \%$. High-resolution hydrodynamical simulations suggest that, while large portions of the combined supernova energy is rapidly radiated away, coupling efficiencies to the gas of this order can easily be achieved \citep{Creasey2013}. Coupling efficiencies of $\lesssim 1\%$ have also been reported for nearby SF outflows observed in the molecular phase \citep{Fluetsch2019}.  Based on ionised gas measurements of SF outflows at higher redshifts, \citet{Swinbank2019} report coupling efficiencies of 0.7 - 3\%, and similarly, coupling efficiencies in the range $\sim 0.4 - 2 \%$ are extracted from \cite{ForsterSchreiber2019}. \par
In Figure \ref{fig:energy_momentum}, we colour-code objects by the relative contribution of the AGN to the total bolometric luminosity $L_{\rm bol}$ which is taken as the combined energy output from star-formation ($L_{\rm SF}\sim \mathrm{SFR} \times L_{\odot}/1.09\times 10^{-10}$; adjusted from \citealt{Kennicutt1998} for a \citealt{Chabrier2003} IMF) and AGN ($L_{\rm AGN}$). In a fraction of systems, where the AGN contribution is high, we find that coupling efficiencies up to $\sim 10 \%$ and above are required in order for star formation to drive the winds.  It is therefore likely that the central AGN is at least partly responsible for the high outflow kinetic power observed.  The bottom panel of Figure \ref{fig:energy_momentum} shows the relation between the outflow momentum rate, given by $\dot{M}_{\rm out} v_{\rm out} $, and the total photon momentum output of the galaxy from both star-formation processes and AGN, estimated as $L_{\rm bol} /c$.  We find a strong correlation between the outflow momentum rate and the total momentum output of the galaxy as expected.  Given that all objects in our MaNGA outflow sample lie on or below the 1:1 line, we conclude that the outflow momentum can be accounted for by the presently observed energy output from star formation and nuclear accretion, where for a large fraction of galaxies the outflow momentum rate is $\lesssim 0.1 L_{\rm bol} /c$.  Consideration of the momentum budget of the ionised gas phase alone therefore does not prompt us to invoke so-called fossil outflows resulting from stronger AGN activity in the past (see \citealt{Fluetsch2019} for context). However, we do find relatively large outflow dynamical times compared to the characteristic timescales probed by our [OIII] AGN luminosity indicator: $t_{\rm AGN} = \frac{R_{\rm 50, [OIII] BPT-AGN}}{c}$, where the numerator corresponds to the half-light radius of the AGN-excited [OIII] emitting gas which we take as the radius containing $50\%$ of the total [OIII] emission measured from the BPT-AGN spaxels (see also Section \ref{LAGN_def.sec}). We find values in the range $t_{\rm AGN} \approx 10^{3.3} - 10^{4.5}$ yr (and possibly extending to $\sim 10^5$ yr when accounting for finite decay times), more than an order of magnitude shorter than the typical dynamical timescale of the outflowing wind gas $t_{\rm dyn, out} = \frac{R_{\rm out}}{V_{\rm out}} \approx 10^{6.6} - 10^{7.4}$ yr. This may hint at AGN variability being a contributor to the scatter between observed outflow properties and the presently observed strength of its potential drivers.\footnote{The scatter due to AGN flickering may be further enhanced if adopting an X-ray AGN luminosity diagnostic as done by Fluetsch et al. (2019), which probes the immediate vicinity of the black hole and thus shorter characteristic timescales.}

\par 
The kinetic energy and the momentum rate of the outflowing gas are interesting physical parameters to compare to feedback models.  We emphasise though that a major caveat of our work is that we only probe one phase of the multi-phase outflowing medium, namely the ionised gas.  This is likely to affect more any conclusions relying on normalisation of outflow strength (such as those presented in this section) rather than conclusions drawn from incidence and/or the relative scaling with galaxy properties (Section\ \ref{results.sec}).  However, the latter may not be free of mono-phase biases either.  For example, \citet{Fluetsch2019} finds the ratio of molecular to ionised gas outflow rates to vary systematically with AGN luminosity, from order unity for SF outflows to values 2 orders of magnitude higher for $L_{\rm AGN} \approx 10^{45}\ \rm{erg}\ \rm{s}^{-1}$, with possible evidence for a turnover beyond this regime presented by \citet{Fiore2017}.  As illustrated in Figure\ \ref{fig:AGN}, the most extreme objects in our sample reach $L_{\rm AGN} \sim 10^{45}\ \rm{erg}\ \rm{s}^{-1}$, but the median AGN luminosity is substantially lower: $\langle L_{\rm AGN}\rangle \sim 10^{43}\ \rm{erg}\ \rm{s}^{-1}$. The plots shown in Figure \ref{fig:energy_momentum} adopt the same style as in \cite{Fluetsch2019} noting that the outflow energy and momentum rates measured for our MaNGA sample fall significantly below those found by these authors.  This is likely related to the fact that their work considers different gas phases and their sample is more biased towards more extreme objects with higher SFRs and $L_{\rm AGN}$ (see also Appendix\ \ref{appendix_literature.sec}).

We conclude that multi-phase observations of outflow activity, for the same set of objects with a sound understanding of how they relate to the underlying galaxy population, remain critical to achieve a meaningful interface with feedback models.  Conversely, while major strides forward have been made in incorporating and analysing stellar and AGN feedback in numerical simulations of galaxy formation that follow the full cosmological context \citep[see, e.g.,]{Nelson2019,Mitchell2020}, much work is still ahead in bridging the gap to the observational realm in terms of radiative transfer and mock observations, and more fundamentally the multi-phase treatment of the ISM and outflowing medium, presently the domain of sub-grid modelling for cosmological simulations that adopt a $10^4$ K cooling floor.

\subsection{Fate and impact of observed winds}
\label{fate.sec}

There is strong evidence from a theoretical perspective that feedback plays an extremely important role in the growth and death of galaxies.  On the observational side, the direct impact of galactic winds is very much debated.  In order to gain insight into the effect of galactic winds in `typical' nearby galaxies, we calculate the fraction of outflows that have sufficient energy to escape the gravitational potential well in which they reside.  \par
We calculate the escape velocity of a galaxy using Eq. 3 from \cite{Swinbank2019}:
\begin{equation}
v_{\rm esc} = [2 v_{\rm rot}^2 \ \rm ln(1 + log(R_{vir} / R_{\rm out})]^{1/2}
\label{eq:vesc}
\end{equation}
where the halo virial radius $R_{\rm vir}$ is computed for each object individually using the stellar-to-halo mass relation of \cite{Moster2013}, and halo mass - halo size relation given by cosmology.  From this, we find that 86 of the 322 galaxies ($\sim 27 \%$) in the MaNGA outflow sample host winds that have $v_{\rm out} > v_{\rm esc}$, i.e., outflow velocities high enough to escape their host halo.  It thus seems that the outflows studied in this work predominantly act as galactic fountains rather than fully expelling the gas reservoirs entrained in the outflowing gas (see, e.g., \citealp{Li2020}). These winds could potentially still contribute to heating the circum-galactic medium and hence reducing the rate at which gas can cool and form stars.  Although outflow activity is often invoked as a mechanism to suppress star formation, we emphasise that in an instantaneous sense what is observed across our sample is a signal of enhanced star formation activity promoting outflow activity, not a direct imprint of negative feedback in the form of an anticorrelation with star formation activity.   \par
As for the $\sim 27 \%$ of galactic winds with $v_{\rm out} > v_{\rm esc}$, roughly half of the galaxies hosting these winds show evidence for AGN activity in their central regions.  We find that the distribution of star-formation driven winds which escape the halo peaks at slightly lower masses $M_{\star}\sim10^{10.1} \ \mathrm{M_{\odot}}$ compared to $M_{\star}\sim10^{10.6} \ \mathrm{M_{\odot}}$ for winds which have AGN as their potential driving mechanism.  Splitting the outflow sample into two halves, below and above the median $M_{\star}$ ($\sigma_{\star \rm{R_{\rm e}}}$), we find the fraction of outflows that escape to be higher among those objects with shallower potential wells ($20\%$ for the lower $M_{\star}$ half and $22\%$ for the lower $\sigma_{\star \rm{R_{\rm e}}}$ half) compared to those objects with deeper potential wells ($7\%$ for the upper $M_{\star}$ half and $5\%$ for the upper $\sigma_{\star \rm{R_{\rm e}}}$ half). \par 
We conclude that in the more typical low-redshift galaxies studied in this paper, star-formation processes are at least equally important in driving winds out of the host galaxy as central accreting supermassive black holes. Our analysis further highlights the critical distinction between on the one hand the effective outflow rates and mass loading factors implied by gas regulator models on the basis of metallicity scaling relations \citep[][quantifying the amount of material lost from the galaxy system altogether]{Lilly2013, Peng2014, Trussler2020}, and on the other hand the outflow scaling relations directly observed on the basis of broad-component line emission, as presented in this paper. The latter primarily assess the amount of warm ionised wind material that is launched. Between the spatial extent reached by the outflow as observed and its parent halo’s estimated virial radius lie factors of $\sim 30$ to 140, during which a host of complex interaction processes in the circum-galactic medium can affect the coasting wind material (e.g., \citealp{Fraternali2016}).  The inference in this Section on wind fate should thus be taken with a grain of salt, and serves predominantly to caution the reader against an interpretation of the observed wind phenomenology as fully ejective outflows.

\section{Summary}
\label{summary.sec}
In this work, we have taken advantage of the spatially resolved, high spectral resolution IFS data from the public data release of the MaNGA galaxy survey ($z \sim 0.04$) to determine the incidence, strength and scaling relations, as well as physical conditions of outflows among galaxies spanning a wide range of physical properties.  We find outflows within 322 galaxies evident as non-gravitational components to the kinematics of the line-emitting gas.  The spatially resolved information allows us to identify nuclear activity, predominantly at a modest level, in 185 of these objects.  We consider the effects of both star formation and AGN activity as wind drivers in low-redshift galaxies.    We summarise our main findings below.

\begin{itemize}
    \item Only a minor fraction ($\sim 10\%$) of MaNGA galaxies exhibits evidence of ionised gas outflows, with their incidence becoming more prevalent among systems of higher star formation activity, stellar mass and surface density, and AGN luminosity.  Within the outflow objects, the winds are centrally concentrated and detected over a spatial extent that scales with a sub-unity power-law slope with galaxy size.
    \item We find strong correlations between the mass outflow rate and the strength of the mechanical drivers, namely SFR and $L_{\rm AGN}$, where $\dot{M}_{\rm out} \propto \rm SFR^{0.97}$ and $\dot{M}_{\rm out} \propto L_{\rm AGN}^{0.55}$.
    \item Given the strong correlations found between the mass outflow rate and numerous galaxy properties, which themselves correlate with one another, we narrow down the key ingredients influencing the strength of the outflow via the derivation of an `all encompassing outflow scaling relation', applicable to SF and AGN outflows alike.  Here, the mass outflow rate of the ionised gas is described with the least possible scatter (0.35 dex) by expressing its dependence jointly on stellar mass ($M_{\star}$), star formation and AGN activity and galaxy size. Although $\sim 90 \%$ of the 2744 MaNGA galaxies considered in our analysis do not show evidence of outflows, we use this relation to show that at least $\sim 32 \%$ of these objects are consistent with hosting weak galactic winds with lower mass outflow rates and broad component intensities too weak to be detected from our method/data (or happen to be inclined such that the outflows are least perceptible in kinematic signatures used for outflow identification). 
    \item The physical conditions of the outflowing gas are distinctly different from the gas within the galactic disks. We show that the outflowing gas component exhibits elevated line ratios compared to the disk gas in both inactive and active galaxies.  We also find evidence for higher dust attenuation in the wind gas, possibly due to an enhanced metal enrichment of the ejected material compared to the average conditions of the ISM in the disk.  Furthermore, the local electron density of ionised gas entrailed within the outflow extends to higher values than the disk gas, albeit with large error on individual measurements of $n_{\rm e}$ determined on an object-by-object basis. In terms of geometry, the observed inclination dependencies are consistent with bi-conical outflows featuring wide wind opening angles.
    \item The energy and momentum of the ionised outflows are consistent with theoretical models of star-formation driven winds with low coupling efficiencies ($\lesssim 1\%$), except for a few objects which have high AGN contributions to the bolometric luminosity.  For these systems, additional energy coupling provided by the central AGN may be required to account for the observed outflow energetics. 
    \item $\sim 27 \%$ of outflows may have sufficient velocity to escape the halo within which they reside.  Half of these outflows are purely SF driven, and half are from systems with central AGN activity.  As expected, we anticipate gas preferentially escaping from systems with shallower potential wells.
    
\end{itemize}
Overall, our results highlight the strong ties between outflow and internal galaxy properties, and the importance of both star formation and AGN as physical drivers of galactic winds in galaxies with evidence of moderate nuclear activity.  Feedback in `typical' nearby galaxies comes mostly in the form of galactic fountains, with escaping winds being more common among systems with shallower potential wells.  The main caveat of our work is that it is limited to the ionised gas phase alone which makes up only a fraction of the total outflowing gas.

\section{Acknowledgements}
We thank the referee for their helpful and insightful comments which have motivated several improvements to this paper.

Funding for the Sloan Digital Sky Survey IV has been provided by the Alfred P. Sloan Foundation, the U.S. Department of Energy Office of Science, and the Participating Institutions. SDSS-IV acknowledges support and resources from the Center for High-Performance Computing at the University of Utah. The SDSS web site is www.sdss.org.

SDSS-IV is managed by the Astrophysical Research Consortium for the  Participating Institutions of the SDSS Collaboration including the 
Brazilian Participation Group, the Carnegie Institution for Science,  Carnegie Mellon University, the Chilean Participation Group, the French Participation Group, Harvard-Smithsonian Center for Astrophysics, Instituto de Astrof\'isica de Canarias, The Johns Hopkins University, Kavli Institute for the Physics and Mathematics of the Universe (IPMU) / University of Tokyo, the Korean Participation Group, Lawrence Berkeley National Laboratory, Leibniz Institut f\"ur Astrophysik Potsdam (AIP), Max-Planck-Institut f\"ur Astronomie (MPIA Heidelberg), Max-Planck-Institut f\"ur Astrophysik (MPA Garching), Max-Planck-Institut f\"ur Extraterrestrische Physik (MPE), National Astronomical Observatories of China, New Mexico State University, New York University, University of Notre Dame, Observat\'ario Nacional / MCTI, The Ohio State University, Pennsylvania State University, Shanghai Astronomical Observatory, United Kingdom Participation Group,
Universidad Nacional Aut\'onoma de M\'exico, University of Arizona,  University of Colorado Boulder, University of Oxford, University of Portsmouth, University of Utah, University of Virginia, University of Washington, University of Wisconsin, Vanderbilt University, and Yale University.

\section{DATA AVAILABILITY}
The data underlying this article were accessed from the publicly available Data Release 15 (DR15) of the Sloan Digital Sky Survey (SDSS-IV).  Specifically we make use of the integral field unit (IFU) spectroscopic observations of nearby galaxies from the SDSS Mapping Nearby Galaxies at APO (MaNGA) survey.  All of the MaNGA data can be accessed using the traditional SDSS Science Archive Server (SAS) at https://data.sdss.org/sas/dr15/manga/spectro/. The NASA-Sloan Atlas can be accessed at http://nsatlas.org/.
The processed datasets underlying this article, which were derived as outlined in Section \ref{method.sec}, will be shared on reasonable request to the corresponding author.

\bibliographystyle{mnras}
\bibliography{library}

\noindent [dataset]* Aguado et al. 2019, Index of /sas/dr15/manga/spectro/, SDSS-IV, https://data.sdss.org/sas/dr15/manga/spectro/

\appendix

\section{Correlation coefficients and linear regression for SF and AGN outflows separately}
\label{appendix_SFAGN.sec}

\begin{figure*}
\centering
\includegraphics[width=\textwidth]{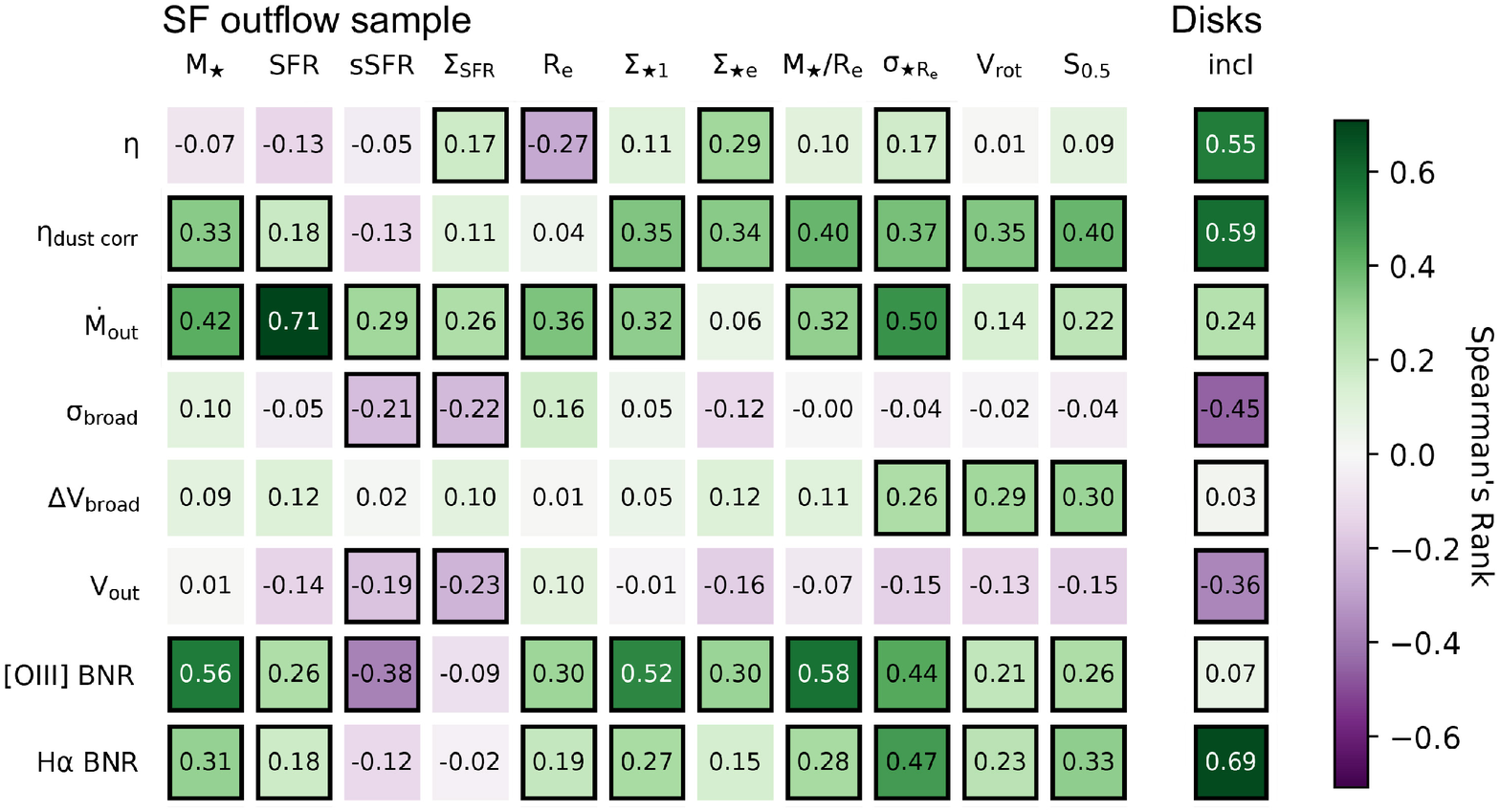}
\caption{Spearman's rank correlation coefficients quantifying the strength of correlations between outflow and host galaxy properties for the star-formation driven outflow sample.  Black outlined boxes indicate a statistically significant correlation, i.e., p-value $< 0.05$.  The dependence on inclination is only investigated for the subsample of morphological disks.  Strong correlations are identified, most notably between the observed outflow rate ($\dot{M}_{\rm out}$) and the intensity of its driver: the rate of star formation (SFR).}
\label{fig:correlation_grid_SF}
\end{figure*}

\begin{figure*}
\centering
\includegraphics[width=\textwidth]{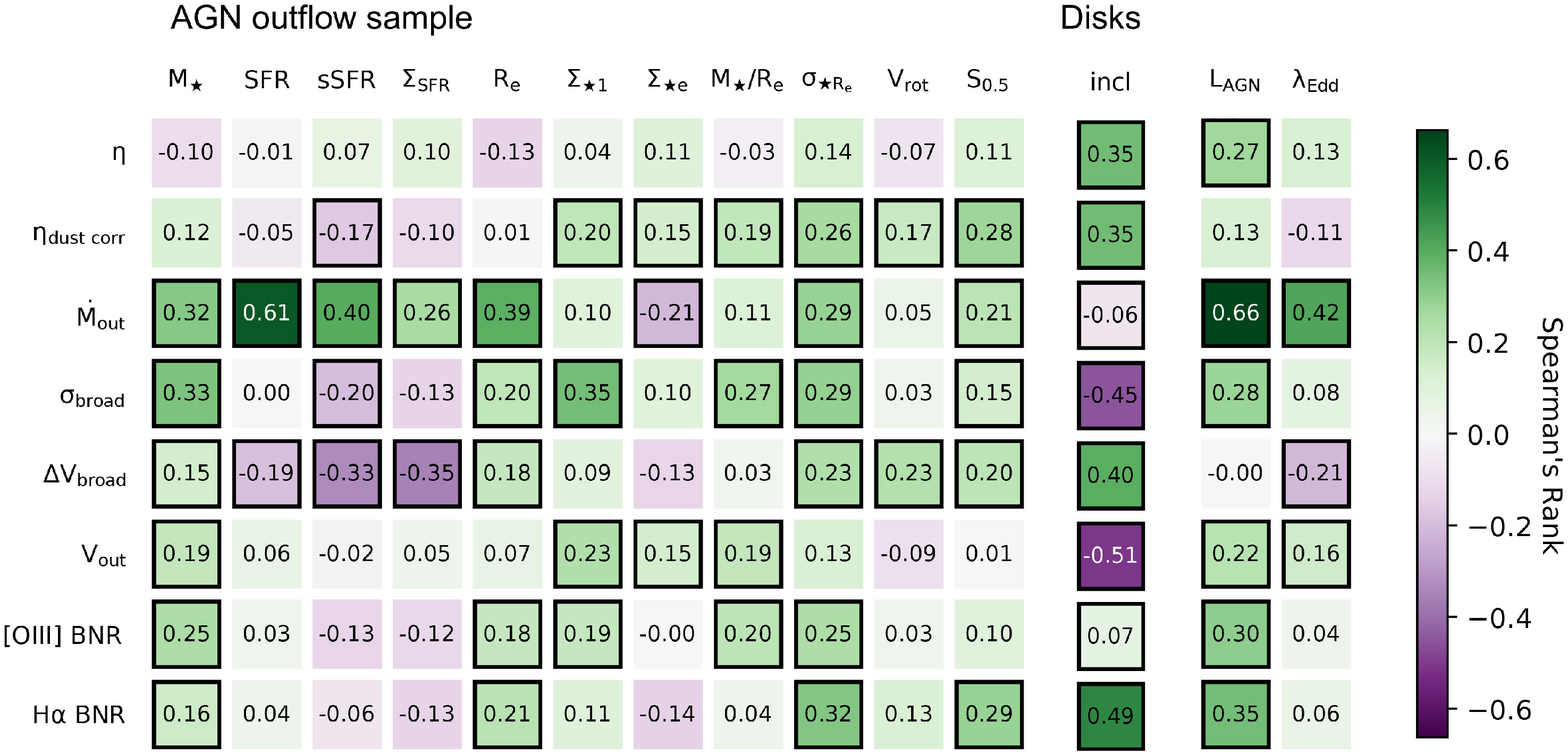}
\caption{Spearman's rank correlation coefficients quantifying the strength of correlations between outflow and host galaxy properties for galaxies with AGN activity.  Black outlined boxes indicate a statistically significant correlation, i.e., p-value $< 0.05$.  The dependence on inclination is only investigated for the subsample of morphological disks.  Strong correlations are identified, most notably between the observed outflow rate ($\dot{M}_{\rm out}$) and the intensity of its potential drivers: the star formation rate and AGN luminosity.}
\label{fig:correlation_grid_AGN}
\end{figure*}

In this Appendix, we provide versions of the grid illustrating correlation strength between outflow and internal galaxy properties for the SF and AGN outflow samples separately (Figures\ \ref{fig:correlation_grid_SF} and\ \ref{fig:correlation_grid_AGN}, respectively).  Tables\ \ref{tab:outflow_scaling_SF} and\ \ref{tab:outflow_scaling_AGN} present the corresponding results of linear regression applied to the outflow - host galaxy relations for SF and AGN samples separately.

% FIG Correlation grid

\begin{table*} %[] 
\centering 
\resizebox{\linewidth}{!}{%
\begin{tabular}{l|rrrrrrrrrrr} 
\hline \hline 
Galaxy property & $A\eta$ & $B\eta$ & $\Delta\eta$ & $A{\dot{M}_{\rm out}}$ & $B{\dot{M}_{\rm out}}$ & $\Delta{\dot{M}_{\rm out}}$ & $A{v_{\rm out}}$ & $B{v_{\rm out}}$ & $\Delta{v_{\rm out}}$ & $x_0$ \\
\hline 
\multicolumn{11}{c}{SF outflow sample} \\ 
\hline 
log($M_{\star} \ [\rm M_{\odot}]$)& $-0.03 \pm 0.07$ & $-1.21 \pm 0.05$ & $0.41$ & $0.36 \pm 0.08$ & $-0.69 \pm 0.06$ & $0.43$ & $0.01 \pm 0.02$ & $2.56 \pm 0.02$ & $0.12$ & $11.0$ \\
log(SFR [$\rm M_{\odot} \ yr^{-1}$])& $-0.03 \pm 0.10$ & $-1.18 \pm 0.06$ & $0.41$ & $0.84 \pm 0.08$ & $-1.29 \pm 0.05$ & $0.33$ & $-0.05 \pm 0.03$ & $2.58 \pm 0.02$ & $0.12$ & $0.0$ \\
log(sSFR \ [yr$^{-1}]$)& $0.05 \pm 0.12$ & $-1.20 \pm 0.04$ & $0.41$ & $0.50 \pm 0.13$ & $-0.93 \pm 0.04$ & $0.45$ & $-0.08 \pm 0.03$ & $2.57 \pm 0.01$ & $0.12$ & $-10.0$ \\
log($\Sigma_{\rm SFR} \ [\rm M_{\odot} \ yr^{-1} \ kpc^{-2}]$)& $0.17 \pm 0.08$ & $-1.10 \pm 0.05$ & $0.40$ & $0.25 \pm 0.09$ & $-0.76 \pm 0.06$ & $0.45$ & $-0.07 \pm 0.02$ & $2.53 \pm 0.02$ & $0.12$ & $-1.0$ \\
log($R_{\rm e}$ [kpc])& $-0.37 \pm 0.13$ & $-1.34 \pm 0.06$ & $0.40$ & $0.69 \pm 0.15$ & $-0.63 \pm 0.07$ & $0.44$ & $0.06 \pm 0.04$ & $2.58 \pm 0.02$ & $0.12$ & $1.0$ \\
log($\Sigma_{\star 1} \ [\rm M_{\odot} \ kpc^{-2}]$)& $0.16 \pm 0.09$ & $-1.20 \pm 0.04$ & $0.41$ & $0.34 \pm 0.10$ & $-0.91 \pm 0.04$ & $0.45$ & $-0.01 \pm 0.03$ & $2.56 \pm 0.01$ & $0.12$ & $9.0$ \\
log($\Sigma_{\rm \star e} \ [\rm M_{\odot} \ kpc^{-2}]$)& $0.29 \pm 0.09$ & $-1.03 \pm 0.06$ & $0.40$ & $-0.00 \pm 0.11$ & $-0.89 \pm 0.07$ & $0.47$ & $-0.04 \pm 0.03$ & $2.54 \pm 0.02$ & $0.12$ & $9.0$ \\
log($M_{\star} / R_{\rm e} \ [\rm M_{\odot} \ kpc^{-1}])$& $0.17 \pm 0.11$ & $-1.17 \pm 0.04$ & $0.41$ & $0.38 \pm 0.12$ & $-0.84 \pm 0.04$ & $0.46$ & $-0.02 \pm 0.03$ & $2.56 \pm 0.01$ & $0.12$ & $10.0$ \\
log($\sigma_{\rm \star R_e} \ [\rm km \ s^{-1}]$)& $1.02 \pm 0.38$ & $-1.23 \pm 0.04$ & $0.42$ & $2.01 \pm 0.33$ & $-0.94 \pm 0.04$ & $0.42$ & $-0.02 \pm 0.12$ & $2.56 \pm 0.01$ & $0.12$ & $2.0$ \\
log($V_{\rm rot} \  [\rm km \ s^{-1}])$& $-0.03 \pm 0.07$ & $-1.19 \pm 0.04$ & $0.41$ & $0.09 \pm 0.08$ & $-0.92 \pm 0.05$ & $0.46$ & $-0.02 \pm 0.02$ & $2.57 \pm 0.01$ & $0.12$ & $2.0$ \\
log($S_{0.5} \ [\rm km \ s^{-1}]$)& $-0.01 \pm 0.10$ & $-1.19 \pm 0.05$ & $0.41$ & $0.14 \pm 0.11$ & $-0.94 \pm 0.05$ & $0.46$ & $-0.02 \pm 0.03$ & $2.57 \pm 0.01$ & $0.12$ & $2.0$ \\
\hline
\multicolumn{11}{c}{SF disks} \\
\hline
${\rm cos(i \ [\deg])}$ & $-1.20 \pm 0.35$ & $-1.72 \pm 0.15$ & $0.34$ & $-0.59 \pm 0.35$ & $-1.22 \pm 0.16$ & $0.37$ & $0.24 \pm 0.13$ & $2.65 \pm 0.06$ & $0.13$ & $2.0$ \\
\hline \hline  
\end{tabular}  
}  
\caption{Scaling relations between star-formation driven ionised gas outflow properties and internal galaxy properties. This table has the same layout as Table \ref{tab:outflow_scaling}, except here we exclude objects where we find evidence of AGN activity.}
\label{tab:outflow_scaling_SF}
\end{table*}

% Table: linear regressions in active galaxies

\begin{table*} %[] 
\centering 
\resizebox{\linewidth}{!}{%
\begin{tabular}{l|rrrrrrrrrrr} 
\hline \hline 
Galaxy property & $A\eta$ & $B\eta$ & $\Delta\eta$ & $A{\dot{M}_{\rm out}}$ & $B{\dot{M}_{\rm out}}$ & $\Delta{\dot{M}_{\rm out}}$ & $A{v_{\rm out}}$ & $B{v_{\rm out}}$ & $\Delta{v_{\rm out}}$ & $x_0$ \\
\hline 
\multicolumn{11}{c}{AGN outflow sample} \\ 
\hline 
log($M_{\star} \ [\rm M_{\odot}]$)& $-0.15 \pm 0.09$ & $-0.96 \pm 0.04$ & $0.48$ & $0.47 \pm 0.11$ & $-0.77 \pm 0.05$ & $0.56$ & $0.06 \pm 0.03$ & $2.64 \pm 0.01$ & $0.14$ & $11.0$ \\
log(SFR [$\rm M_{\odot} \ yr^{-1}$])& $-0.05 \pm 0.16$ & $-0.91 \pm 0.06$ & $0.49$ & $1.21 \pm 0.14$ & $-1.26 \pm 0.07$ & $0.60$ & $0.09 \pm 0.05$ & $2.60 \pm 0.02$ & $0.15$ & $0.0$ \\
log(sSFR \ [yr$^{-1}]$)& $0.21 \pm 0.21$ & $-1.05 \pm 0.14$ & $0.50$ & $0.94 \pm 0.20$ & $-1.46 \pm 0.14$ & $0.66$ & $-0.00 \pm 0.06$ & $2.63 \pm 0.04$ & $0.15$ & $-11.0$ \\
log($\Sigma_{\rm SFR} \ [\rm M_{\odot} \ yr^{-1} \ kpc^{-2}]$)& $0.15 \pm 0.13$ & $-0.94 \pm 0.04$ & $0.49$ & $0.37 \pm 0.16$ & $-0.93 \pm 0.05$ & $0.59$ & $0.04 \pm 0.04$ & $2.62 \pm 0.01$ & $0.15$ & $-2.0$ \\
log($R_{\rm e}$ [kpc])& $-0.26 \pm 0.14$ & $-1.00 \pm 0.05$ & $0.48$ & $0.96 \pm 0.17$ & $-0.62 \pm 0.06$ & $0.54$ & $0.03 \pm 0.04$ & $2.63 \pm 0.02$ & $0.15$ & $1.0$ \\
log($\Sigma_{\star 1} \ [\rm M_{\odot} \ kpc^{-2}]$)& $0.09 \pm 0.11$ & $-0.95 \pm 0.04$ & $0.49$ & $0.20 \pm 0.13$ & $-0.94 \pm 0.05$ & $0.58$ & $0.12 \pm 0.03$ & $2.60 \pm 0.01$ & $0.14$ & $9.0$ \\
log($\Sigma_{\rm \star e} \ [\rm M_{\odot} \ kpc^{-2}]$)& $0.12 \pm 0.13$ & $-0.87 \pm 0.08$ & $0.49$ & $-0.50 \pm 0.15$ & $-1.14 \pm 0.09$ & $0.57$ & $0.08 \pm 0.04$ & $2.67 \pm 0.02$ & $0.14$ & $9.0$ \\
log($M_{\star} / R_{\rm e} \ [\rm M_{\odot} \ kpc^{-1}])$& $-0.05 \pm 0.17$ & $-0.92 \pm 0.04$ & $0.49$ & $0.28 \pm 0.20$ & $-0.91 \pm 0.04$ & $0.58$ & $0.17 \pm 0.05$ & $2.62 \pm 0.01$ & $0.14$ & $10.0$ \\
log($\sigma_{\rm \star R_e} \ [\rm km \ s^{-1}]$)& $1.44 \pm 0.42$ & $-1.05 \pm 0.05$ & $0.50$ & $2.66 \pm 0.48$ & $-1.12 \pm 0.06$ & $0.59$ & $0.40 \pm 0.13$ & $2.59 \pm 0.02$ & $0.15$ & $2.0$ \\
log($V_{\rm rot} \  [\rm km \ s^{-1}])$& $-0.10 \pm 0.09$ & $-0.88 \pm 0.05$ & $0.49$ & $-0.01 \pm 0.10$ & $-0.89 \pm 0.06$ & $0.59$ & $-0.00 \pm 0.03$ & $2.63 \pm 0.02$ & $0.15$ & $2.0$ \\
log($S_{0.5} \ [\rm km \ s^{-1}]$)& $0.00 \pm 0.11$ & $-0.93 \pm 0.06$ & $0.49$ & $0.12 \pm 0.13$ & $-0.94 \pm 0.07$ & $0.58$ & $0.03 \pm 0.03$ & $2.62 \pm 0.02$ & $0.15$ & $2.0$ \\
\hline
\multicolumn{11}{c}{AGN disks} \\
\hline
${\rm cos(i \ [\deg])}$ & $-0.63 \pm 0.31$ & $-0.66 \pm 0.16$ & $0.38$ & $0.42 \pm 0.38$ & $-1.04 \pm 0.20$ & $0.50$ & $0.37 \pm 0.12$ & $2.43 \pm 0.06$ & $0.15$ & $2.0$ \\
\hline \hline  
\end{tabular}  
}  
\caption{Scaling relations between ionised gas outflow properties and internal galaxy properties for galaxies with AGN activity only. This table has the same layout as table \ref{tab:outflow_scaling}.}
\label{tab:outflow_scaling_AGN}
\end{table*}

\section{Comparison to literature samples}
\label{appendix_literature.sec}
\begin{figure*}
\includegraphics[width=0.49\linewidth]{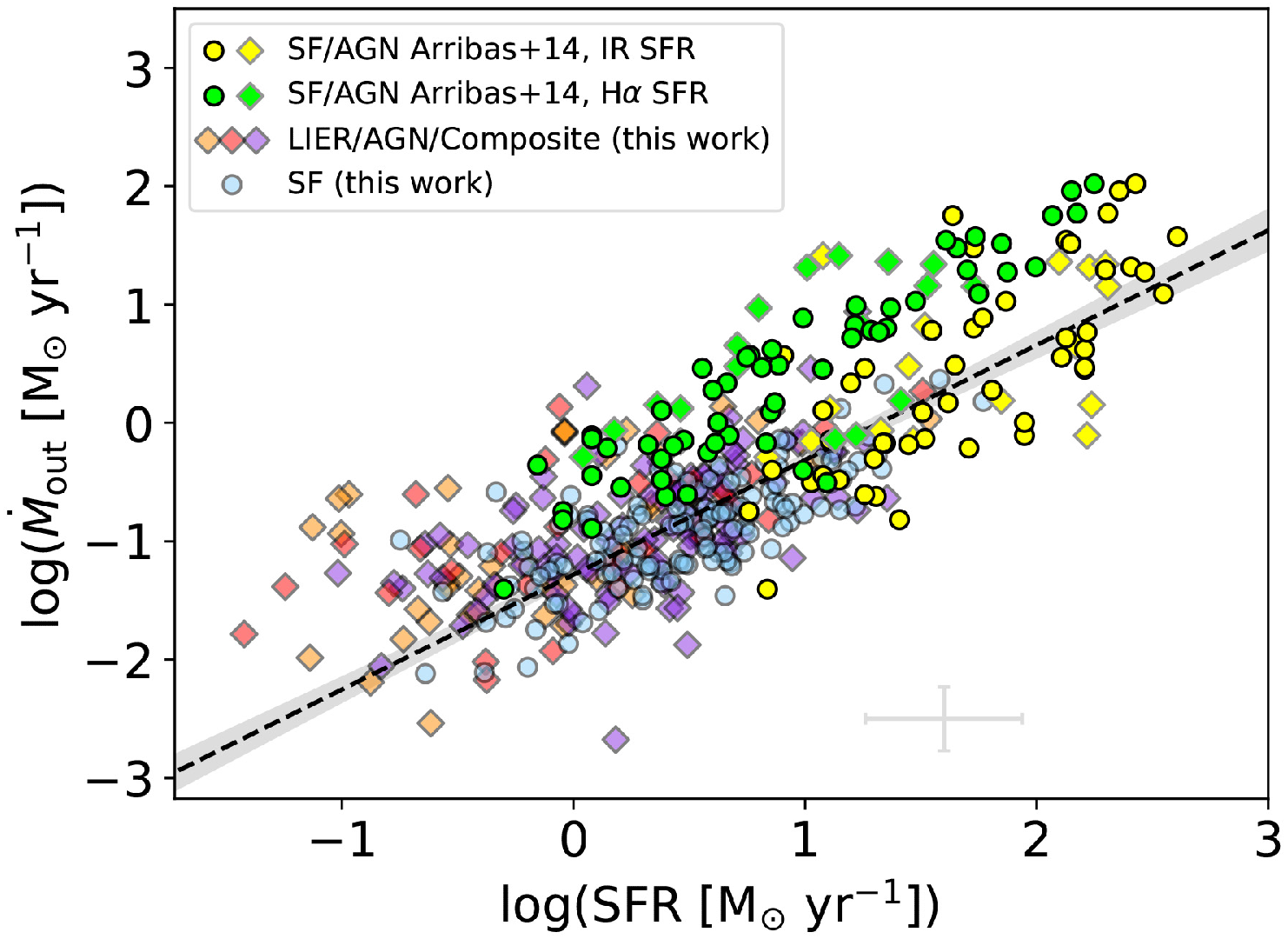}
\includegraphics[width=0.49\linewidth]{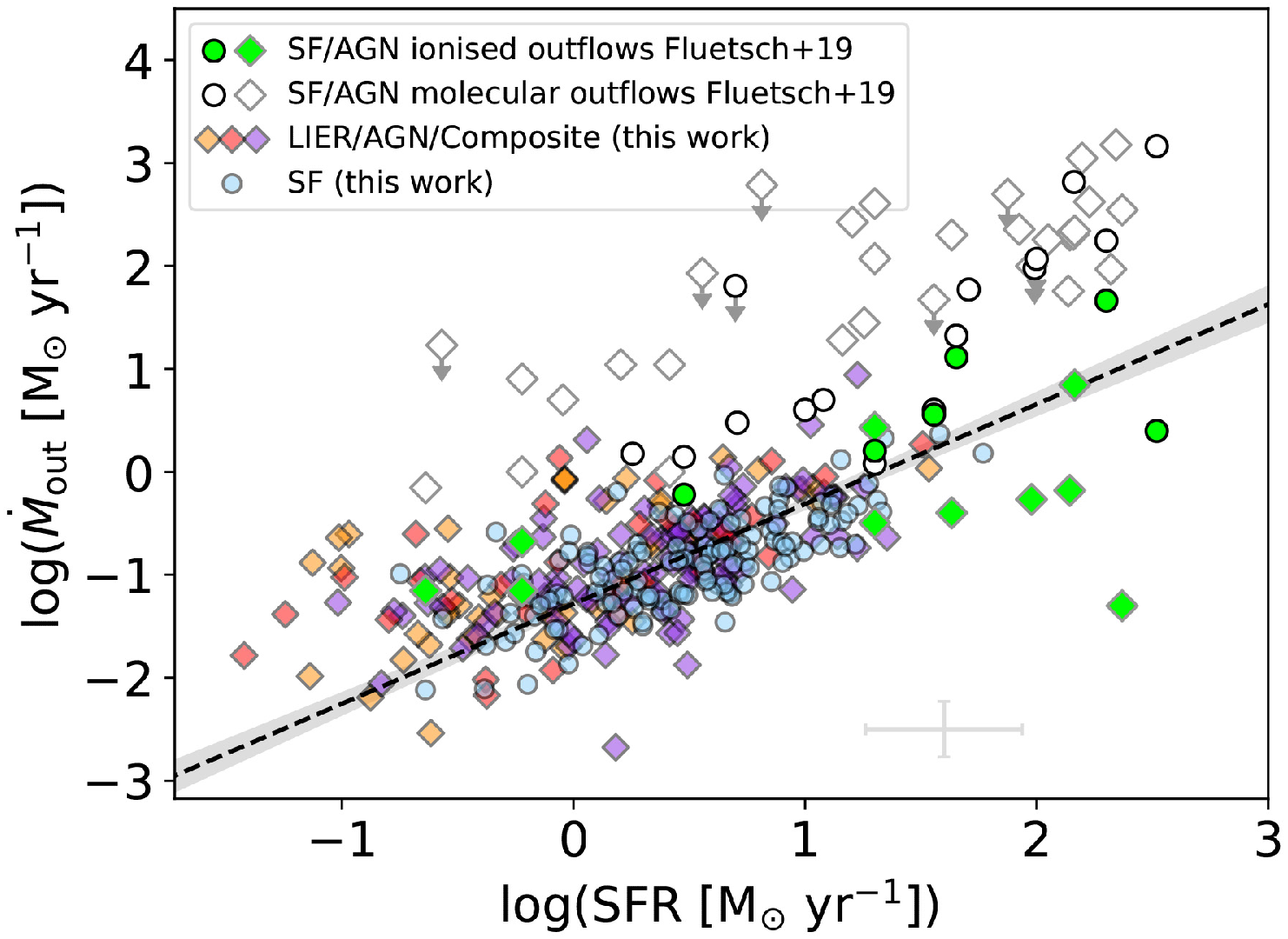}
\caption{{\it Left:} SFR - $\dot{M}_{\rm out}$ relation from \protect\citet{Arribas2014} overplotted on the MaNGA relation presented in Figure \protect\ref{fig:SFR}.  The grey polygon and dashed black line reproduce the linear regression to the outflow sample analysed in this work.  {\it Right:} \protect\cite{Fluetsch2019} results of molecular and ionised gas outflows overplotted on Figure \protect\ref{fig:SFR}.}
\label{fig:SFR_literature}
\end{figure*}

\begin{figure*}
\includegraphics[width=0.7\linewidth]{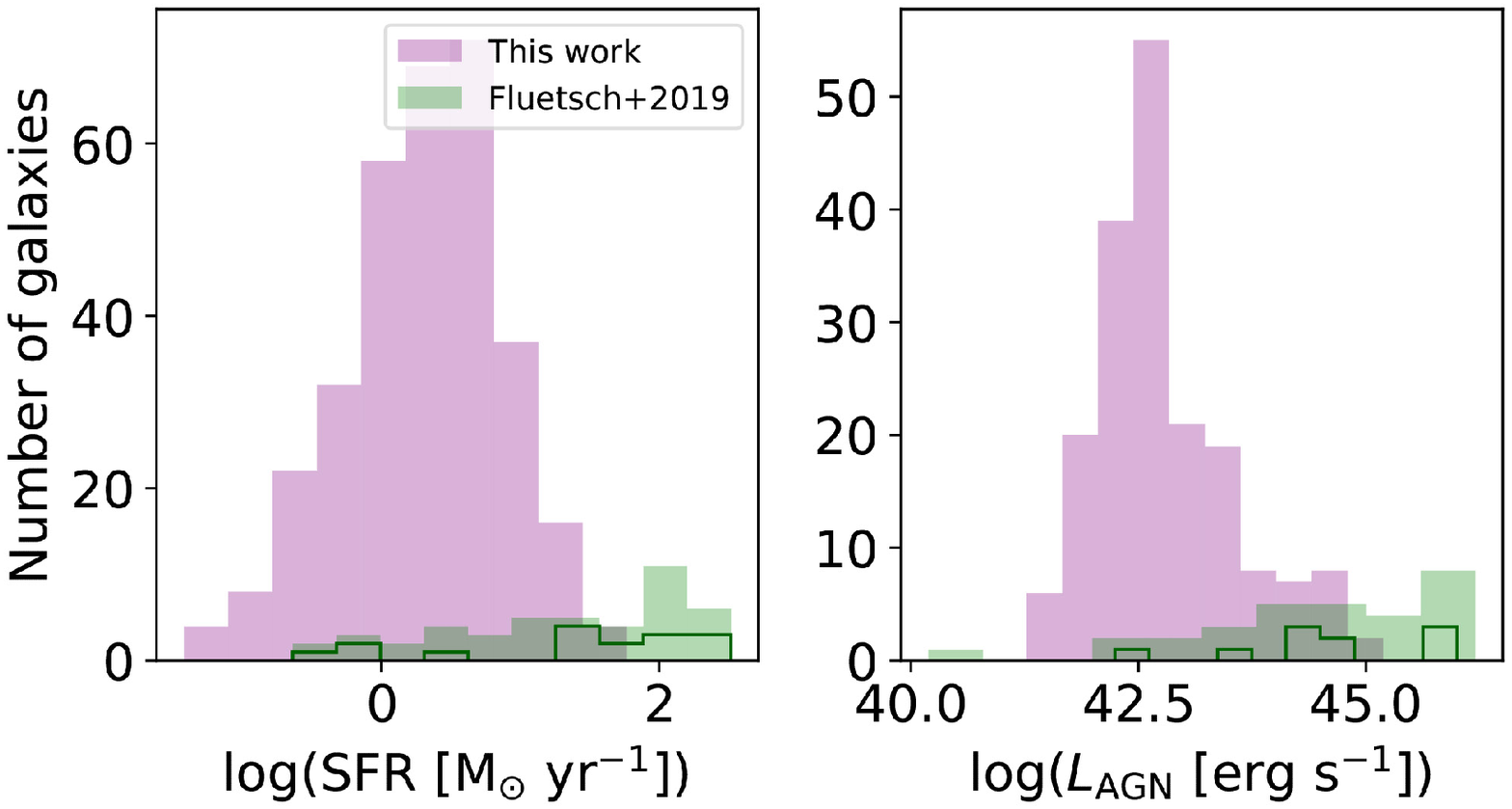}
\caption{SFR and $L_{\rm AGN}$ distributions of the full outflow sample ({\it left}) and AGN outflow sample ({\it right}) presented in this work ({\it purple}), contrasted to that probed in the molecular phase ({\it filled green}) and ionised gas phase ({\it open green}) by \protect\cite{Fluetsch2019}.}
\label{fig:hist_F19}
\end{figure*}

In Figure\ \ref{fig:SFR_literature} we compare the $\dot{M}_{\rm out} - {\rm SFR}$ relation derived in our work with the (U)LIRG sample analysed by \cite{Arribas2014}, and the molecular gas outflows (plus subset with ionised gas measurements) of the on average more luminous galaxies studied by \cite{Fluetsch2019}.  The distribution of star formation rates and AGN luminosities for the latter are contrasted to our MaNGA outflow sample and the AGN subset, respectively, in Figure\ \ref{fig:hist_F19}. A better agreement with \citet{Arribas2014} is found when adopting the IR-based SFR values for their (U)LIRGs, with the two samples covering a complementary dynamic range.  Alternatively, adopting their H$\alpha$-based SFRs, their ${\rm SFR} - \dot{M}_{\rm out}$ relation may be offset to higher outflow strengths because their targets are more compact, thus featuring higher star formation surface densities at a given SFR.  The ionised gas outflow rates presented by \citet{Fluetsch2019} lie roughly along the best-fit linear relation obtained on the basis of our MaNGA outflow sample, with most of them extending the dynamic range into the higher star formation activity regime, whereas the scaling relation for molecular gas outflows follows a similar slope but is offset to higher amplitudes.

\end{document}